\documentclass{jfm}
\usepackage{graphicx}
\usepackage{mathrsfs}
\usepackage{amsmath}
\usepackage{placeins}
\usepackage{epstopdf, epsfig}
\usepackage{subcaption}
\usepackage[export]{adjustbox}
\usepackage[%
    colorlinks=true,
    pdfborder={0 0 0},
    linkcolor=blue,
 citecolor=blue,
urlcolor=Blue
]{hyperref}

\shorttitle{The influence of Reynolds and Froude number}
\shortauthor{M. Momenifar, R. Dhariwal and A.D. Bragg}

\title{The influence of Reynolds and Froude number on the motion of settling, bidisperse inertial particles in turbulence}

\author{Mohammadreza Momenifar,\aff{1}
  ,
  Rohit Dhariwal\aff{1}\\
 \and Andrew D. Bragg\aff{1}\corresp{\email{andrew.bragg@duke.edu}}}

\affiliation{\aff{1}Department of Civil and Environmental Engineering, Duke University,
Durham, NC 27708, USA
}

\begin{document}

\maketitle

\begin{abstract}
	Using Direct Numerical Simulations (DNS), we examine the effects of Taylor Reynolds number, $R_\lambda$, and Froude number, $Fr$, on the motion of settling, bidisperse inertial particles in isotropic turbulence. Particle accelerations play a key role in the relative motion of bidisperse particles, and we find that reducing $Fr$ leads to an enhancement of the accelerations, but a suppression of their intermittency. For Stokes numbers $St>1$, the effect of $R_\lambda$ on the accelerations is enhanced by gravity, since settling causes the particle accelerations to be affected by a larger range of flow scales. The results for the Probability Density Function (PDF) of the particle relative velocities show that for bidisperse particles, decreasing $Fr$ leads to an enhancement of their relative velocities in both the vertical (parallel to gravity) and horizontal directions. Importantly, our results show that even when the particles are settling very fast, turbulence continues to play a key role in their vertical relative velocities, and increasingly so as $R_\lambda$ is increased. This occurs because although the settling velocity may be much larger than typical velocities of the turbulence, due to intermittency, there are significant regions of the flow where the turbulence contribution to the particle motion is of the same order as that from gravitational settling. Increasing $R_\lambda$ enhances the non-Gaussianity of the relative velocity PDFs, while reducing $Fr$ has the opposite effect, and for fast settling particles, the PDFs become approximately Gaussian. Finally, we observe that low-order statistics such as the Radial Distribution Function (RDF) and the particle collision kernel, are strongly affected by $Fr$ and $St$, and especially by the degree of bidispersity of the particles. Indeed, even when the difference in the value of $St$ of the two particles is $\ll1$, the results can differ strongly from the monodisperse case, especially when $Fr\ll1$. However, we also find that these low-order statistics are very weakly affected by $R_\lambda$ when $St\leq O(1)$, irrespective of the degree of bidispersity. Therefore,  although the mechanisms controlling the collision rates of monodisperse and bidisperse particles are different, they share the property of a weak sensitivity to $R_\lambda$ when $St\leq O(1)$.
\end{abstract}


\section{Introduction}
Multiphase turbulent flows have attracted substantial interest for many years both because of the intellectually stimulating challenges associated with understanding them, and also because of their practical importance to a wide range of applications. Examples include the interaction between multiple flowing fluids such as is found in oil-refrigerant mixtures for refrigeration systems (\cite{borghi2013turbulent,kolev2005multiphase,momenifar2015effect}), and the transport of bubbles in liquids such as occur in airlift pumps (\cite{brennen2005fundamentals,prosperetti2009computational,hanafizadeh2014void}). The class of multiphase flows involving the motion of small suspended particles in turbulent flows is of particular interest to the present work, being important for environmental sciences (atmospheric pollution transport, sea spray, and cloud formation), astrophysics (protoplanetary disks, and the atmospheres of planets and dwarf stars) and industrial processes (turbulent combustion, spray nozzles, fluidized bed reactors). In many of these applications, the particle motion is not only affected by the fluid turbulence, but also by particle inertia and gravitational settling. 

The motion of particles with inertia can differ profoundly from that of inertialess fluid particles in turbulent flows \citep{toschi2009lagrangian}. For example, even in incompressible flows, inertial particles can spatially cluster across a range of length scales \citep{maxey87,bec07,balachandar10,bragg14b,ireland2016effecta,gustavsson16}, and their trajectories in configuration space can intersect \citep{wilkinson05,gustavsson12,bragg14c}. Gravity not only leads to finite settling velocities for inertial particles, but it also modifies the way the particles interact with the turbulent flow \citep{maxey87,wang93}. The latter is in fact quite subtle, and recent results have shown that it can lead to non-trivial effects on the multiscale motion of inertial particles in turbulence \citep{bec14b,gustavsson14,parishani2015effects,ireland2016effectb,dhariwal2018small}.


When considering problems such as particle mixing and collision rates, it is the relative motion of particles in turbulence that is of importance. This relative motion is often studied by considering the relative motion of two particles (``particle-pairs'') in the turbulent flow \citep{salazar09,toschi2009lagrangian}. The motion of inertial particle-pairs can differ substantially depending upon whether the two particles have the same (monodisperse) or different (bidisperse) inertia. Several studies show that in the absence of gravity, bidispersity enhances the relative velocities and suppresses the spatial clustering of inertial particles compared with the monodisperse case \citep{chun05,pan10,pan14,dhariwal2018small}.

In a recent study, we considered the effect of gravity (characterized by the Froude number, $Fr\equiv a_\eta/g$, where $a_\eta$ is the Kolmogorov acceleration, and $g$ is the acceleration due to gravity) on the relative motion of bidisperse particles \citep{dhariwal2018small}, and found that the combined effects of gravity and turbulence lead to some interesting effects which were absent in the monodisperse case. Using Direct Numerical Simulations (DNS), statistics of the particle relative velocities in the directions parallel and perpendicular to the direction of gravity were computed. It was observed that at the small scales of the turbulent flow, decreasing $Fr$ leads to an enhancement of the particle relative velocities, not only in the direction of gravity, but even in the plane normal to its action. This is quite unlike the monodisperse case where it has been shown that decreasing $Fr$ leads to a uniform suppression of the inertial particle relative velocities in all directions \citep{bec14b,ireland2016effectb}. The results in \cite{dhariwal2018small} also showed that unlike the monodisperse case \citep{bec14b,ireland2016effectb}, the clustering of bidisperse inertial particles is always suppressed as $Fr$ is decreased (except when the bidisipersity is very weak, i.e. the monodisperse limit). The theoretical analysis in \cite{dhariwal2018small} explained these differences between bidisperse and monodisperse particles as being due to the fact that at the small scales, the relative motion of bidisperse particles is dominated by a term in their equation of relative motion that depends upon the acceleration of the particles, and the accelerations are enhanced in the presence of gravity \citep{parishani2015effects,ireland2016effectb}. For monodisperse particles, this acceleration contribution vanishes, and their relative motion is dominated by the particle interaction with the fluid relative velocity field.
 
Since most of these studies are based on DNS at low to moderate Reynolds numbers, it is important to understand how representative the results are of the real problems of interest, since in nature the flows have much larger Reynolds numbers \citep{toschi2009lagrangian}. Reynolds number affects turbulence in two distinct, but related ways, namely through the classical effect of enhanced scale separations with increasing Reynolds number \citep{pope}, and enhanced internal intermittency at the small scales of the flow \citep{frisch}. Given the current limitations of the Reynolds numbers accessible with DNS, one way to explore the effect of Reynolds number on particle motion in turbulence would be to use theoretical models. However, current (fully predictive) theoretical models of inertial particle motion at the small scales of turbulence are only accurate for weak particle inertia, can fail dramatically for moderate to strong particle inertia, and do not account for the effects of internal intermittency in the turbulence \citep[e.g. see][]{bragg14b,bragg14c}. An alternative method is to use DNS over a range of Reynolds numbers to look for trends in the behavior. This can provide insight regarding the extent to which results obtained at low/moderate Reynolds numbers might be extrapolated the real problems of interest where the Reynolds numbers are much larger.

In  \cite{ireland2016effecta} \&  \cite{ireland2016effectb}, the effect of Reynolds number on the motion of monodisperse inertial particles with and without gravity was explored using DNS over the range $88\leq R_\lambda\leq 598$, where $R_\lambda$ is the Taylor Reynolds number \citep{pope}. As might be expected, the higher-order statistics of the particle relative velocities (e.g. kurtosis) showed a strong dependence on $R_\lambda$. However, they also showed that collision rates (which depend on low-order statistics) of monodisperse particles with Stokes numbers $St\lesssim 1$ are essentially independent of $R_\lambda$. This result implies that the essential physics governing particle collisions in atmospheric clouds where $R_\lambda=O(10^4)$ and typically $St<1$ \citep{shaw2003particle} might in fact be captured by DNS studies with $R_\lambda=O(10^2) $. For $St>1$, the results in \cite{ireland2016effecta} \&  \cite{ireland2016effectb} show that the collision rates are sensitive to $R_\lambda$, and it was argued this is most likely due to the fact that particles with sufficient inertia are affected by the increasing range of scales as $R_\lambda$ is increased, owing to the fact that they posses sufficient memory to be affected by their past interaction with scales outside the dissipation range, even when their current separation lies in the dissipation range (though it is possible that enhanced intermittency also plays a role).


 To the best of our knowledge, no previous study has explored the effect of $R_\lambda$ on the motion of bidisperse particles in turbulence. Previous DNS studies of settling bidisperse particles considered only single (and low) Reynolds numbers ($R_\lambda=84.9$ in \cite{woittiez2009combined}, $R_\lambda=143$ in \cite{parishani2015effects}, and $R_\lambda=90 $ in \cite{dhariwal2018small}). As explained earlier, the relative motion of bidisperse particles differs substantially from that of monodisperse particles, being dominated by different effects and mechanisms. It is therefore possible that the $R_\lambda$ dependence of bidisperse particles could differ from that of monodisperse particles. Another key difference is that in the monodisperse case, gravity only affects the particle motion implicitly through the way it modifies the particle interactions with the turbulence, since the relative motion induced by the gravitational settling is zero (both particles have the same settling speeds). By contrast, in the bidisperse case, gravity has an explicit effect, and in the direction of gravity one might expect gravity to dominate the particle motion when $Fr\ll1$ (if the difference in the Stokes numbers of the two particles is $\geq O(1)$. But as discussed in \cite{dhariwal2018small}, this is not guaranteed since fluid accelerations are highly intermittent in turbulence \citep{laporta01}, such that even if the average Froude number is $\ll1$, there may be significant regions of the flow where the instantaneous Froude number, ${Fr}'\equiv \|\boldsymbol{a}\|/g$ (where $\boldsymbol{a}$ is the instantaneous fluid acceleration), is $\geq O(1)$. Since intermittency increases with increasing $R_\lambda$, then at large $R_\lambda$, turbulence may continue to play a key role in the bidisperse particle relative motion in the direction of gravity even when $Fr\ll1$.
 
 In order to address these issues, the present paper considers the effect of both $R_\lambda$ and $Fr$ on the motion of bidisperse particles in turbulence. Our previous study considered $Fr=\infty,0.3,0.052$ and $R_\lambda=90$ \citep{dhariwal2018small}. The present study significantly extends the parameter space by considering $Fr=\infty,0.3,0.052$ and $R_\lambda =90, 224, 398$.
\section{Theoretical Considerations}\label{TC}
We are concerned with dilute suspensions of particles in turbulent flows, with the particle mass loading sufficiently small so the back-reaction of the particles on the underlying flow can be ignored, corresponding to the one-way coupled regime \citep{balachandar10}. Such a regime is applicable for the motion of droplets in atmospheric clouds \citep{shaw2003particle}, where particles may also be assumed to be small (i.e $d/\eta \ll 1$ where $d$ is the particle diameter and $\eta$ is the Kolmogorov length scale) and dense ($\rho_p/\rho_f \gg 1$, where $\rho_p$ and $\rho_f$ represent the particle density and fluid density, respectively). These assumptions justify the use of a point-particle approach where the inertial particle motion is governed by a simplified version of the equation of \cite{maxey1983equation}
\begin{equation}
\ddot{\boldsymbol{x}}^p(t)\equiv\dot{\boldsymbol{v}}^p(t)=\frac{\boldsymbol{u}(\boldsymbol{x}^p(t),t)-\boldsymbol{v}^p(t)}{\tau_p}+\boldsymbol{g},
\label{MR_inertial}
\end{equation}
where $\boldsymbol{u}(\boldsymbol{x}^p(t),t)$ denotes fluid velocity at the particle position $\boldsymbol{x}^p(t)$, $\boldsymbol{v}^p(t)$ is the particle velocity, $\tau_p\equiv\rho_pd^2/18\rho_f\nu $ is the particle response time, $\nu$ is the fluid kinematic viscosity, and $\boldsymbol{g}$ is the gravitational acceleration vector. We also consider fluid particles whose motion obeys $\dot{\boldsymbol{x}}^p(t)\equiv\boldsymbol{u}(\boldsymbol{x}^p(t),t)$. 

Particle inertia can be characterized by the Stokes number, $St\equiv\tau_p/\tau_\eta$, where $\tau_\eta$ is the Kolmogorov time scale. Equation (\ref{MR_inertial}) assumes a linear drag force on the particles, which is thought to be valid for $St\le O(1)$ \citep{good14,ireland2016effectb}, and this is the range we restrict attention to. The effect of gravity on the particle motion may be characterized through the settling parameter, $Sv$, defined as the ratio of particle's settling velocity in a quiescent flow, $\tau_p g$ (where $g\equiv \|\boldsymbol{g}\|$), to the Kolmogorov velocity scale, $u_\eta\equiv\eta/\tau_\eta$. 
The effect of gravity on the flow can be characterized by the Froude number, $ Fr\equiv a_\eta/{g}=\epsilon^{3/4}/(\nu^{1/4}{g})$, where $a_\eta\equiv u_\eta/\tau_\eta$ is the Kolmogorov acceleration, and $\epsilon$ is the mean turbulent kinetic energy dissipation rate. Note that from these definitions we also have $Fr\equiv St/Sv$.

To consider the relative motion between two particles, we consider the motion of a ``satellite'' particle relative to a ``primary'' particle. When each particle is governed by (\ref{MR_inertial}), the equation describing their relative motion (in non-dimensional form) is \citep{dhariwal2018small} 
\begin{equation}\label{eq:ND_RME}
\widetilde{\boldsymbol{\ddot{r}}^p}(\tilde{t})=
\widetilde{\boldsymbol{\dot{w}}^p}(\tilde{t})=
\frac{\Delta \tilde{\boldsymbol{u}} ( \widetilde{\boldsymbol{x}^p}(\tilde{t}),\widetilde{\boldsymbol{r}^p}(\tilde{t}),\tilde{t})-\widetilde{{\boldsymbol{w}^p}}(\tilde{t})  }{St_2}
+
\frac{\Delta St ( \widetilde{{\boldsymbol{a}}^p}(\tilde{t})- \boldsymbol{e}_g Fr^{-1})}{St_2}
\end{equation}
where $\widetilde{\cdot}$ denotes a quantity non-dimensionalized using the Kolmogorov scales, $\Delta {\boldsymbol{u}}$ is the difference in the fluid velocity at the two particle positions, ${\boldsymbol{x}^p}({t})$ is the position of the primary particle, ${\boldsymbol{x}^p}({t})+{\boldsymbol{r}^p}({t})$ is the position of the satellite particle, ${\boldsymbol{w}^p}({t})$ is their relative velocity, $St_1$ and  $St_2$ are the Stokes numbers of the primary and satellite particles, respectively, $\Delta St\equiv St_1-St_2$, ${{\boldsymbol{a}}^p}({t})$ is the primary particle acceleration, and $\boldsymbol{e}_g\equiv \boldsymbol{g}/g$ is the unit vector in the direction of gravity. 

The formal solution of (\ref{eq:ND_RME}) is (we drop the $\widetilde{\cdot}$ for notational ease, and assume $t\gg St_2$)

\begin{equation} \label{eq:w_p_fs}
\boldsymbol{w}^p(t)=\frac{1}{St_2} \int_0^{t} \mathrm e^{-(t-s)/St_2} \Delta \boldsymbol{u}^p(s) \mathrm{d} s
-
\frac{\Delta St}{Fr}\boldsymbol{e}_g
+
\frac{\Delta St}{St_2} \int_0^{t} \mathrm e^{-(t-s)/St_2} \boldsymbol{a}^p(s) \mathrm{d} s,
\end{equation}
and the particle acceleration $\boldsymbol{a}^p$ may be expressed as \citep{dhariwal2018small} 
\begin{equation}\label{eq:a_p}
\boldsymbol{a}^p(t)=\frac{1}{St_1^2} \int_0^{t} \mathrm e^{-(t-t^\prime)/St_1}(\boldsymbol{u}^p (t)-\boldsymbol{u}^p(t^\prime))\mathrm{d} t^\prime.
\end{equation}
In the following, we discuss the implications of the above equations and summarize the key findings of \cite{dhariwal2018small} regarding the effects of bidispersity and gravity on the relative motion of inertial particles in turbulence.   

In (\ref{eq:w_p_fs}), only the first term survives for monodisperse particles, and the effect of gravity appears implicitly, through the way it affects the particle interaction with $\Delta\boldsymbol{u}$. The first integral of (\ref{eq:w_p_fs}) reveals that the particle-pairs are influenced by their past interaction with the turbulent flow over the time-span $t-s\le O(St_2)$ along their trajectory history. The impact of this path-history effect on the statistics of $\boldsymbol{w}^p(t)$ depends upon both $St_2$, and the timescale of $\Delta\boldsymbol{u}^p$. In the presence of gravity, the particles fall through the flow, and the timescale of $\Delta\boldsymbol{u}^p$ is reduced compared to the case without gravity \citep{ireland2016effectb}. Consequently, gravity reduces the path-history effect, and in the case of monodisperse particles, this leads to a suppression of the particle relative velocities \citep{ireland2016effectb}.

The second term on the rhs of (\ref{eq:w_p_fs}) describes the explicit effect of gravity on the particle motion; it acts only in the direction of gravity and represents the difference in the settling velocity of the two particles (the ``differential settling velocity''). The third term depends upon the particle acceleration, which is implicitly affected by gravity. This third term causes the relative velocities of bidisperse particles to be greater than those of monodisperse particles in the absence of gravity. In \cite{ireland2016effectb}, it was argued that because gravity causes particles to fall through the fluid velocity field, the fluid velocity changes more rapidly along their trajectory than in the absence of gravity (i.e. larger values of $ \boldsymbol{u}^p (t)-\boldsymbol{u}^p(t^\prime)$ for a given $t-t'$ in (\ref{eq:a_p})), such that gravity enhances the inertial particle accelerations. This enhancement occurs for both the horizontal and vertical components of $\boldsymbol{a}^p$, and is in fact stronger for the horizontal component \citep{ireland2016effectb}. Gravity therefore enhances both the vertical and horizontal components of $\boldsymbol{w}^p(t)$.

We now turn to consider the effect of $R_\lambda$ on the motion of settling, bidisperse inertial particles in turbulence. As is well-known, turbulent flows become increasingly intermittent at the small scales as $R_\lambda$ is increased \citep{pope,frisch}. This would then lead to the expectation that $\boldsymbol{w}^p(t)$ should exhibit increasingly intermittent fluctuations due to the first and third terms on the rhs (\ref{eq:w_p_fs}). While this will strongly affect the higher-order moments of $\boldsymbol{w}^p(t)$, the effect on lower-order moments, such as those relevant to particle collision rates, is not immediately clear. Indeed, in \cite{ireland2016effectb} it was shown that for monodisperse particles, the lower-order moments of $\boldsymbol{w}^p(t)$ are almost independent of $R_\lambda$ for $St\lesssim 1$.

For typical (r.m.s.) values of $\Delta\boldsymbol{u}^p$ and $\boldsymbol{a}^p(t)$, we expect that the differential settling contribution to (\ref{eq:w_p_fs}) will completely dominate the behavior of $\boldsymbol{w}^p(t)$ in the vertical direction when $Fr\ll1$ if $|\Delta St|\gg Fr$. However, owing to intermittency, as $R_\lambda$ is increased, regions where the first and third terms  on the rhs (\ref{eq:w_p_fs}) become $O(|\Delta St |/Fr )$ are increasingly probable. Therefore, for a given value of $Fr$, turbulence is expected to play an increasingly important role on the verticle relative motion of the particles as $R_\lambda$ is increased, even when $|\Delta St|/Fr\gg1$, where simple dimensional arguments would suggest the effect of turbulence could be neglected.

In \cite{ireland2016effectb}, we derived the following asymptotic prediction for the particle accelerations in the vertical direction in the regime $St_1\gg (u'/u_\eta)Fr$
\begin{equation}
\frac{\langle a_3^p(t)a_3^p(t) \rangle}{a_\eta^{2}}=
\frac{1}{Fr}\Bigg(\frac{u^\prime}{u_\eta}\Bigg)^2 
\Bigg(\frac{1}{St_1^2 Fr^{-1}+l\eta^{-1}} \Bigg),\label{a3model}
\end{equation}
where $u^\prime$ is the fluid r.m.s. velocity, and $l$ is the integral length scale of the flow. In \cite{ireland2016effectb}, (\ref{a3model}) was found to agree very well with DNS over the regime for which it was derived. Using the scaling $u'/u_\eta\sim Re^{1/4}$, $l/\eta\sim Re^{3/4}$, where $Re\equiv u' l/\nu$, and $R_\lambda=\sqrt{15 Re}$, we find 
\begin{equation}\label{a1asymp1}
\frac{\langle a_3^p(t)a_3^p(t) \rangle}{a_\eta^{2}} \sim 15^{1/4} /(Fr R_\lambda ^{1/2}),\quad\textrm{for}\,(St_1^2 /Fr)^{2/3}\ll R_\lambda \ll (St_1 /Fr)^{2},
\end{equation}
whereas
\begin{equation}\label{a1asymp2}
\frac{\langle a_3^p(t)a_3^p(t) \rangle}{a_\eta^{2}} \sim R_\lambda / (15^{1/2} St_1^{2}) ,\quad\textrm{for}\,R_\lambda\ll \min[(St_1^2 /Fr)^{2/3},(St_1 /Fr)^{2}].
\end{equation}
Consequently, two very different asymptotic behaviors are predicted depending upon the parameter regimes of the system, although likely only \eqref{a1asymp1} would be obtainable in realistic applications where $R_\lambda\gg 1$. The asymptotic model for the horizontal accelerations given in \cite{ireland2016effectb} leads to similar predictions for that direction. 
\section{Computational Details}\label{CompD}
We perform Direct Numerical Simulations (DNS) of the incompressible Navier-Stokes equation on a triperiodic cube of length $\mathscr {L}$, using a pseudo-spectral method on a uniform mesh with $N^3$ grid points. The fluid velocity of isotropic turbulence flow field $\boldsymbol{u}(\boldsymbol{x},t)$ is obtained by solving the incompressible Navier-Stokes equation
\begin{eqnarray}\label{eq:NSE}
\partial_t\boldsymbol{u} + {\boldsymbol{\omega}} \times \boldsymbol{u} 
+ \boldsymbol{\nabla}\left ( \frac{p}{\rho_f} + \frac{\|\boldsymbol{u}\|^{2}}{2}  \right )  = \nu {\nabla}^2 \boldsymbol{u} + \boldsymbol{f}, \,\boldsymbol{\nabla}\bcdot\boldsymbol{u}=0 \label{eq:ns}
\end{eqnarray}
where $\boldsymbol{\omega} \equiv \boldsymbol{\nabla} \times \boldsymbol{u}$ is the vorticity, $\rho_f$ is the fluid density, $p$ is the pressure (determined by using $\boldsymbol{\nabla}\bcdot\boldsymbol{u}=0$), $\nu$ is the kinematic viscosity and  $\boldsymbol{f}$ is the external forcing term to numerically generate statistically stationary homogeneous turbulence flow field.
A deterministic forcing scheme was used for $\boldsymbol{f}$, where the energy dissipated during one time step is resupplied to the low wavenumbers (large scales) with magnitude $\kappa \in (0, \sqrt{2}]$. Time integration is performed through a second-order, explicit Runge-Kutta scheme and alias control is achieved through a combination of spherical truncation and phase-shifting.

\begin{table}
	\centering
	\renewcommand{\arraystretch}{1.2}
	\setlength{\tabcolsep}{12pt}
\begin{tabular}{ccccc}
    $\mathrm{Parameter}$ & $\mathrm{DNS} \,1 $ & $\mathrm{DNS} \,2$  & $\mathrm{DNS} \,3$  & $\mathrm{DNS} \,4$   \\
	$N$ & 128 & 128 & 1024 & 512  \\
	$R_\lambda$ &  93 &  94 &  90 &  224  \\
	$Fr$ &  $ \infty $ &  0.3 &  0.052 &  $ \infty $  \\
	$\mathscr{L}$ & 2$\upi$ & 2$\upi$ & 16$\upi$ & 2$\upi$   \\       
	$\nu$ & 0.005 & 0.005 & 0.005 & 0.0008289  \\
	$\epsilon$ & 0.324 & 0.332 & 0.257 & 0.253  \\
	$l$ & 1.48 & 1.49 & 1.47 & 1.40 \\
	$l/\eta$ & 59.6 & 60.4 & 55.6 & 204 \\
	$u'$ & 0.984 & 0.996 & 0.912 & 0.915  \\
	$u'/u_\eta$ & 4.91 & 4.92 & 4.82 & 7.60  \\
	$T_L$ & 1.51 & 1.50 & 1.61 & 1.53  \\
	$T_L/\tau_\eta$ & 12.14 & 12.24 & 11.52 & 26.8 \\
	$\kappa_{{\rm max}}\eta$ & 1.5 & 1.48 & 1.61 & 1.66 \\
	$N_p$ & 262,144 & 262,144 & 16,777,216 & 2,097,152  \\
\end{tabular}
\begin{tabular}{ccccc}
	\hline
	$\mathrm{Parameter}$ & $\mathrm{DNS} \,5$ & $\mathrm{DNS} \,6$  & $\mathrm{DNS}\,7 $  & $\mathrm{DNS} \,8$   \\
	$N$ & 512 & 1024 & 1024 & 1024  \\
	$R_\lambda$ &  237 &  230 &  398 &  398 \\
	$Fr$ &  0.3 &  0.052 & $ \infty $  & 0.052 \\
	$\mathscr{L}$ & 2$\upi$ & 4$\upi$ & 2$\upi$ & 2$\upi$  \\       
	$\nu$ & 0.0008289  & 0.0008289 & 0.0003 & 0.0003 \\
	$\epsilon$ & 0.2842 & 0.239 & 0.223 & 0.223  \\
	$l$ & 1.43 & 1.49 & 1.45 & 1.45  \\
	$l/\eta$ & 214 & 213 & 436 & 436  \\
	$u'$ & 0.966 & 0.914 & 0.915 & 0.915  \\
	$u'/u_\eta$ & 7.82 & 7.7 & 10.1 & 10.1  \\
	$T_L$ & 1.48 & 1.63 & 1.58 & 1.58  \\
	$T_L/\tau_\eta$ & 27.36 & 27.66 & 43.0 & 43.0  \\
	$\kappa_{{\rm max}}\eta$ & 1.62 & 1.68 & 1.60 & 1.60 \\
	$N_p$ & 2,097,152 & 16,777,216 & 2,097,152 & 2,097,152  \\
\end{tabular}

	\caption{Simulation parameters for the DNS study of isotropic turbulence (arbitrary units).
		$N$ is the number of grid points in each direction, 
		$R_\lambda \equiv u'\lambda/\nu$ is the Taylor micro-scale
		Reynolds number ($R_\lambda \equiv \sqrt{15Re}$ for homogeneous and isotropic flows), $\lambda\equiv u'/\langle(\boldsymbol{\nabla} \boldsymbol{u})^2\rangle^{1/2} $ is the Taylor micro-scale,
		$\mathscr{L} $ is the box size, $\nu$ is the fluid kinematic viscosity, $\epsilon \equiv 2\nu \int_0^{\kappa_{\rm max}}\kappa^2 E(\kappa) {\rm d}\kappa $ is the mean
		turbulent kinetic energy dissipation rate,
		$l \equiv 3\upi/(2k)\int_0^{\kappa_{\rm max}}E(\kappa)/\kappa {\rm d}\kappa $  is the integral length scale, $\eta \equiv \nu^{3/4}/\epsilon^{1/4}$ is the Kolmogorov length scale, 
		$u' \equiv \sqrt{(2k/3)}$ is the fluid r.m.s. fluctuating 
		velocity, $k$ is the turbulent kinetic energy, 
		$u_\eta$ is the Kolmogorov velocity scale, 
		$T_L \equiv l/u^\prime$ is the large-eddy turnover
		time, $\tau_\eta \equiv \sqrt{(\nu/\epsilon)}$ is the Kolmogorov time scale, 
		$\kappa_{\rm max}=\sqrt{2}N/3$ is the maximum
		resolved wavenumber, and $N_p$ is  
		the number of particles per Stokes number.}
	{\label{tab:parameters}}
\end{table}
\FloatBarrier

Inertial particles governed by \eqref{MR_inertial} were tracked in the turbulent flow, assuming the influence of particles on the flow is negligible (a good approximation for the motivating applications). Fifteen different particle classes are simulated with Stokes numbers and settling parameters in the range of $0\le St \le 3$ and $0\le Sv \lesssim 58$, respectively, with $N^3/64$ particles simulated per $St$. Once the flow field has become statistically stationary, particles are uniformly introduced into the flow with the local fluid velocity. The particle statistics were computed after the particle distributions and velocities had reached a statistically stationary state, independent of their initial conditions.

The solution of \eqref{MR_inertial}, depends upon the fluid velocity at the particle position $\boldsymbol{u}(\boldsymbol{x}^p(t),t)$, and this must be evaluated by interpolating the grid values of fluid velocity at the surrounding points to particle centers $\boldsymbol{x}^p(t)$. In this study we apply an $8^{th}$-order, B-spline interpolation (from the Eulerian grid) which provides a good balance between high-accuracy and efficiency \citep[see][]{ireland2013highly}. Further details on all aspects of the computational methods can be found in \cite{ireland2013highly}. 

In this study, eight different simulations are performed in which $90\leq R_\lambda \leq 398$. The primary object of this study is to explore the effect of gravity and $R_\lambda$ on the small-scale motion of the bidisperse particles, particularly at conditions representative of those in cumulus clouds. Therefore, in addition to the zero gravity case $Fr=\infty$, we follow \cite{ireland2016effectb} and consider $Fr= 0.3, 0.052$, which characterize strongly turbulent cumulonimbus clouds and weakly turbulent stratiform clouds, respectively \citep{pinsky2007collisions}. 

As reported in our recent works (\cite{ireland2016effectb}  \& \cite{dhariwal2018small}), in the case of strong gravity ($Fr<1$) the use of periodic boundary conditions in the DNS can artificially influence the motion of inertial particles if the box length $\mathscr{L}$ is too small. In particular, periodic boundary conditions become problematic when the time it takes the settling particles to traverse the distance $\mathscr{L}$ is smaller than the large eddy turnover time, $\mathscr{L}/\tau_p{g}< O(T_L)$. As discussed in \cite{ireland2016effectb}, if $\mathscr{L}/\tau_p{g}< O(T_L)$, particles can artificially re-encounter the same large eddy as they are periodically looped through the domain. To resolve this issue, larger domain sizes must be used to ensure $\mathscr{L}/\tau_p{g}>O(T_L)$ for each particle class simulated, and this places significant limitations on the value of $R_\lambda$ that can be reliably simulated when $Fr\ll1$.

Our DNS satisfy the requirement $\mathscr{L}/\tau_p{g}>O(T_L)$ for the range of $St$ and $Fr$ considered, and are designed to keep both the small and large scales of the flow approximately constant while extending the domain size. Details of the DNS are summarized in Table~\ref{tab:parameters}.

\section{Results and discussion}
\subsection{Acceleration}
\begin{figure}
  \centering
  \begin{subfigure}[b]{0.5\linewidth}
    \includegraphics[max size={\textwidth}{\textheight}]{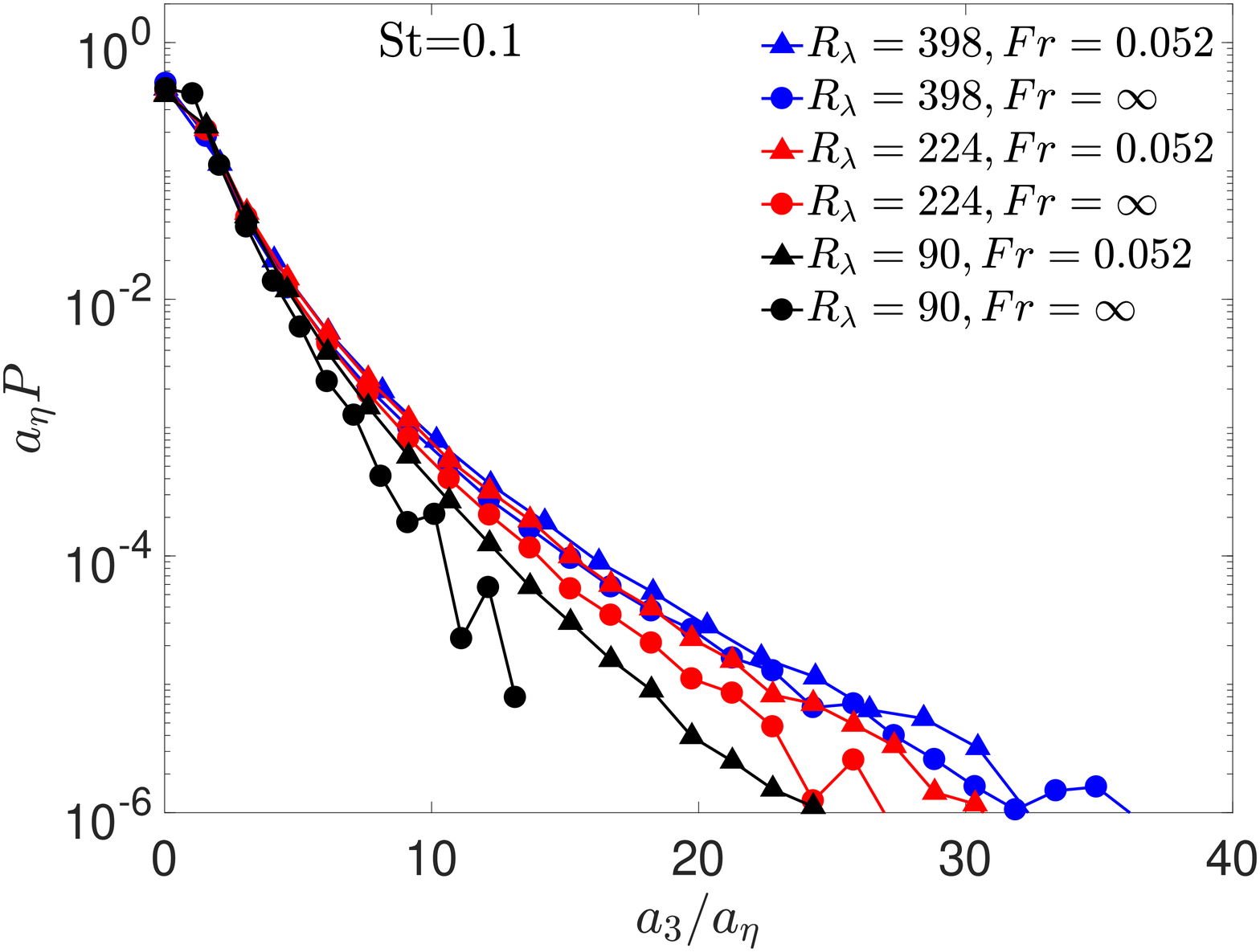}
    \caption{}
  \end{subfigure}%
    \begin{subfigure}[b]{0.5\linewidth}
    \includegraphics[max size={\textwidth}{\textheight}]{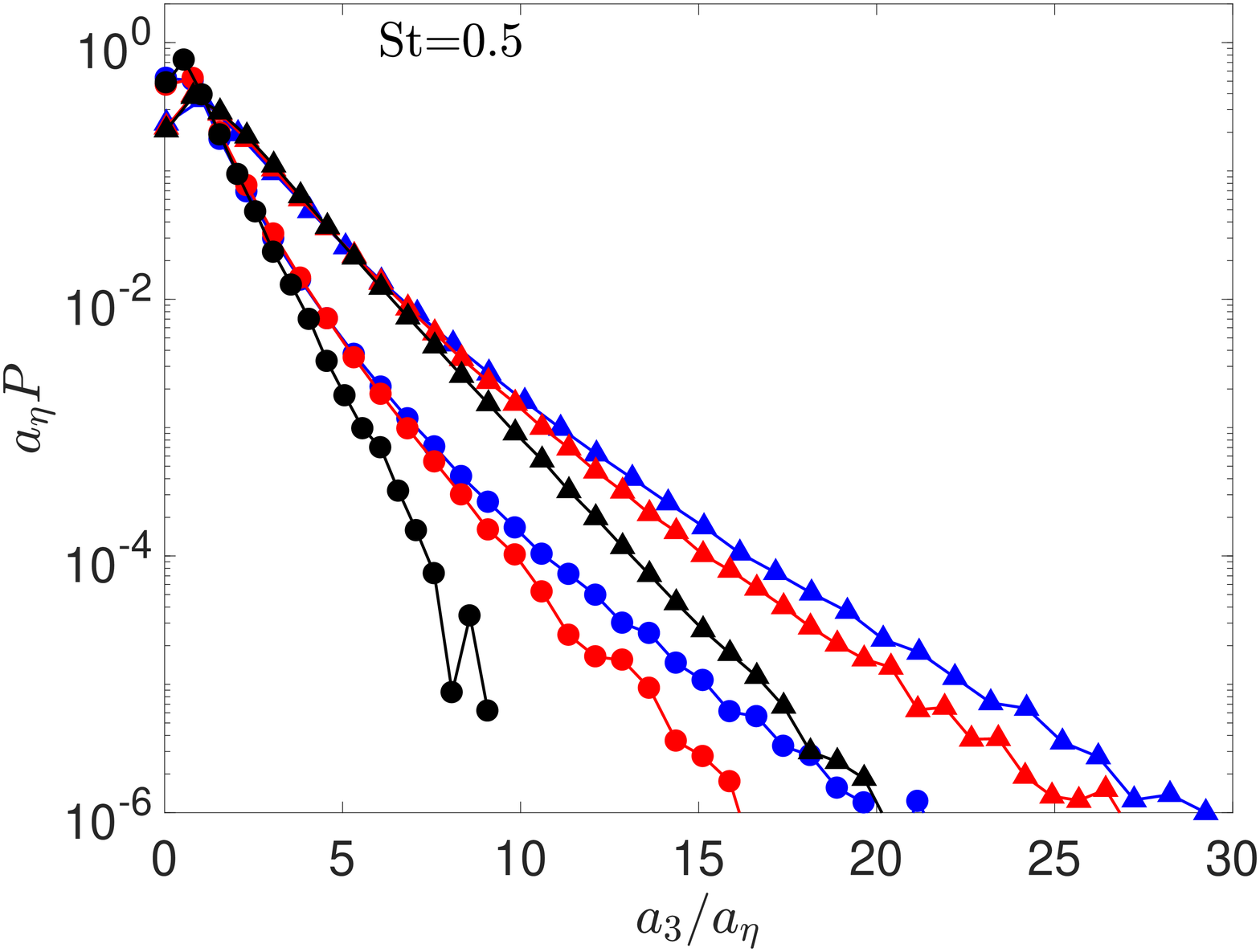}
    \caption{}
  \end{subfigure}
    \begin{subfigure}[b]{0.5\linewidth}
    \includegraphics[max size={\textwidth}{\textheight}]{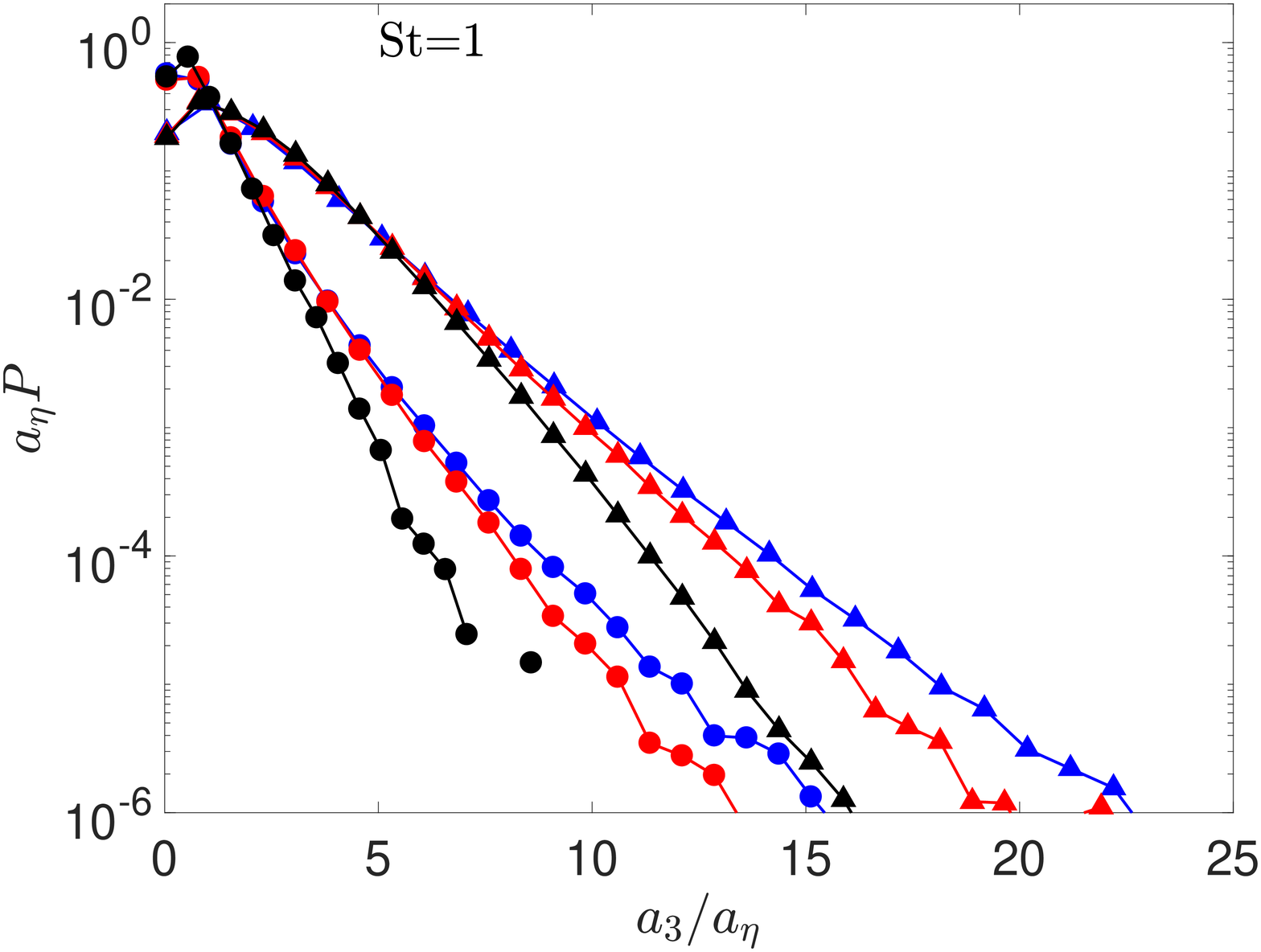}
    \caption{}
  \end{subfigure}%
    \begin{subfigure}[b]{0.5\linewidth}
    \includegraphics[max size={\textwidth}{\textheight}]{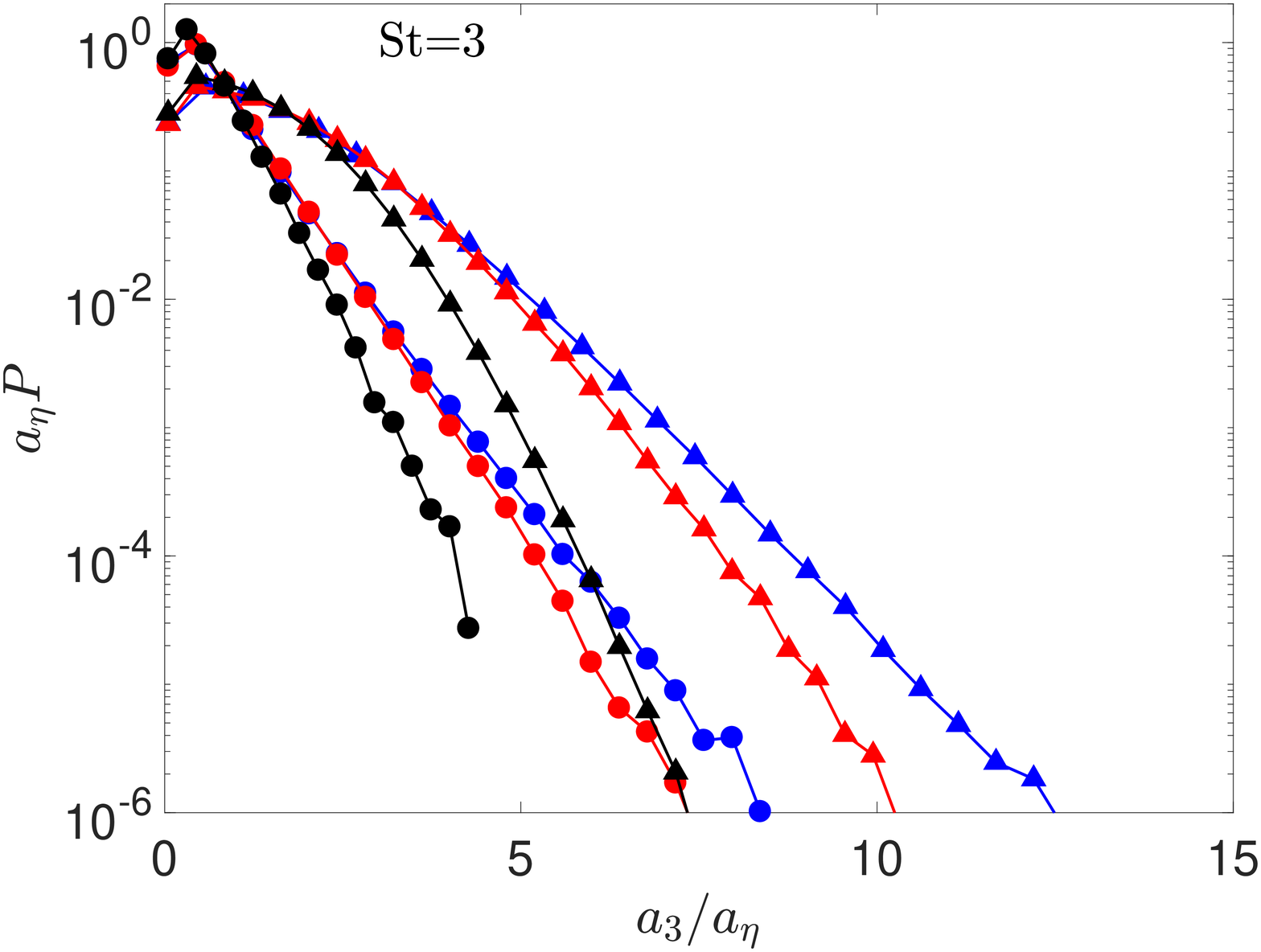}
    \caption{}
  \end{subfigure}
  \caption{PDF of the vertical particle acceleration, normalized by $a_\eta$, for different $St$, $Fr$, and $R_\lambda$ combinations. Black, red and blue lines correspond to $R_\lambda=90$, $R_\lambda=224$ and $R_\lambda=398$, respectively, and circle and triangle symbols denote $Fr=\infty$ and $Fr=0.052$, respectively.}\label{fig:Acceleration_PDF_gDir}
\end{figure}

\begin{figure}
  \centering
  \begin{subfigure}[b]{0.5\linewidth}
    \includegraphics[width=\linewidth]{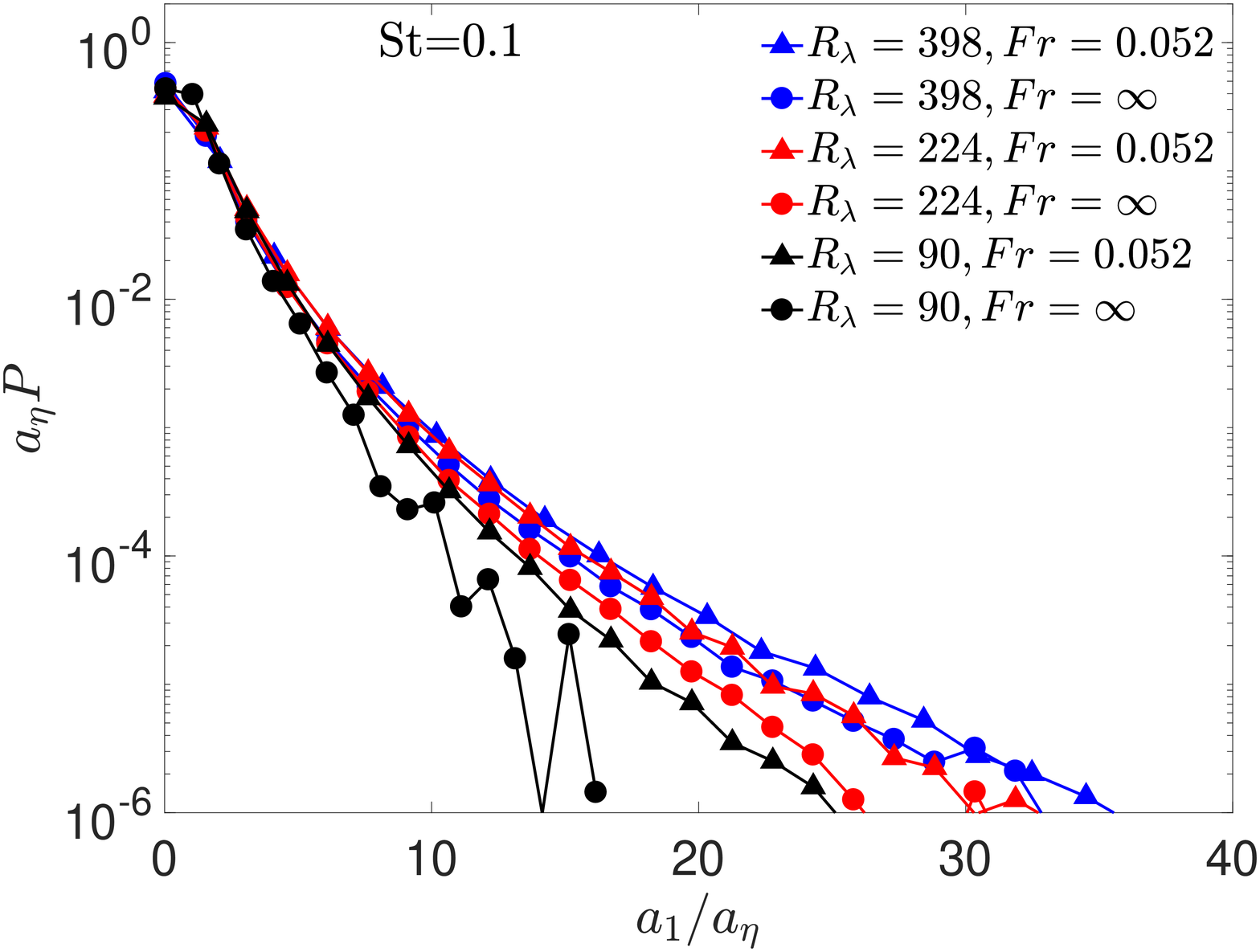}
    \caption{}
  \end{subfigure}%
    \begin{subfigure}[b]{0.5\linewidth}
    \includegraphics[width=\linewidth]{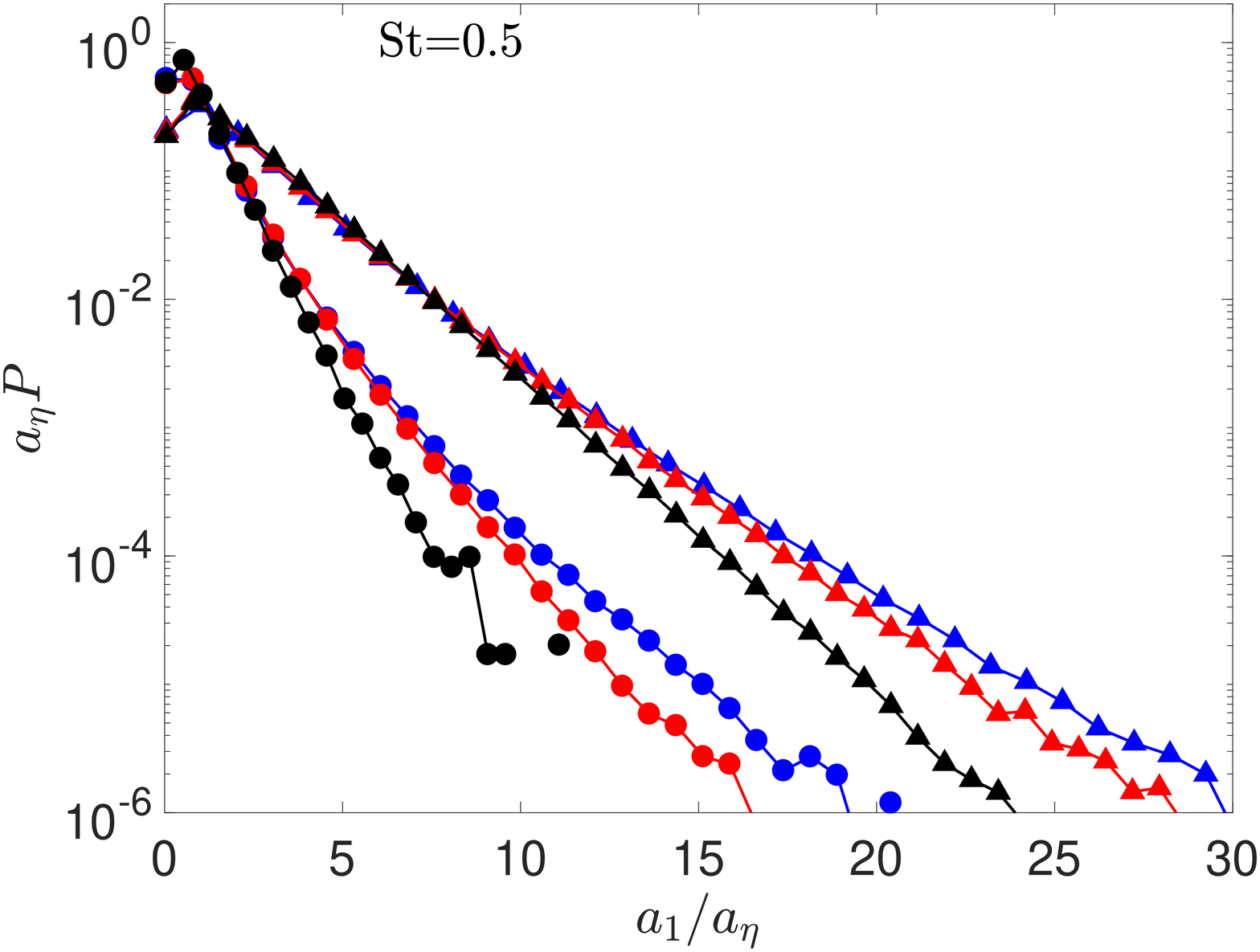}
    \caption{}
  \end{subfigure}
    \begin{subfigure}[b]{0.5\linewidth}
    \includegraphics[width=\linewidth]{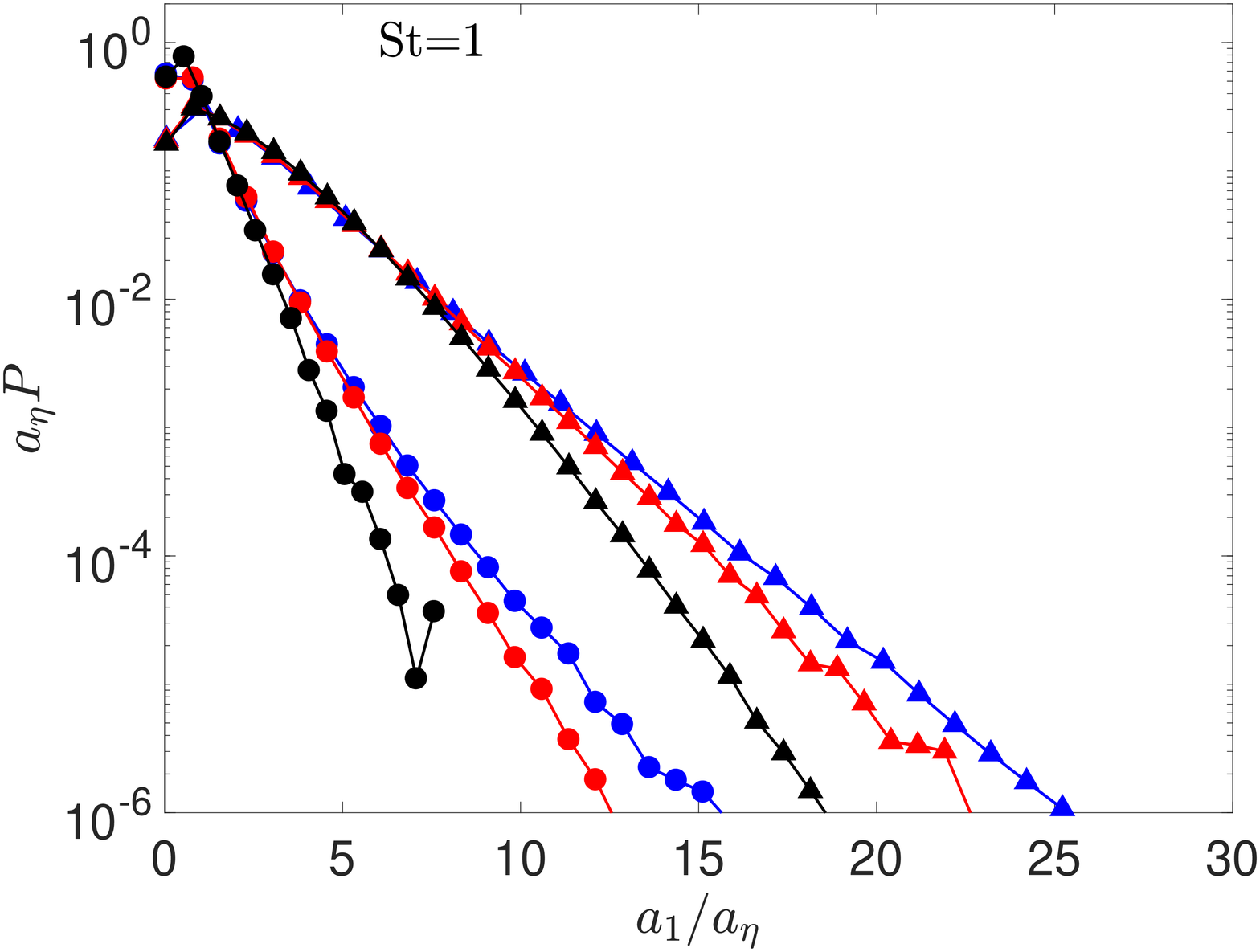}
    \caption{}
  \end{subfigure}%
    \begin{subfigure}[b]{0.5\linewidth}
    \includegraphics[width=\linewidth]{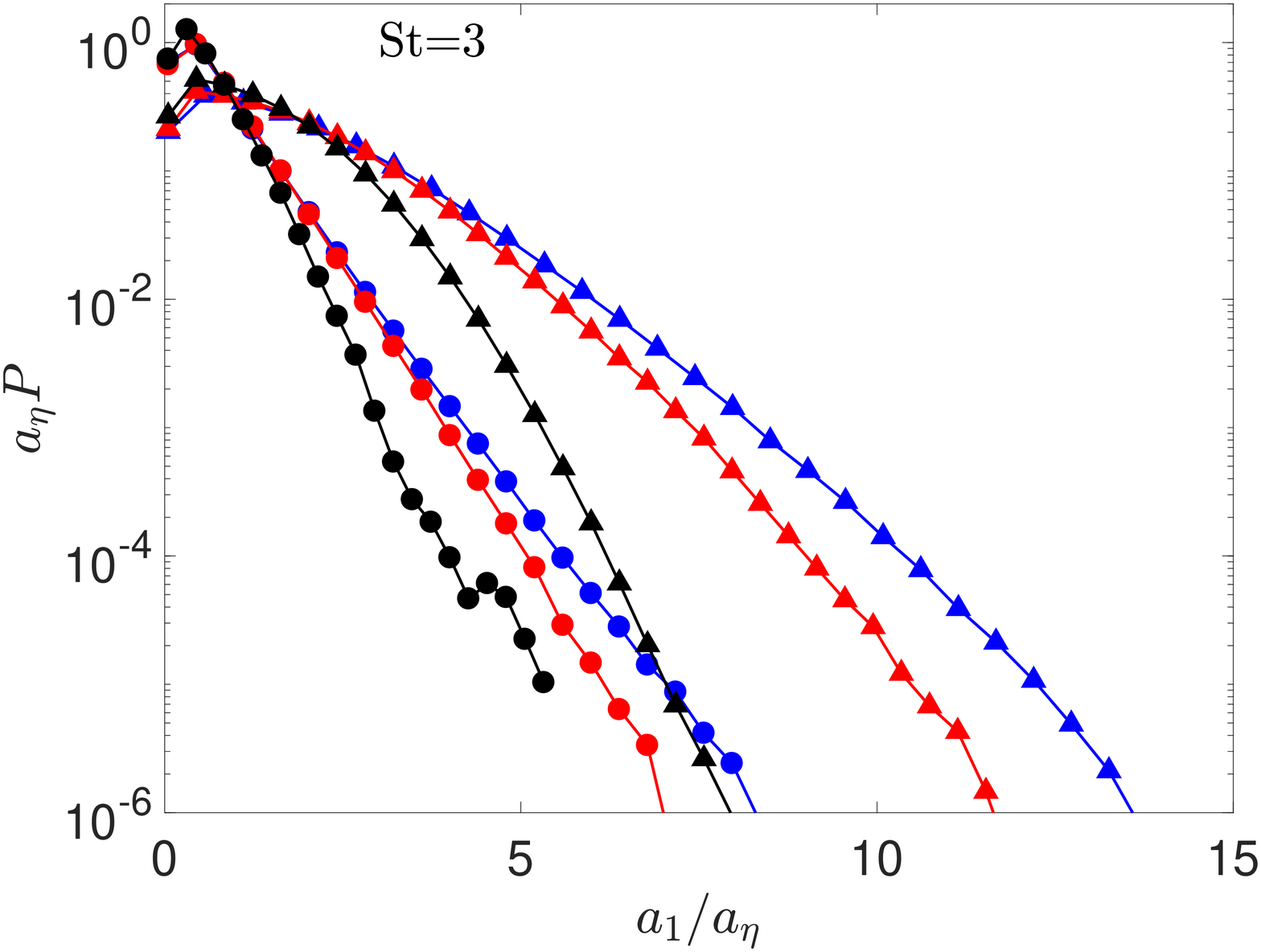}
    \caption{}
  \end{subfigure}

  \caption{PDF of the horizontal particle acceleration, normalized by $a_\eta$, for different $St$, $Fr$, and $R_\lambda$ combinations. Black, red and blue lines correspond to $R_\lambda=90$, $R_\lambda=224$ and $R_\lambda=398$, respectively, and circle and triangle symbols denote $Fr=\infty$ and $Fr=0.052$, respectively}\label{fig:Acceleration_PDF_NtgDir}
\end{figure}

Figures \ref{fig:Acceleration_PDF_gDir} \& \ref{fig:Acceleration_PDF_NtgDir} show the DNS results for the acceleration PDFs, at three different $R_\lambda$ both with gravity and without gravity, in the vertical and horizontal directions, respectively. We observe that the $R_\lambda$ dependency can be stronger for the case with gravity than without gravity, especially for $St>1$. A possible explanation for this is as follows. For particles moving according to \eqref{MR_inertial}, the acceleration (normalized by the Kolmogorov scales) may be written as \citep{dhariwal2018small} 
\begin{equation}\label{eq:a_p2}
\boldsymbol{a}^p(t)=\frac{1}{St^2} \int_0^{t} \mathrm e^{-(t-t^\prime)/St}(\boldsymbol{u}^p (t)-\boldsymbol{u}^p(t^\prime))\mathrm{d} t^\prime.
\end{equation}
Equation  \eqref{eq:a_p2} shows that since $\mathrm e^{-(t-t^\prime)/St}$ decays on the scale $St$, $\boldsymbol{a}^p(t)$ is only affected by $\boldsymbol{u}^p (t)-\boldsymbol{u}^p(t^\prime)$ for times $t-t'\leq O(St)$. Further, the quantity $\boldsymbol{u}^p (t)-\boldsymbol{u}^p(t^\prime)\equiv \boldsymbol{u}(\boldsymbol{x}^p(t),t)-\boldsymbol{u}(\boldsymbol{x}^p(t'),t')$ depends in part on $\| \boldsymbol{x}^p(t)-\boldsymbol{x}^p(t')\|$. If $St\ll1$, $\| \boldsymbol{x}^p(t)-\boldsymbol{x}^p(t')\|$ is small (compared to the integral lengthscale of the flow) for $t-t'\leq O(St)$, and so $\boldsymbol{u}^p (t)-\boldsymbol{u}^p(t^\prime)$, and hence $\boldsymbol{a}^p(t)$, will be dominated by the small-scales. As $St$ is increased, $\| \boldsymbol{x}^p(t)-\boldsymbol{x}^p(t')\|$ can be significant for $t-t'\leq O(St)$, and so the accelerations of these particles are increasingly affected by larger scales in the flow. However, in the regime $Fr\ll1$ and $St\geq O(1)$, gravity significantly enhances the particle displacement $\| \boldsymbol{x}^p(t)-\boldsymbol{x}^p(t')\|$ over the time span $t-t'\leq O(St)$, due to the fast settling of the particles. As a result, the particle accelerations for a given $St$ can become increasingly affected by larger scales in the flow as $Fr$ is decreased, and hence the accelerations of inertial particles can be more sensitive to $R_\lambda$ with gravity than without, since gravity causes the particles accelerations to be affected by a wider range of flow lengthscales.
\FloatBarrier

\begin{figure}
  \centering
  \begin{subfigure}[b]{0.5\linewidth}
    \includegraphics[width=\linewidth]{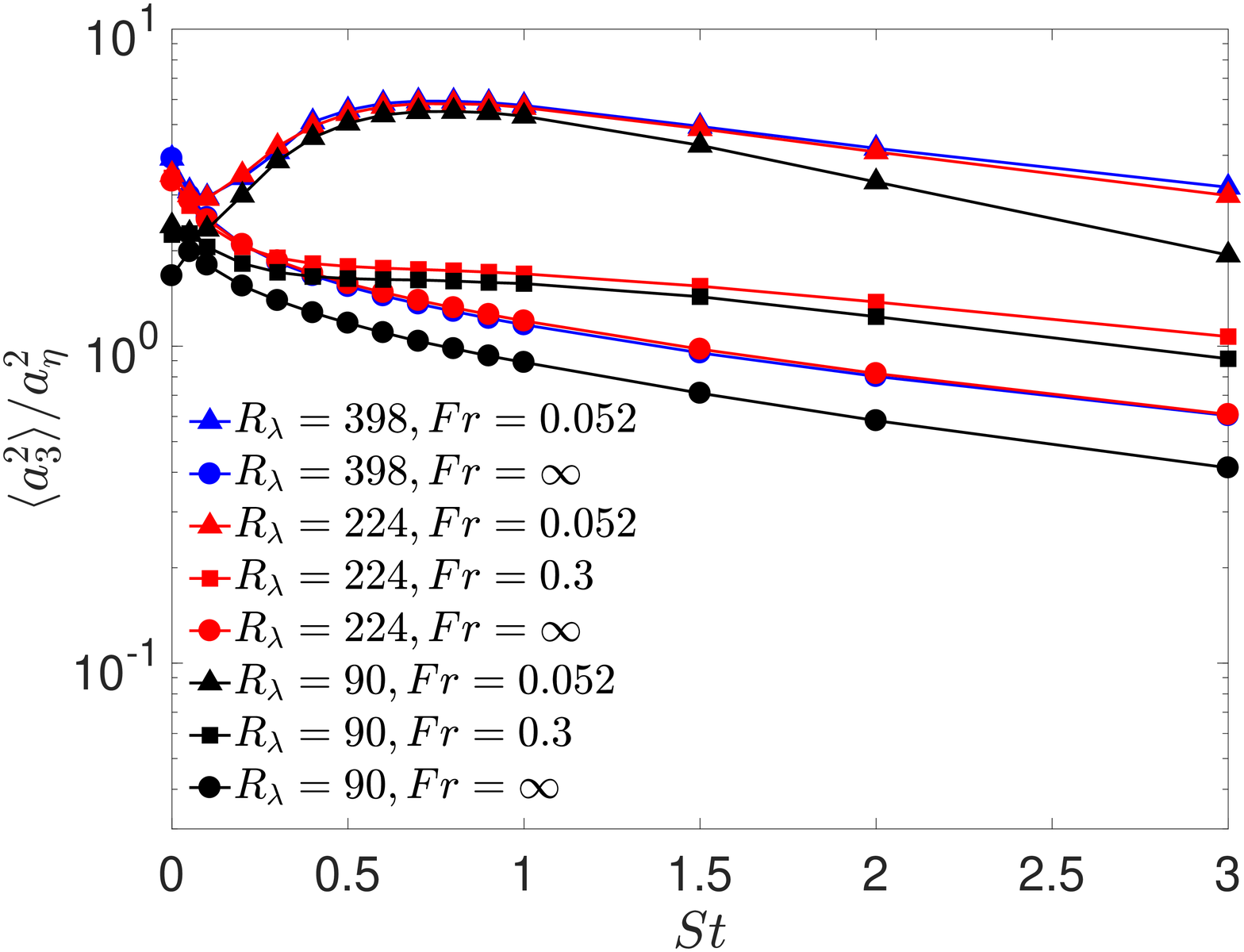}
    \caption{Vertical}
  \end{subfigure}%
    \begin{subfigure}[b]{0.5\linewidth}
    \includegraphics[width=\linewidth]{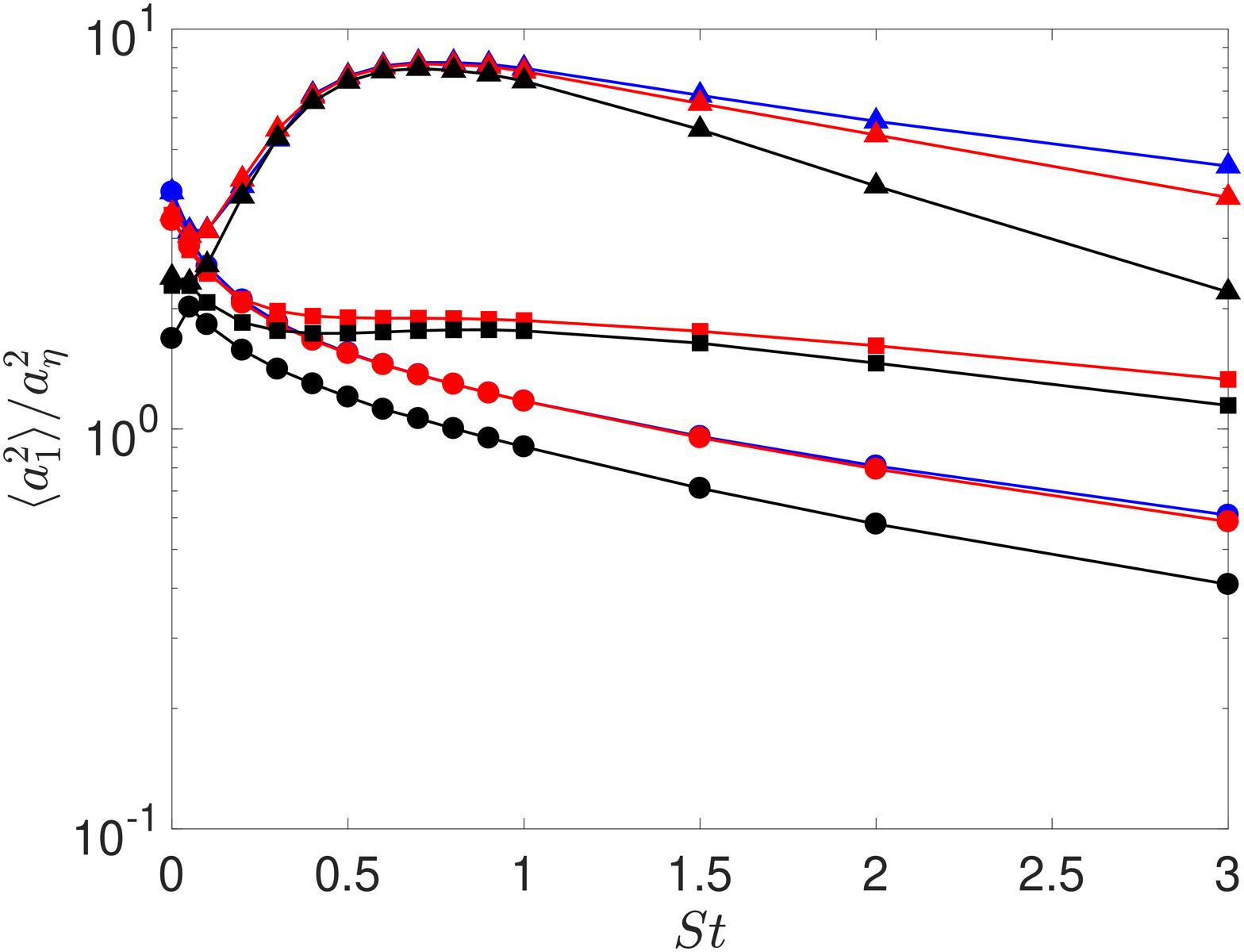}
    \caption{Horizontal}
    \end{subfigure}
  \caption{Variance of (a) vertical and (b) horizontal particle accelerations, as a function of $St$ for different $Fr$ and $R_\lambda$ combinations. Black, red and blue lines correspond to $R_\lambda=90$, $R_\lambda=224$ and $R_\lambda=398$, respectively, and circle, square and triangle symbols denote $Fr=\infty$, $Fr=0.3$ and $Fr=0.052$, respectively.}\label{fig:Acceleration_Variance_direct_computation}
\end{figure}

In figure \ref{fig:Acceleration_Variance_direct_computation}, we show results for the variance of the particle accelerations, and the ratio of their values for the case with gravity to the case without gravity. The results indicate that for $Fr=\infty$, increasing $St$ monotonically decreases the particle accelerations. This decrease occurs both due to the effect of preferential sampling, whereby the inertial partcles avoid strongly vortical regions where there is rapid fluid acceleration, and also due to the filtering effect, whereby with increasing $St$, the particles become sluggish and have a modulated response to the fluid accelerations along their trajectory \citep{bec06a,sathya08a}. The results for $Fr<1$ show a non-monotonic dependence of the fluid acceleration variances on $St$. The initial enhancement of the particle accelerations with increasing $St$ is explained by the arguments in \S\ref{TC}, namely, that the fast settling of the particles causes them to experience rapid changes in the fluid velocity along their trajectory, leading to large particle accelerations. However, as $St$ is increased, the filtering effect begins to take over, and the accelerations begin to reduce. 

It is interesting to note that the results for $Fr=0.052$ show that as $R_\lambda$ is increased, the dependence of the acceleration variances on $St$ become weaker for $St\gtrsim 1$. This is because as $R_\lambda$ is increased, the behavior approaches (although the data indicates it does not reach) the asymptotic regime described by \eqref{a1asymp1} in which the acceleration variances become independent of $St$.


\begin{figure}
  \centering
  \begin{subfigure}[b]{0.5\linewidth}
    \includegraphics[width=\linewidth]{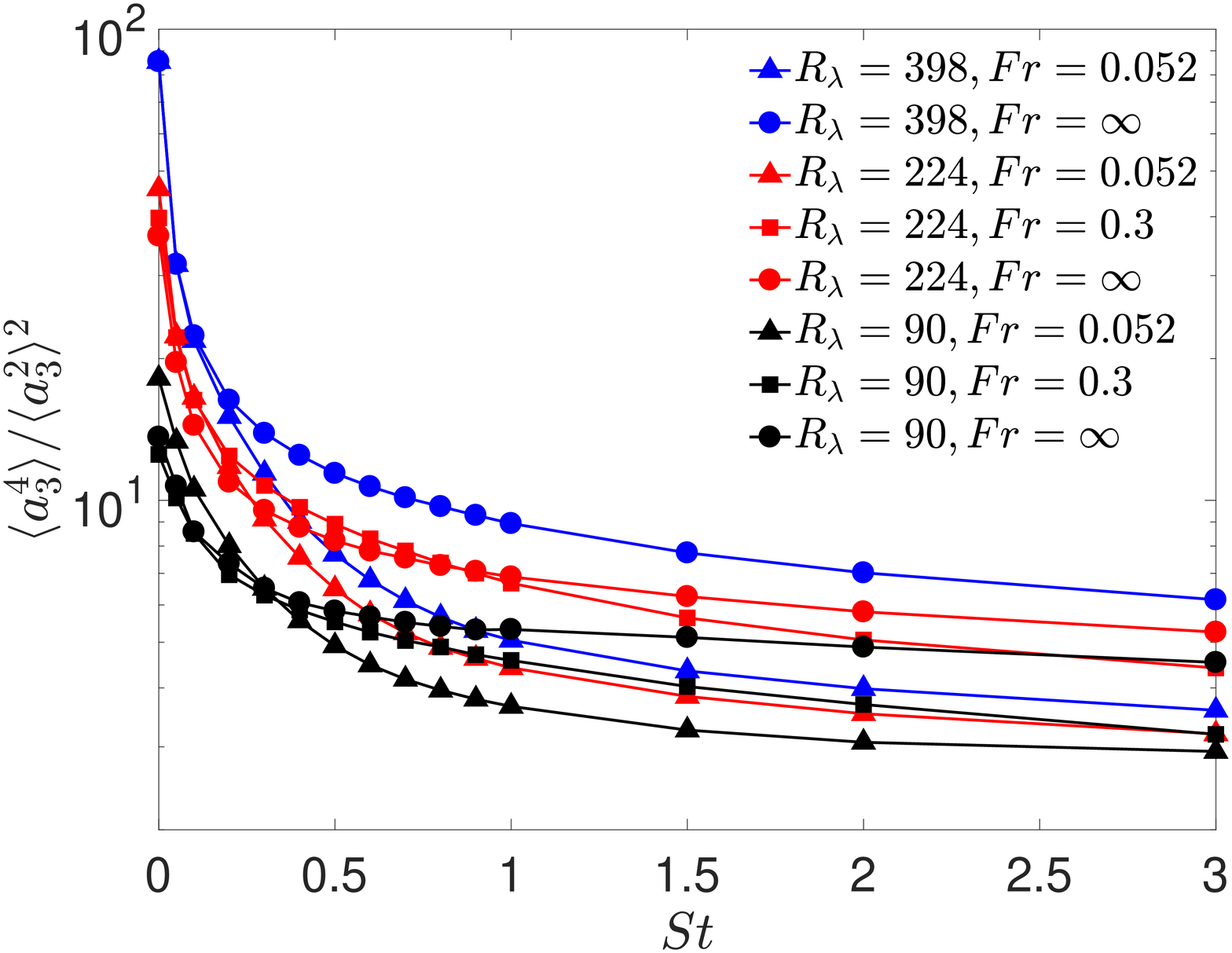}
    \caption{Vertical}
  \end{subfigure}%
    \begin{subfigure}[b]{0.5\linewidth}
    \includegraphics[width=\linewidth]{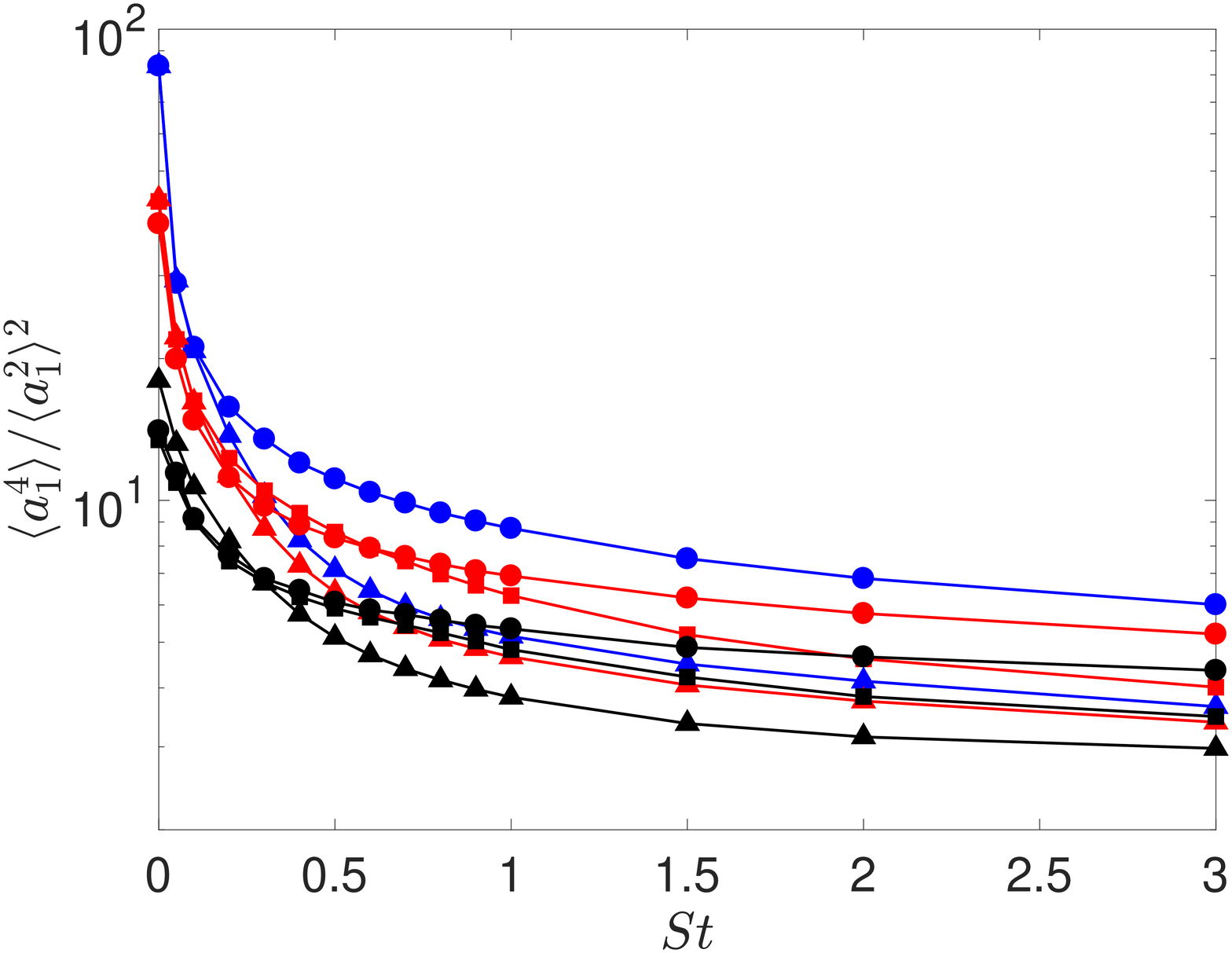}
    \caption{Horizontal}
    \end{subfigure}
  \caption{Kurtosis of (a) vertical and (b) horizontal particle accelerations, as a function of $St$ for different $Fr$ and $R_\lambda$ combinations. Black, red and blue lines correspond to $R_\lambda=90$, $R_\lambda=224$ and $R_\lambda=398$, respectively, and circle, square and triangle symbols denote $Fr=\infty$, $Fr=0.3$ and $Fr=0.052$, respectively.}\label{fig:Acceleration_Kurtosis_direct_computation}
\end{figure}

In figure \ref{fig:Acceleration_Kurtosis_direct_computation} we plot the kurtosis of the inertial particle accelerations to explore intermittency in the accelerations. The results show that while increasing $R_\lambda$ enhances the kurtosis for all $St$, increasing $St$ monotonically suppresses the kurtosis of the particle accelerations relative to the fluid particle case $St=0$, as previously observed in \cite{ireland2016effecta}. However, we also find that decreasing $Fr$ significantly suppresses the kurtosis further, producing values approaching those of a Gaussian PDF. Therefore, the effect of gravity on the inertial particle accelerations is to enhance the size of the fluctuations, but also to suppress intermittency in the fluctuations. Note that since we are considering homogeneous, stationary turbulence, for which $\langle\boldsymbol{a}^p(t)\rangle=0$, our results for the kurtosis imply that for $Fr\ll1$ and $St\gtrsim 1$ (i.e. $Sv\gg 1$), the particle acceleration PDFs could be modeled as a Gaussian distribution with variance given by the asymptotic models in \cite{ireland2016effectb}. 


\subsection{Relative Velocities}

\begin{figure}
  \centering
  \begin{subfigure}[b]{0.5\linewidth}
    \includegraphics[width=\linewidth]{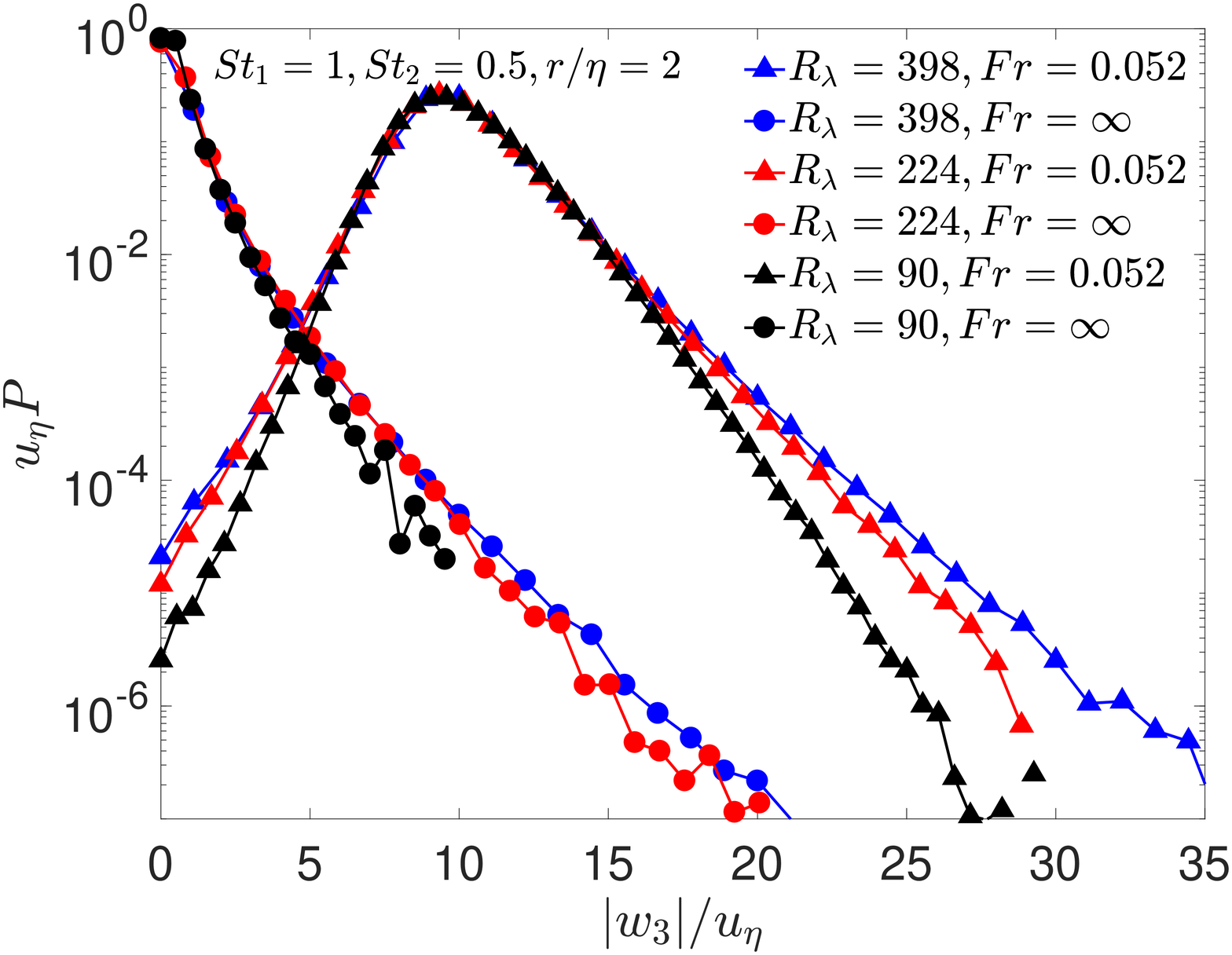}
    \caption{Vertical }
  \end{subfigure}%
    \begin{subfigure}[b]{0.5\linewidth}
    \includegraphics[width=\linewidth]{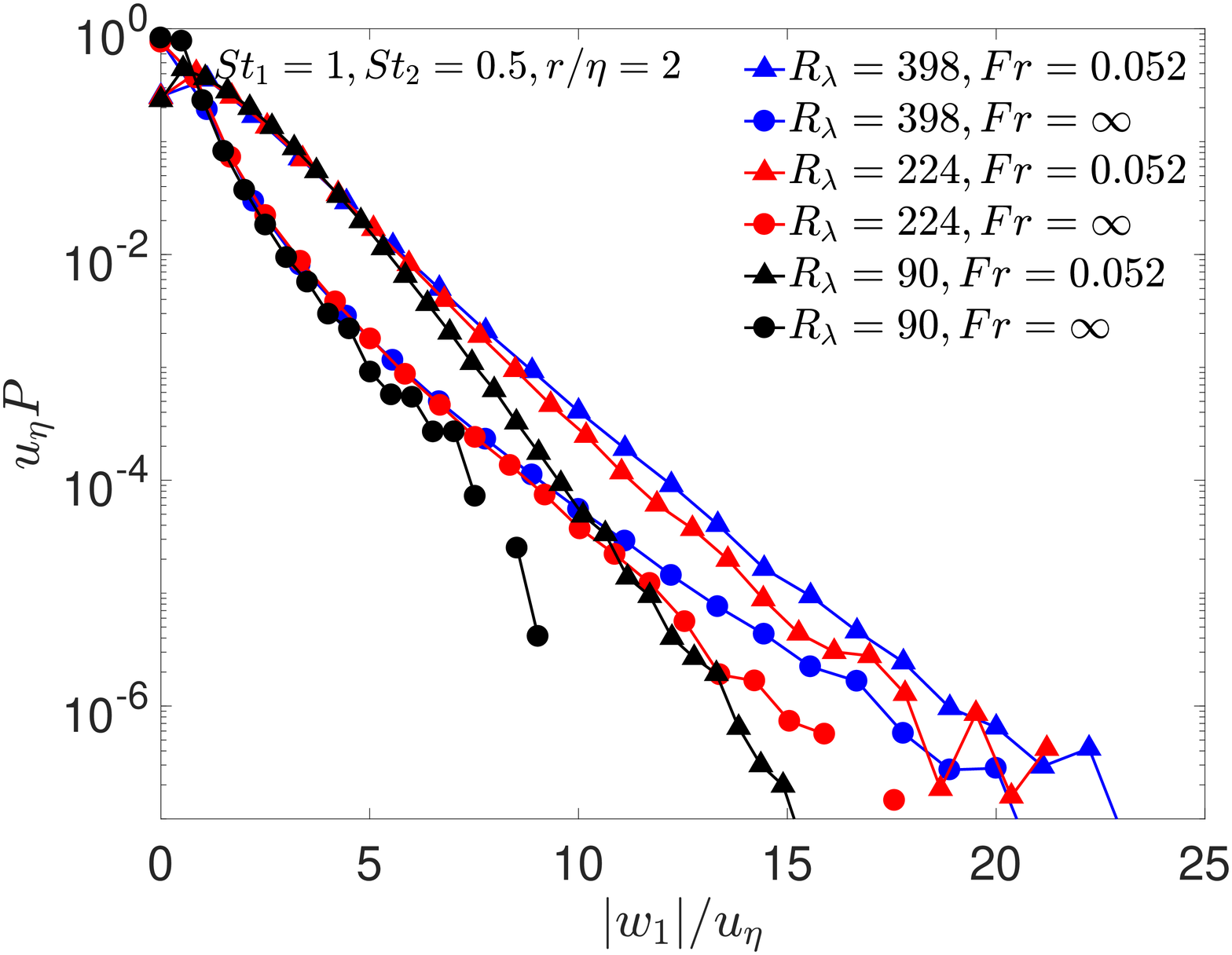}
    \caption{Horizontal }
    \end{subfigure}

      \begin{subfigure}[b]{0.5\linewidth}
    \includegraphics[width=\linewidth]{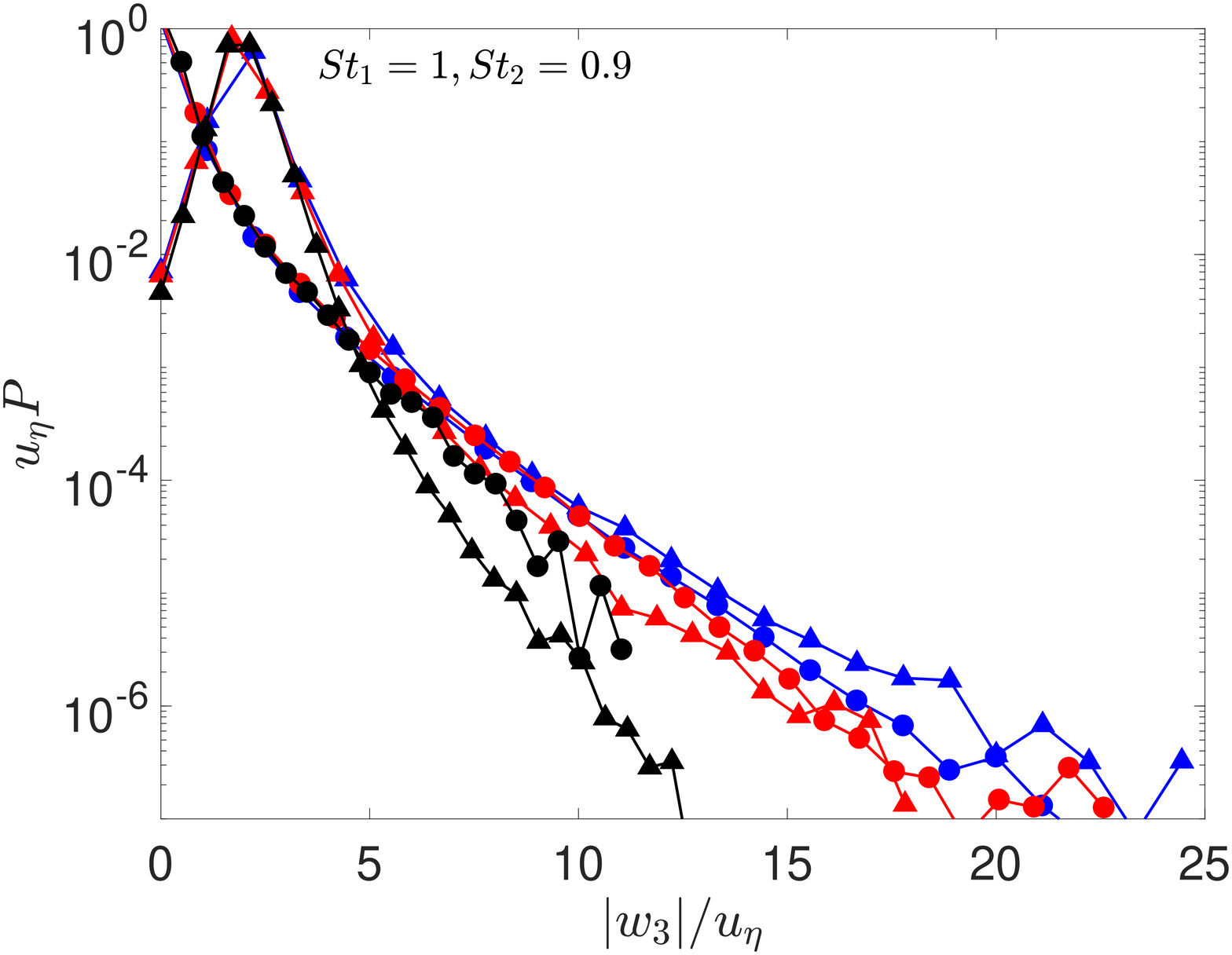}
    \caption{Vertical }
  \end{subfigure}%
    \begin{subfigure}[b]{0.5\linewidth}
    \includegraphics[width=\linewidth]{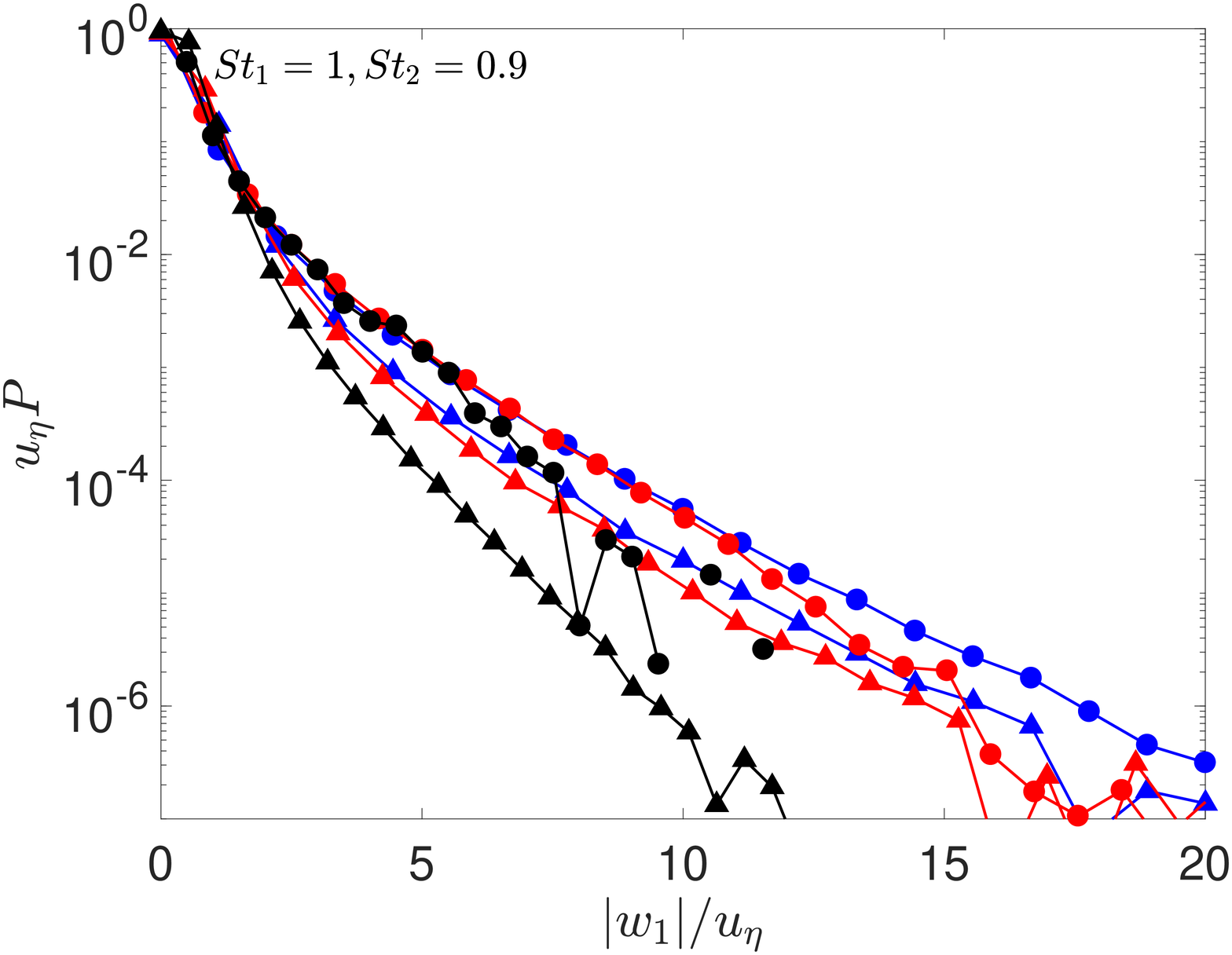}
    \caption{Horizontal }
    \end{subfigure}
      \begin{subfigure}[b]{0.5\linewidth}
    \includegraphics[width=\linewidth]{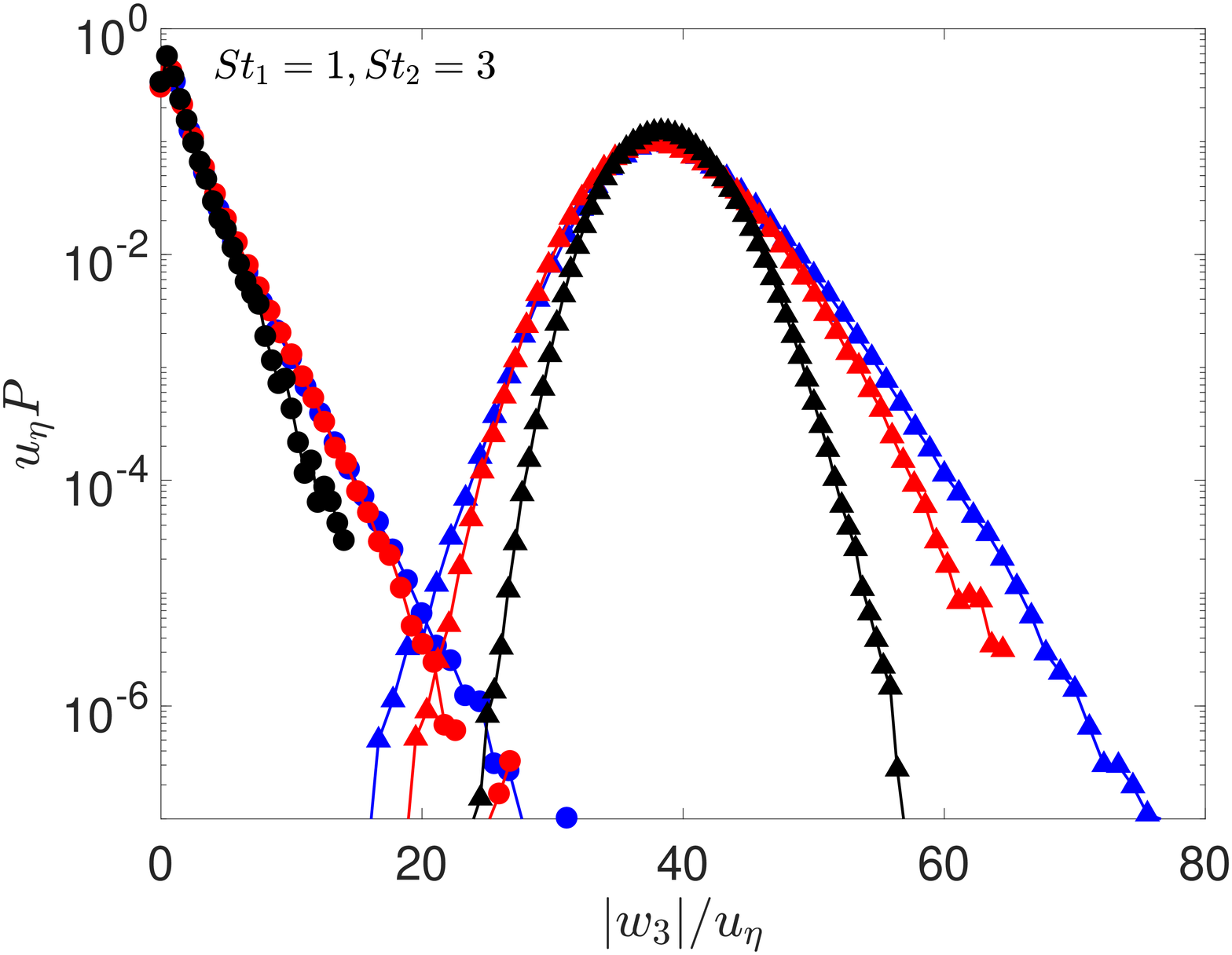}
    \caption{Vertical }
  \end{subfigure}%
    \begin{subfigure}[b]{0.5\linewidth}
    \includegraphics[width=\linewidth]{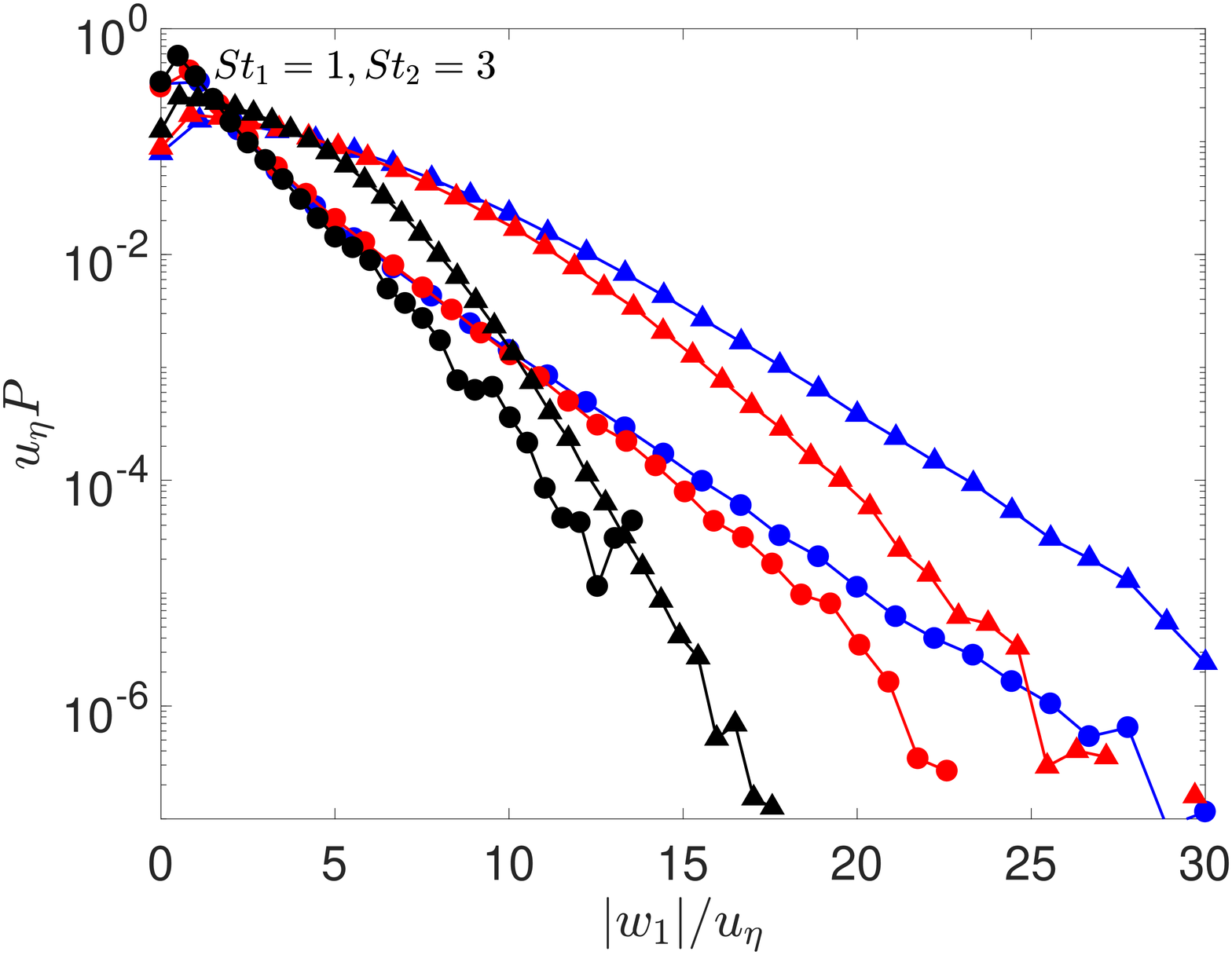}
    \caption{Horizontal }
    \end{subfigure}
  \caption{PDF of (a),(c),(e) vertical, and (b),(d),(f) horizontal relative velocity for $St_1=1$, and different $St_2$, $Fr$ and $R_\lambda$ combinations, and for particles with separation $r \in [0,2]\eta $. Black, red and blue lines correspond to $R_\lambda=90$, $R_\lambda=224$ and $R_\lambda=398$, respectively, and circle and triangle symbols denote $Fr=\infty$ and $Fr=0.052$, respectively.}\label{fig:RelativeVelocity_PDF_very_near_separation}
\end{figure}

\begin{figure}
  \centering
  \begin{subfigure}[b]{0.5\linewidth}
    \includegraphics[width=\linewidth]{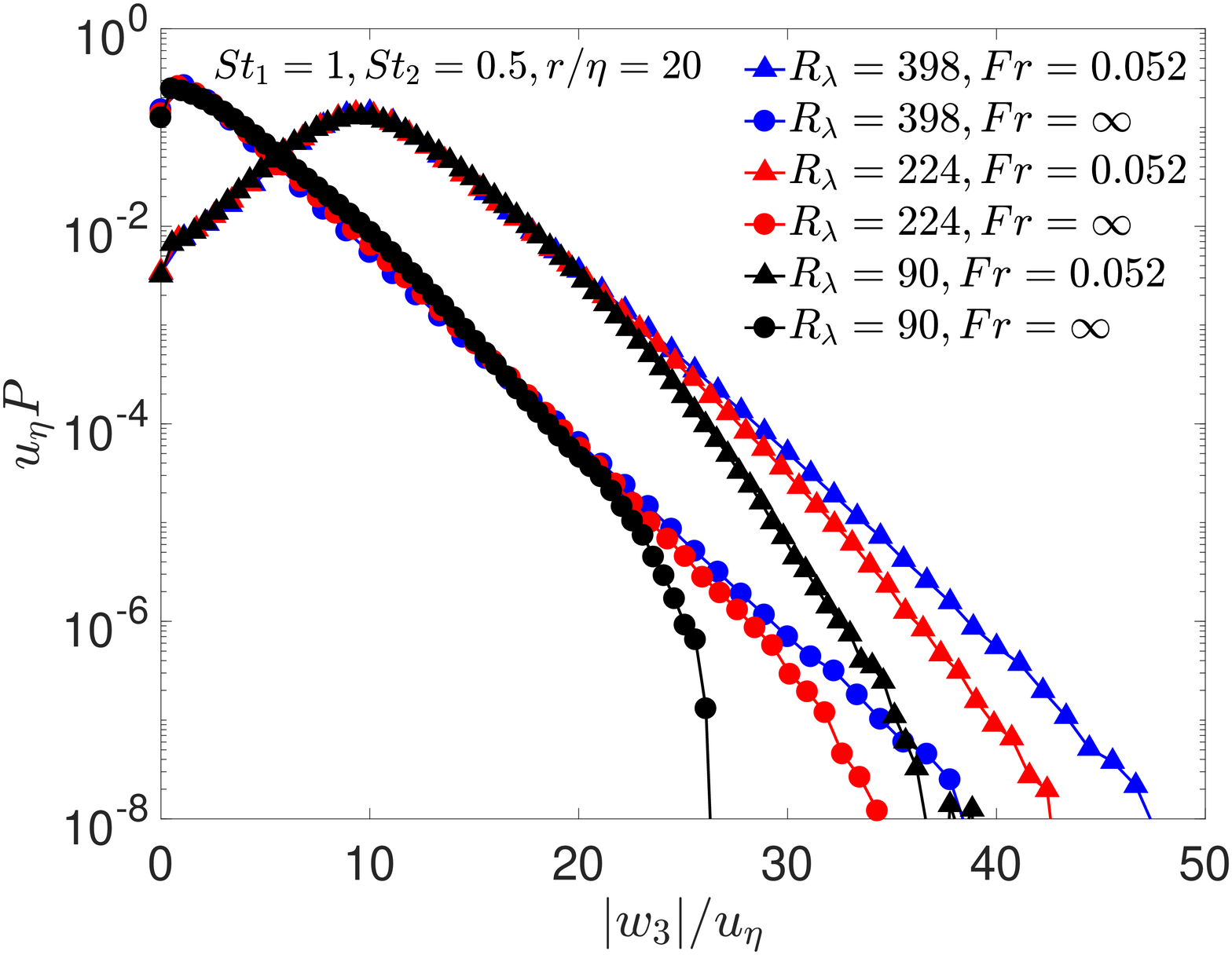}
    \caption{Vertical }
  \end{subfigure}%
    \begin{subfigure}[b]{0.5\linewidth}
    \includegraphics[width=\linewidth]{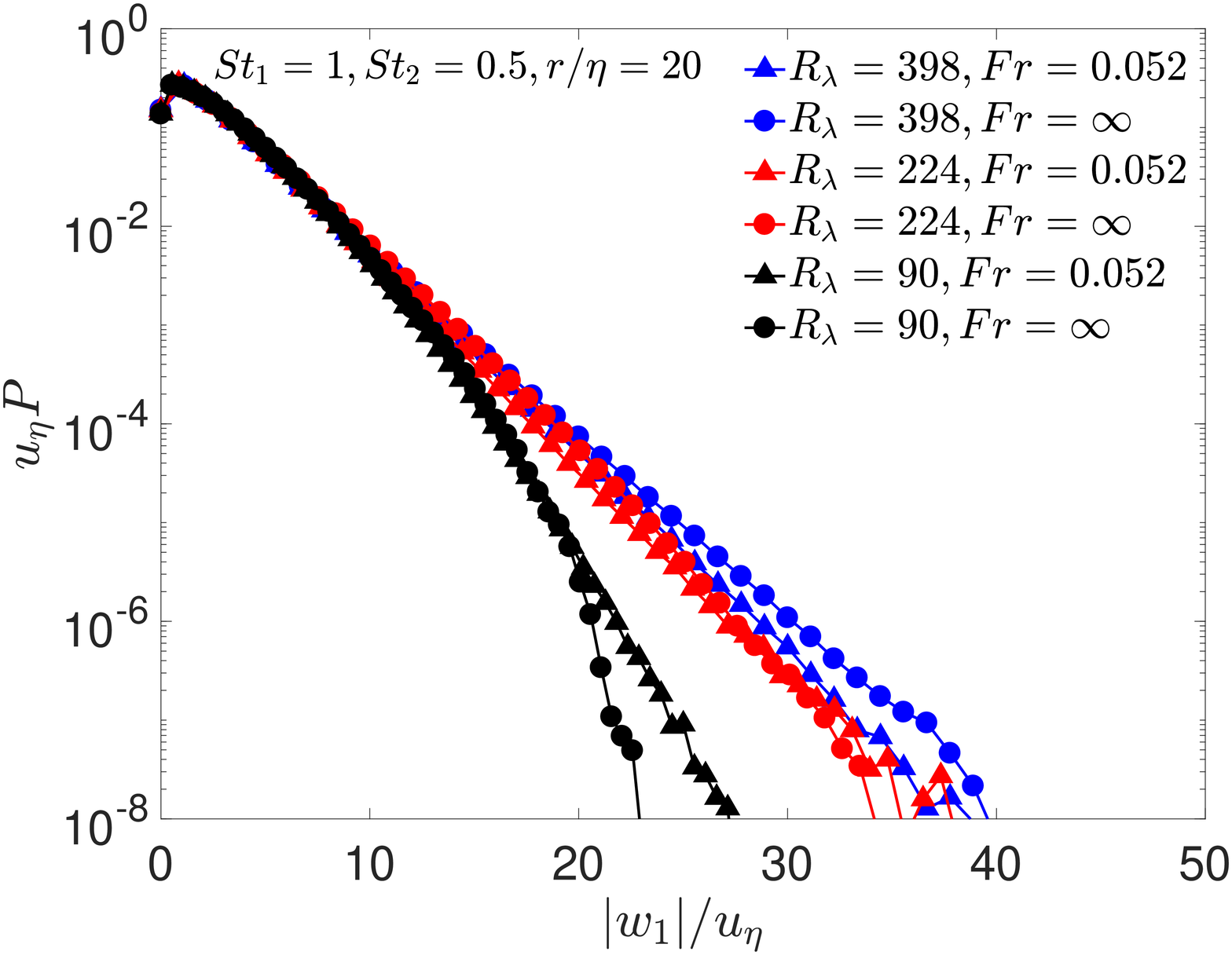}
    \caption{Horizontal }
    \end{subfigure}
      \begin{subfigure}[b]{0.5\linewidth}
    \includegraphics[width=\linewidth]{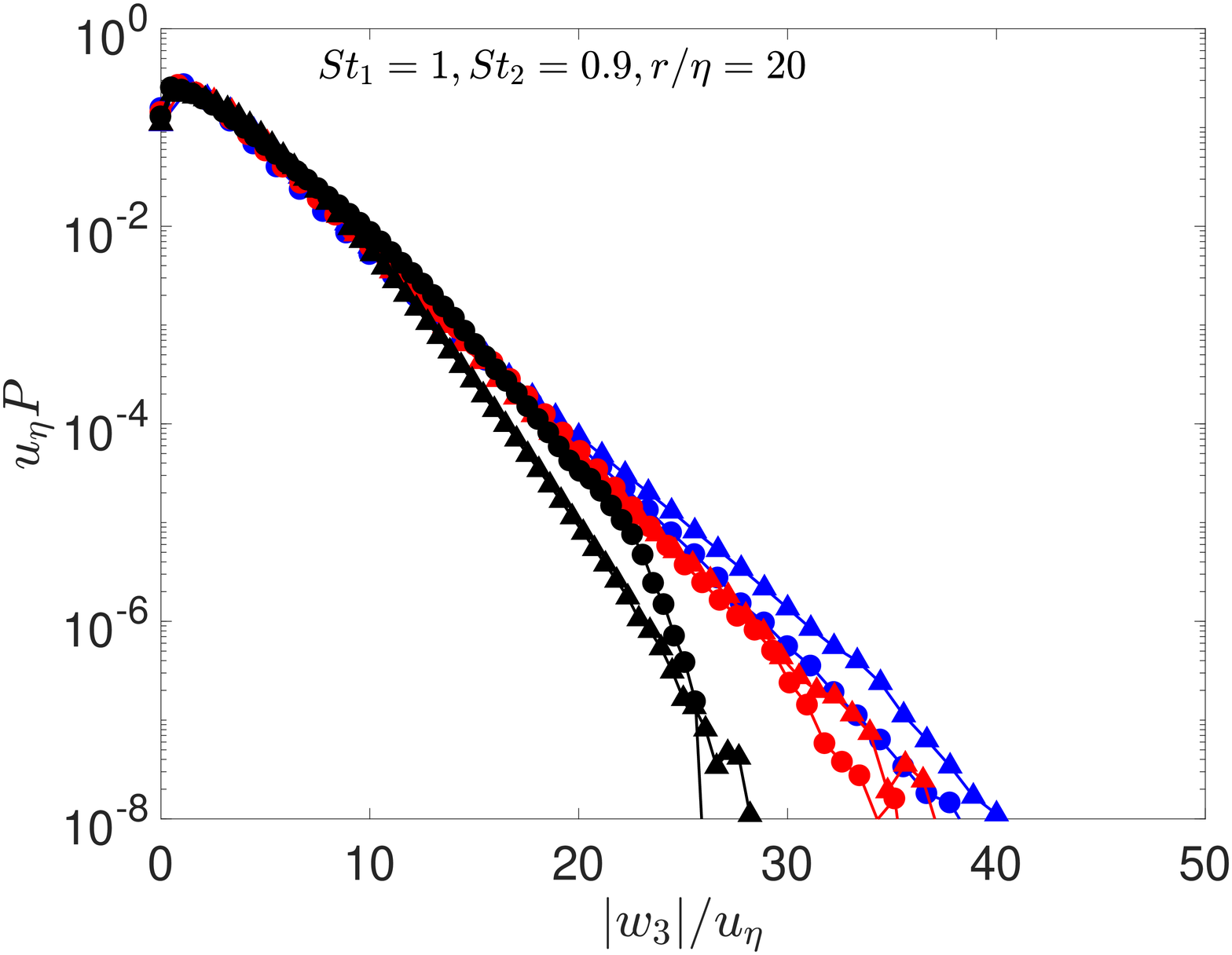}
    \caption{Vertical }
  \end{subfigure}%
    \begin{subfigure}[b]{0.5\linewidth}
    \includegraphics[width=\linewidth]{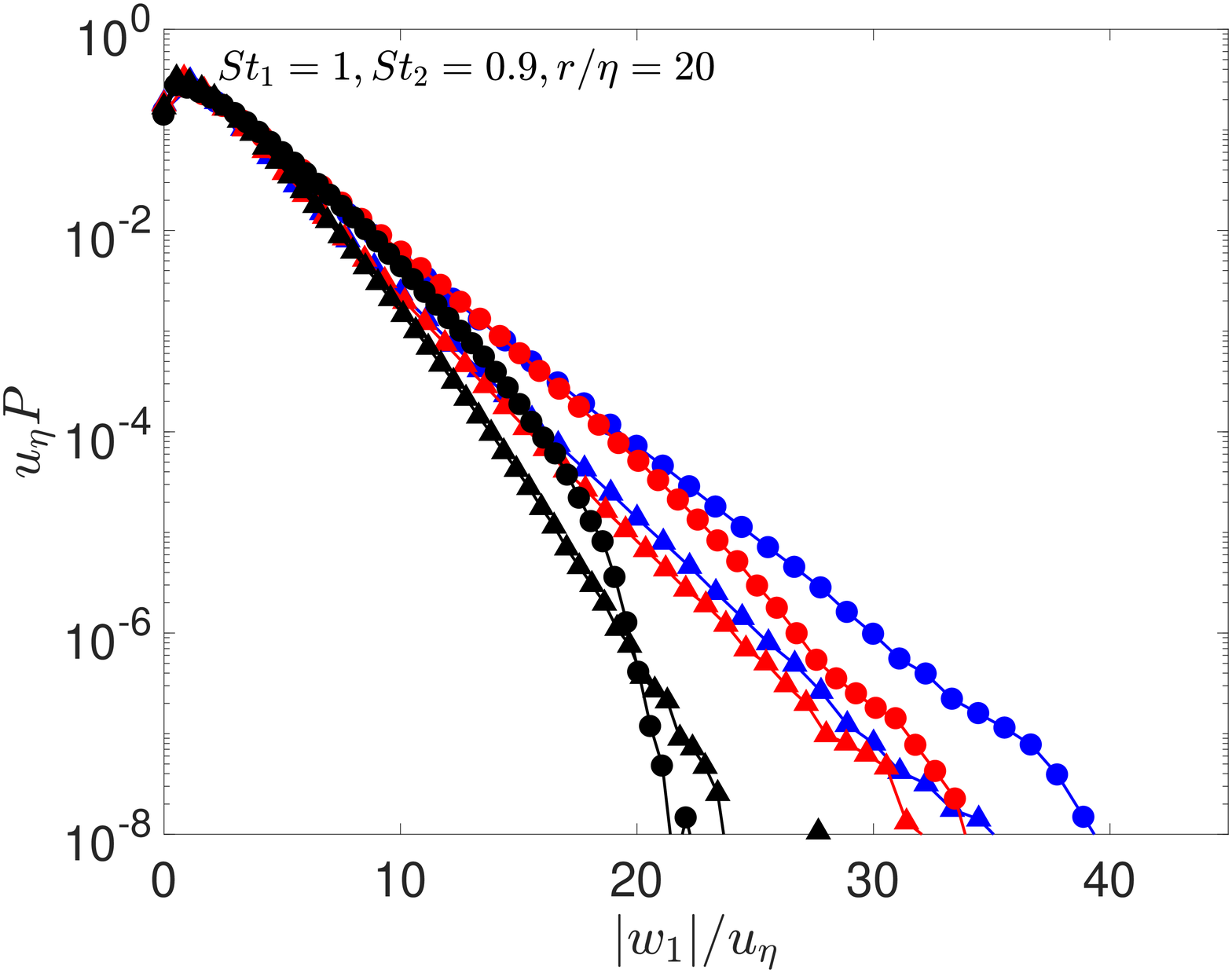}
    \caption{Horizontal }
    \end{subfigure}

      \begin{subfigure}[b]{0.5\linewidth}
    \includegraphics[width=\linewidth]{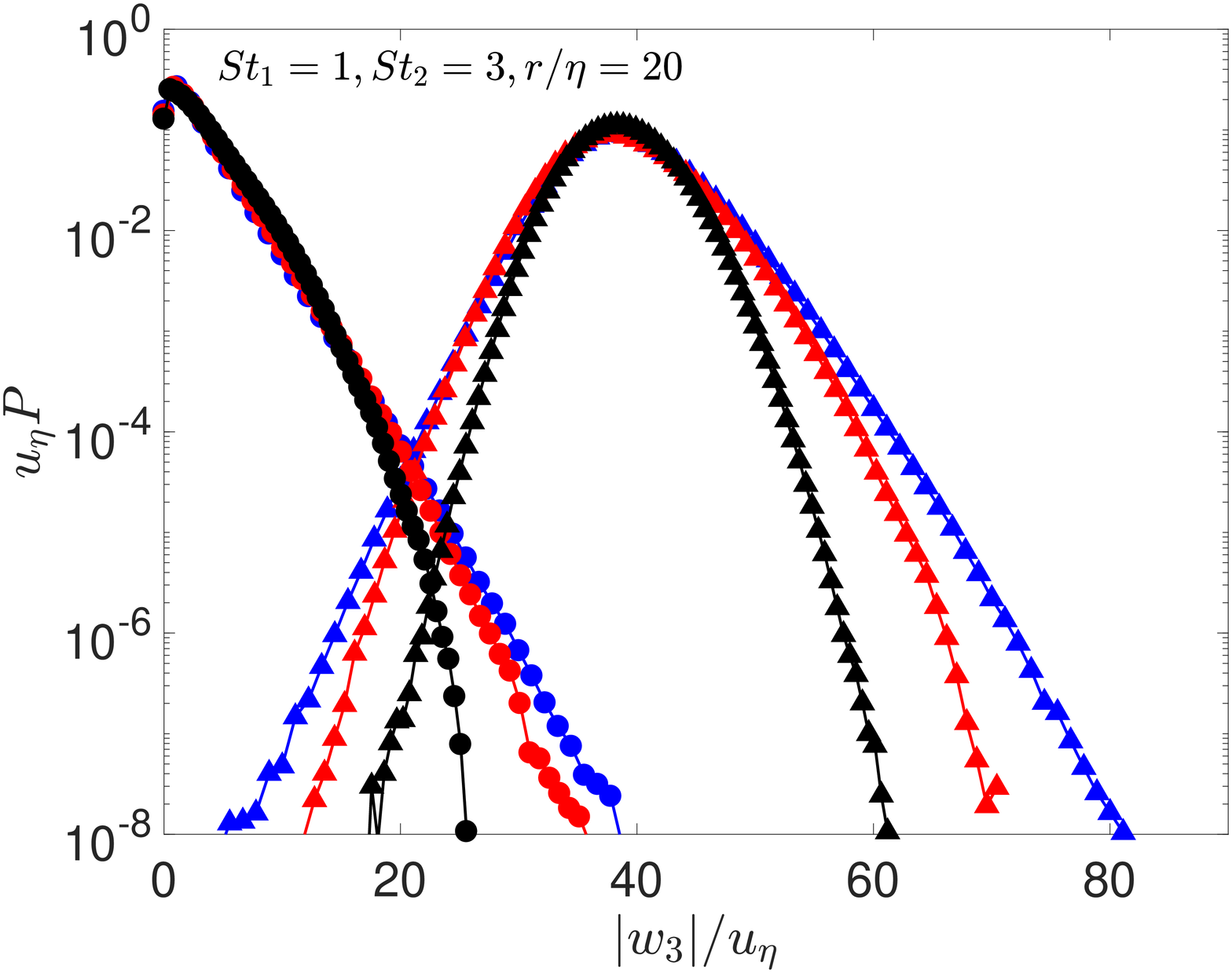}
    \caption{Vertical }
  \end{subfigure}%
    \begin{subfigure}[b]{0.5\linewidth}
    \includegraphics[width=\linewidth]{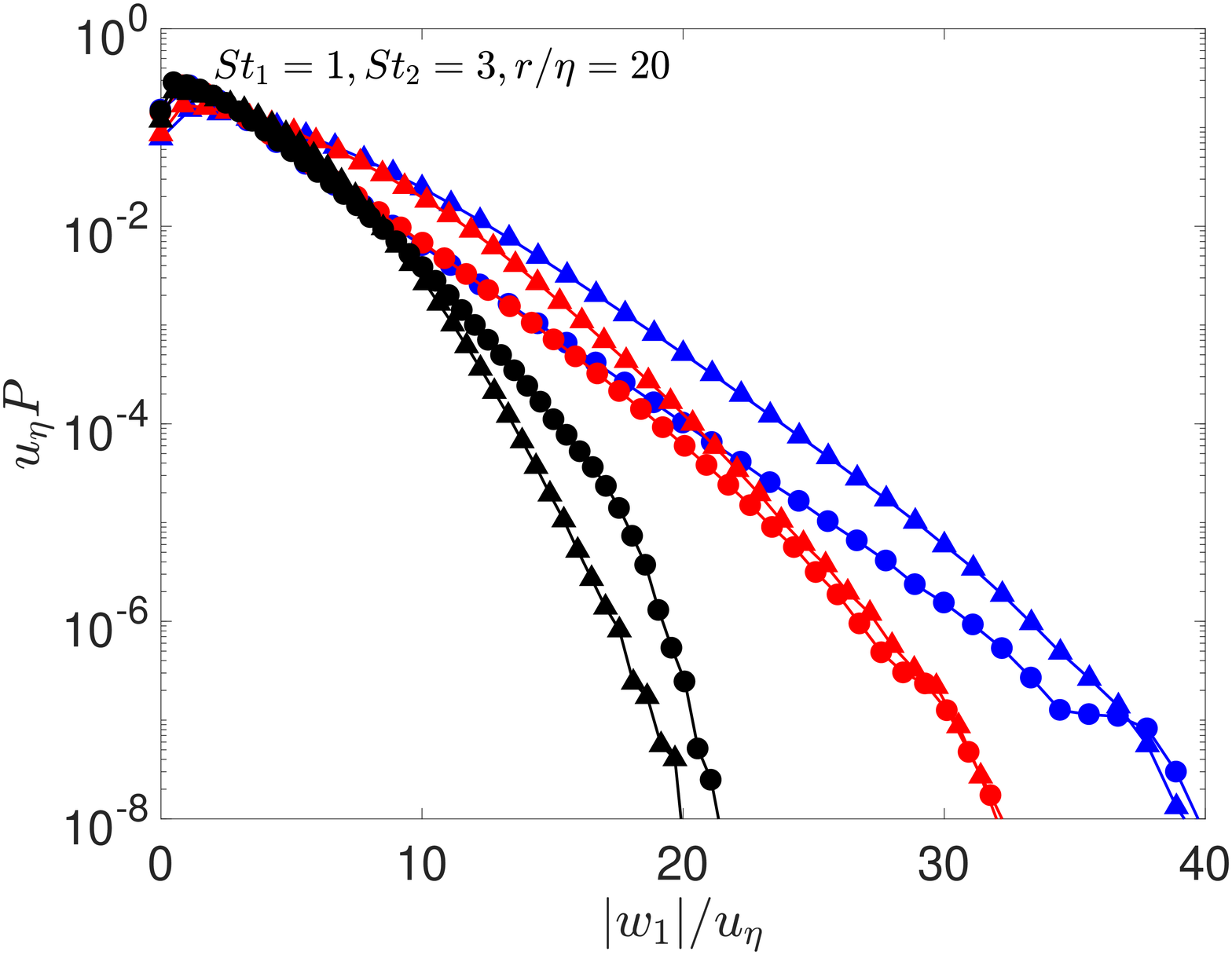}
    \caption{Horizontal }
    \end{subfigure}
  \caption{PDF of (a),(c),(e) vertical, and (b),(d),(f) horizontal relative velocity for $St_1=1$, and different $St_2$, $Fr$ and $R_\lambda$ combinations, and for particles with separation $r \in [18,20]\eta $. Black, red and blue lines correspond to $R_\lambda=90$, $R_\lambda=224$ and $R_\lambda=398$, respectively, and circle and triangle symbols denote $Fr=\infty$ and $Fr=0.052$, respectively.}
  \label{fig:RelativeVelocity_PDF_far_separation}
\end{figure}

We now turn our attention to the PDFs of the vertical and horizontal components of the particle relative velocities. In figures  \ref{fig:RelativeVelocity_PDF_very_near_separation} and \ref{fig:RelativeVelocity_PDF_far_separation}, the values of the Stokes numbers for the particle-pairs are chosen to represent weak ($|\Delta St|=0.1$), moderate ($|\Delta St|=0.5$) and strong bidispersity ($|\Delta St|=2$). Figure \ref{fig:RelativeVelocity_PDF_very_near_separation} shows the PDFs for particles with separation in the dissipation range ($0\le r/\eta \le 2$), and figure \ref{fig:RelativeVelocity_PDF_far_separation}  shows the PDFs for particles with separation in the lower end of the inertial range ($18\le r/\eta \le 20$). 

\FloatBarrier
  \FloatBarrier
An important finding from these results is that, as anticipated in \S\ref{TC}, even when $|\Delta St|/Fr\gg1$, the effect of turbulence on the vertical relative motion of the particles cannot be ignored. Indeed, when gravity dominates the vertical relative velocities (i.e. for $|\Delta St|/Fr\to\infty$), the PDF of the vertical relative velocity is a delta function centered on $u_\eta|\Delta St|/Fr$. However, the results in figure  \ref{fig:RelativeVelocity_PDF_very_near_separation}(e) show that even when $|\Delta St|/Fr\approx 39$ (i.e. $\gg 1$), the PDF is far from such a delta function, and that departures of the PDF from a delta function are becoming stronger as $R_\lambda$ is increased. This is despite the fact that a standard scaling analysis of the equation of relative motion suggests that the effect of turbulence should be negligible compared to the differential sedimentation velocity when $|\Delta St|/Fr\gg1$. In general, arguments based on scaling analysis of the equations can be misleading in turbulent flows as they do not accurately characterize the behavior of the system during large fluctuations about the mean-field behavior. Our results imply that extremely large values of $|\Delta St|/Fr$ would be required to observe a regime where the vertical relative motion of the particles is completely dominated by gravity. This has important implications for modeling since in many applications, $|\Delta St|/Fr $ may never be large enough to fully neglect the role of turbulent fluctuations on their motion. However, our results do show that the mode of the vertical relative velocity PDFs are close to the gravity-dominated prediction $u_\eta|\Delta St|/Fr$ for each of the values of $R_\lambda$ considered, as was also observed in \cite{dhariwal2018small} for $R_\lambda\approx 90$. Therefore, if only low-order moments of the relative velocities need to be predicted, the effects of turbulence could be ignored when $|\Delta St|/Fr\gg1$.

The results in figures  \ref{fig:RelativeVelocity_PDF_very_near_separation} and \ref{fig:RelativeVelocity_PDF_far_separation} also show that the effect of $R_\lambda$ is generally stronger as $Fr$ is decreased, similar to what was observed with the accelerations. This is largely due to the fact that at these separations, the particle acceleration contribution to $\boldsymbol{w}^p(t)$ is stronger than that associated with $\Delta\boldsymbol{u}$, and as explained earlier, gravity can enhance the dependency of the accelerations on $R_\lambda$ as it increases the range of scales affecting the particle accelerations.

\begin{figure}
  \centering
  \begin{subfigure}[b]{0.5\linewidth}
    \includegraphics[width=\linewidth]{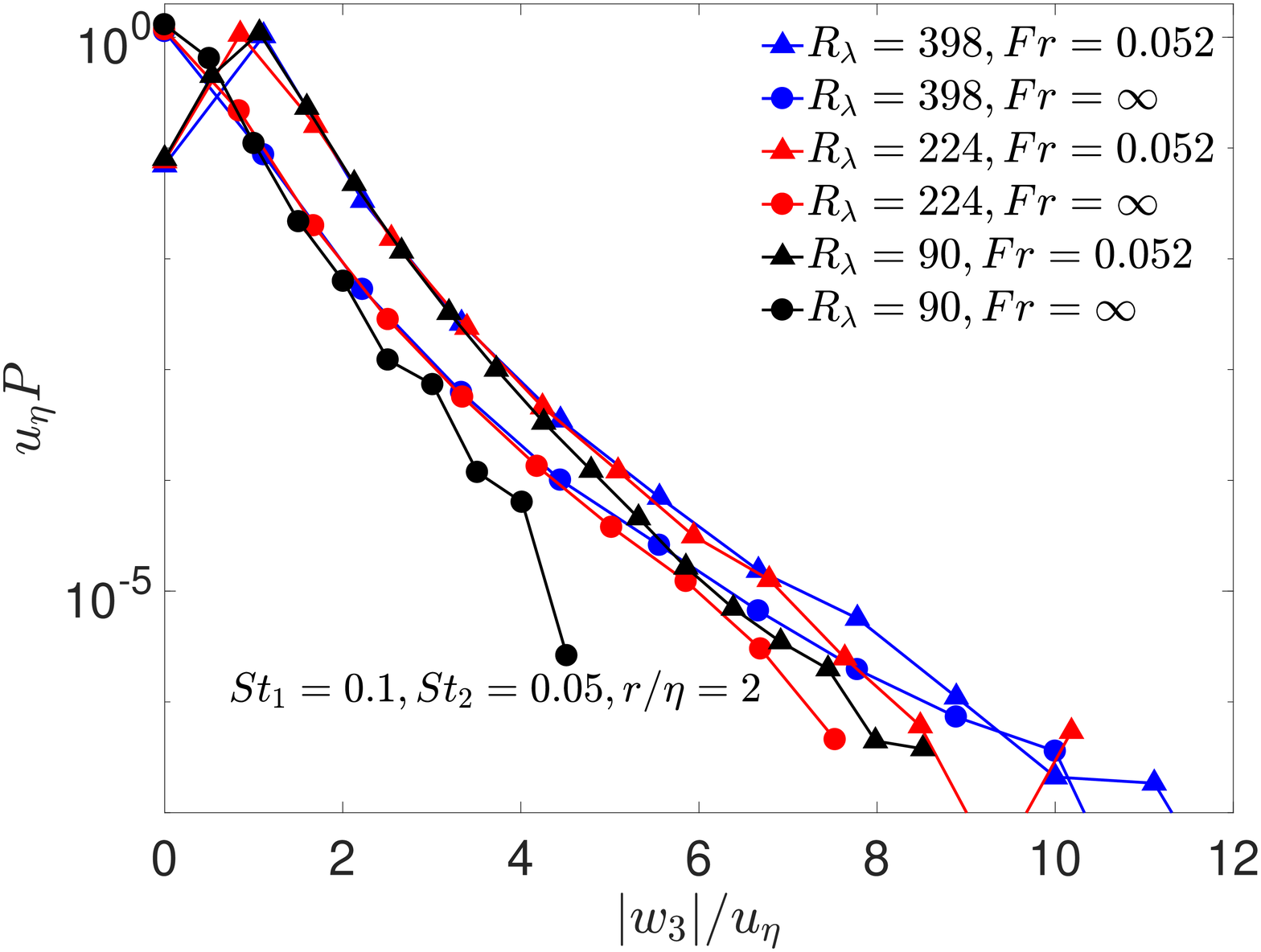}
    \caption{Vertical }
  \end{subfigure}%
    \begin{subfigure}[b]{0.5\linewidth}
    \includegraphics[width=\linewidth]{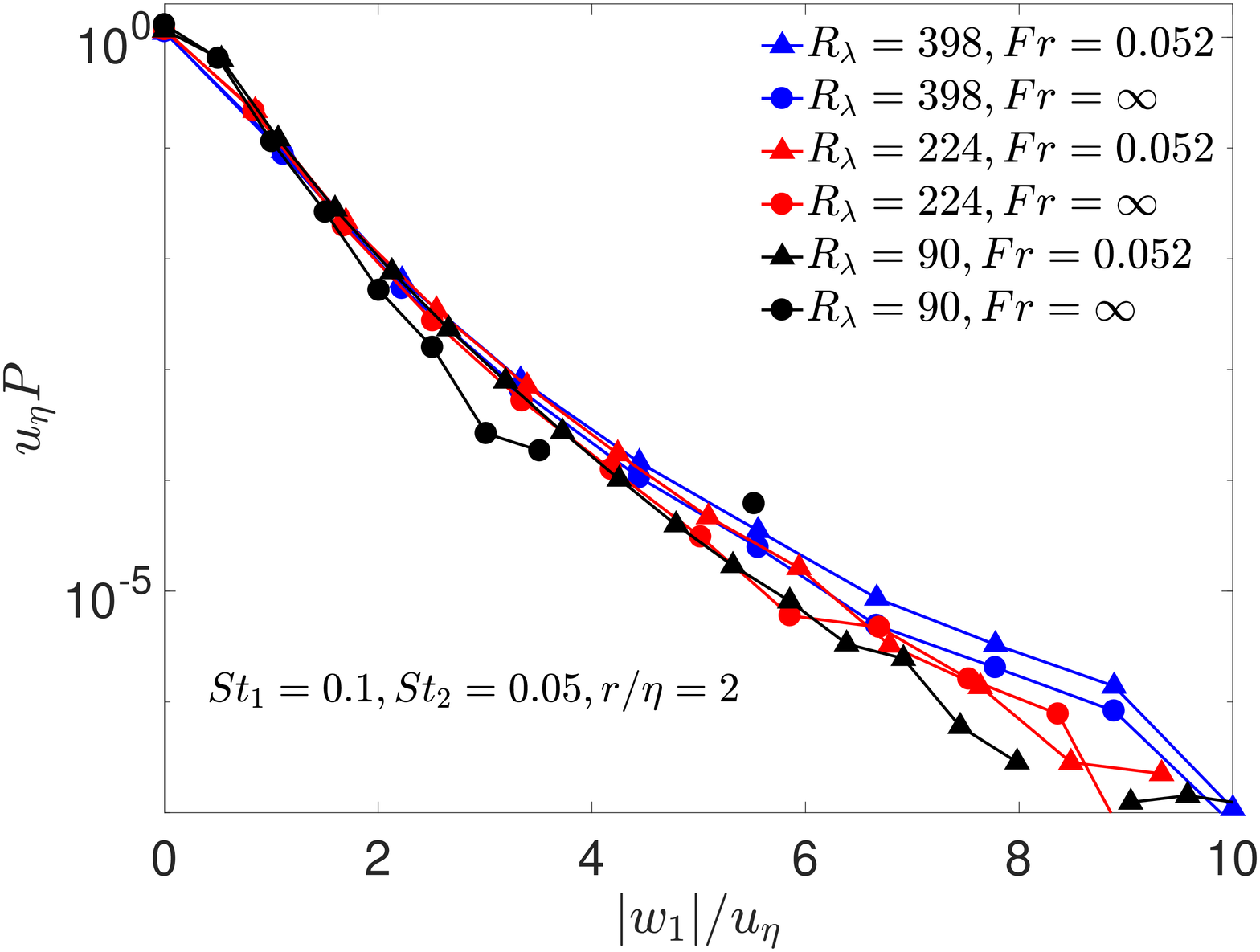}
    \caption{Horizontal }
    \end{subfigure}
      \begin{subfigure}[b]{0.5\linewidth}
    \includegraphics[width=\linewidth]{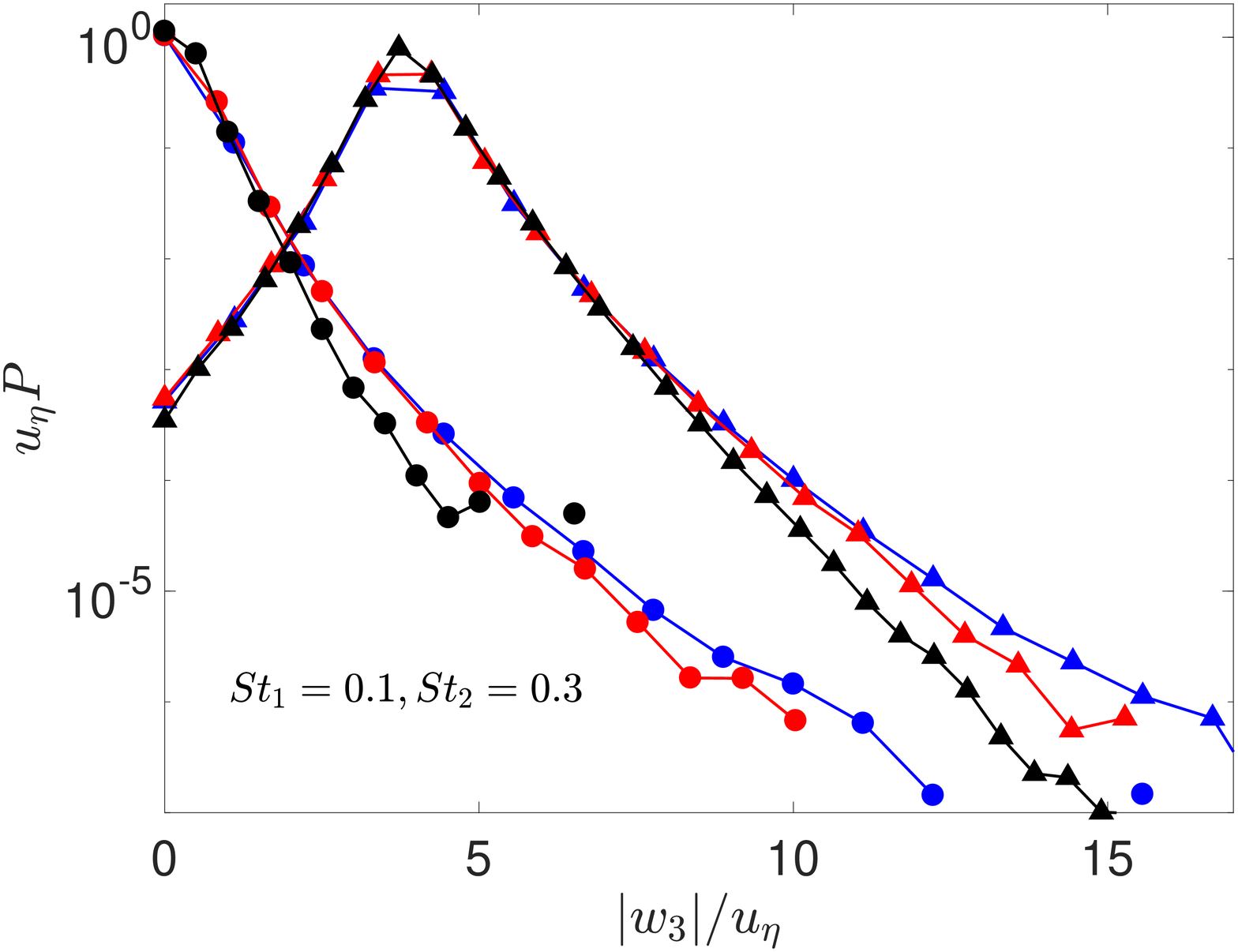}
    \caption{Vertical }
  \end{subfigure}%
    \begin{subfigure}[b]{0.5\linewidth}
    \includegraphics[width=\linewidth]{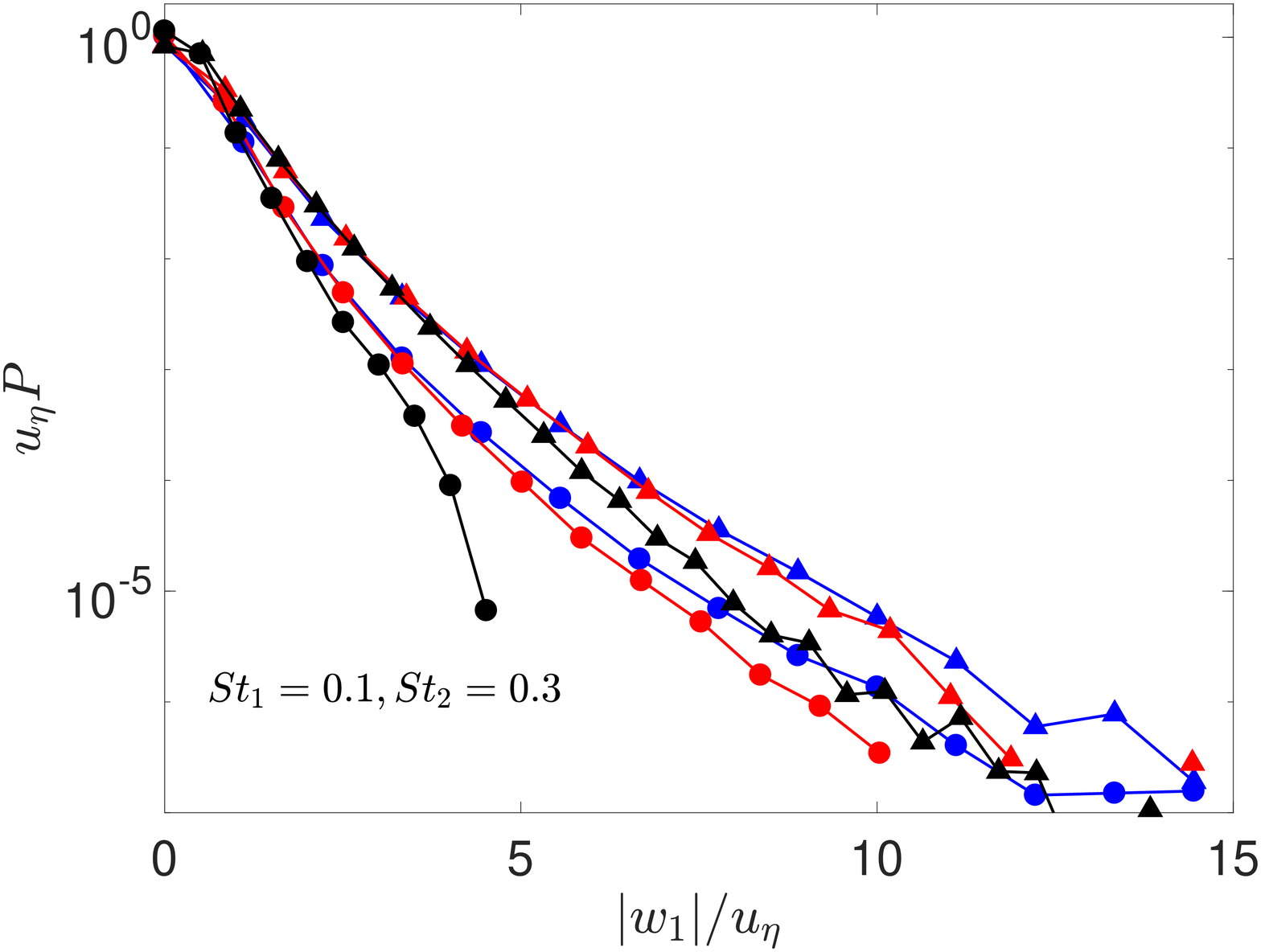}
    \caption{Horizontal }
    \end{subfigure}
  \caption{PDF of (a),(c),(e) vertical, and (b),(d),(f) horizontal relative velocity for $St_1=0.1$, and different $St_2$, $Fr$ and $R_\lambda$ combinations, and for particles with separation $r \in [0,2]\eta $. Black, red and blue lines correspond to $R_\lambda=90$, $R_\lambda=224$ and $R_\lambda=398$, respectively, and circle and triangle symbols denote $Fr=\infty$ and $Fr=0.052$, respectively.}\label{fig:LowSt_RelativeVelocity_PDF_very_near_separation}
\end{figure}

In figure \ref{fig:LowSt_RelativeVelocity_PDF_very_near_separation} we show results for the relative velocities of bidisperse particles with weak inertia. For $St_1=0.1, St_2=0.05$, the vertical velocities are slightly enhanced by gravity, while the horizontal velocities are not affected by $Fr$. However, for $St_1=0.1, St_2=0.3$, the effect of $Fr$ is appreciable, with gravity noticably enhancing both the vertical and horizontal velocities. 

\begin{figure}
  \centering
  \begin{subfigure}[b]{0.5\linewidth}
    \includegraphics[width=\linewidth]{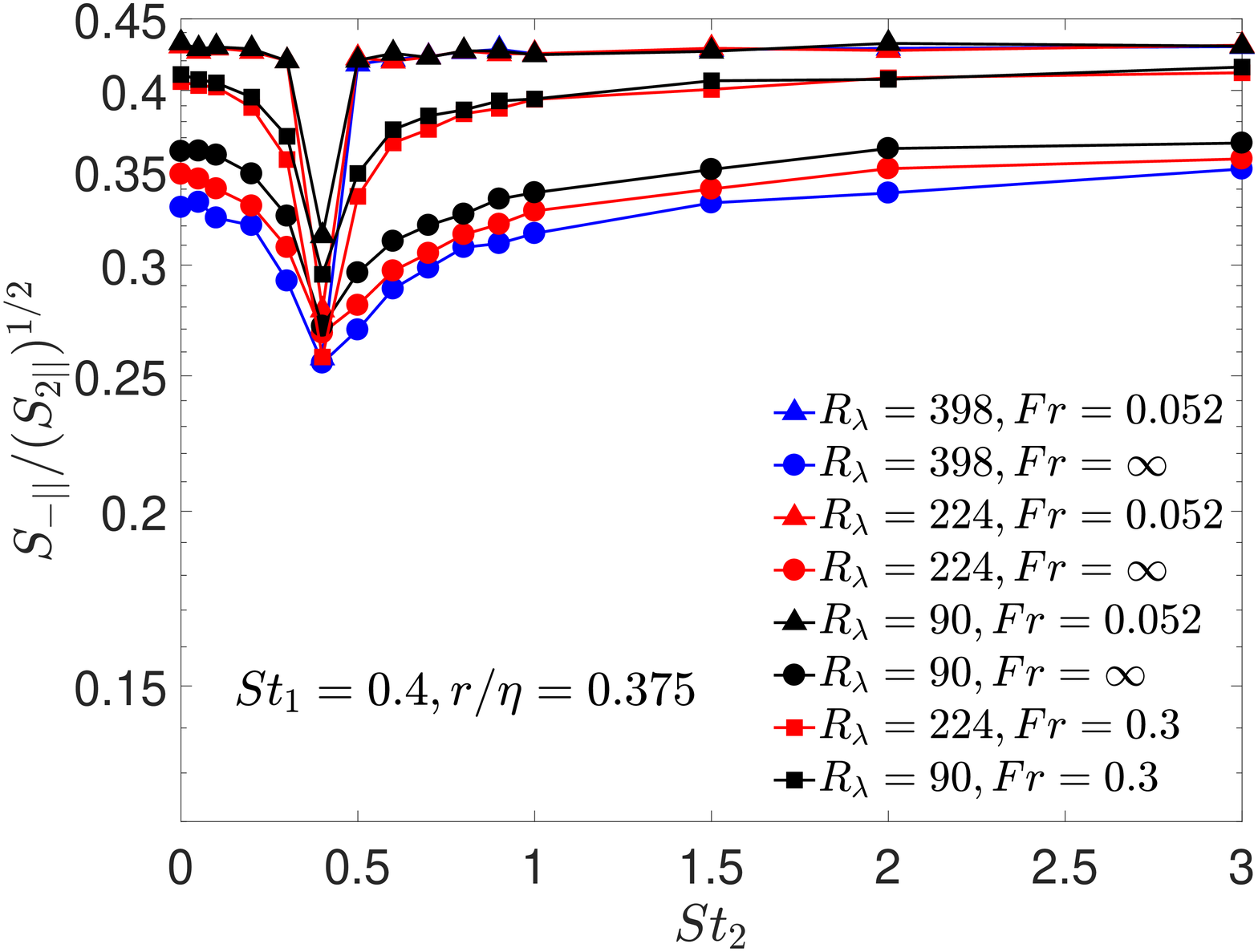}
    \caption{}
  \end{subfigure}%
    \begin{subfigure}[b]{0.5\linewidth}
    \includegraphics[width=\linewidth]{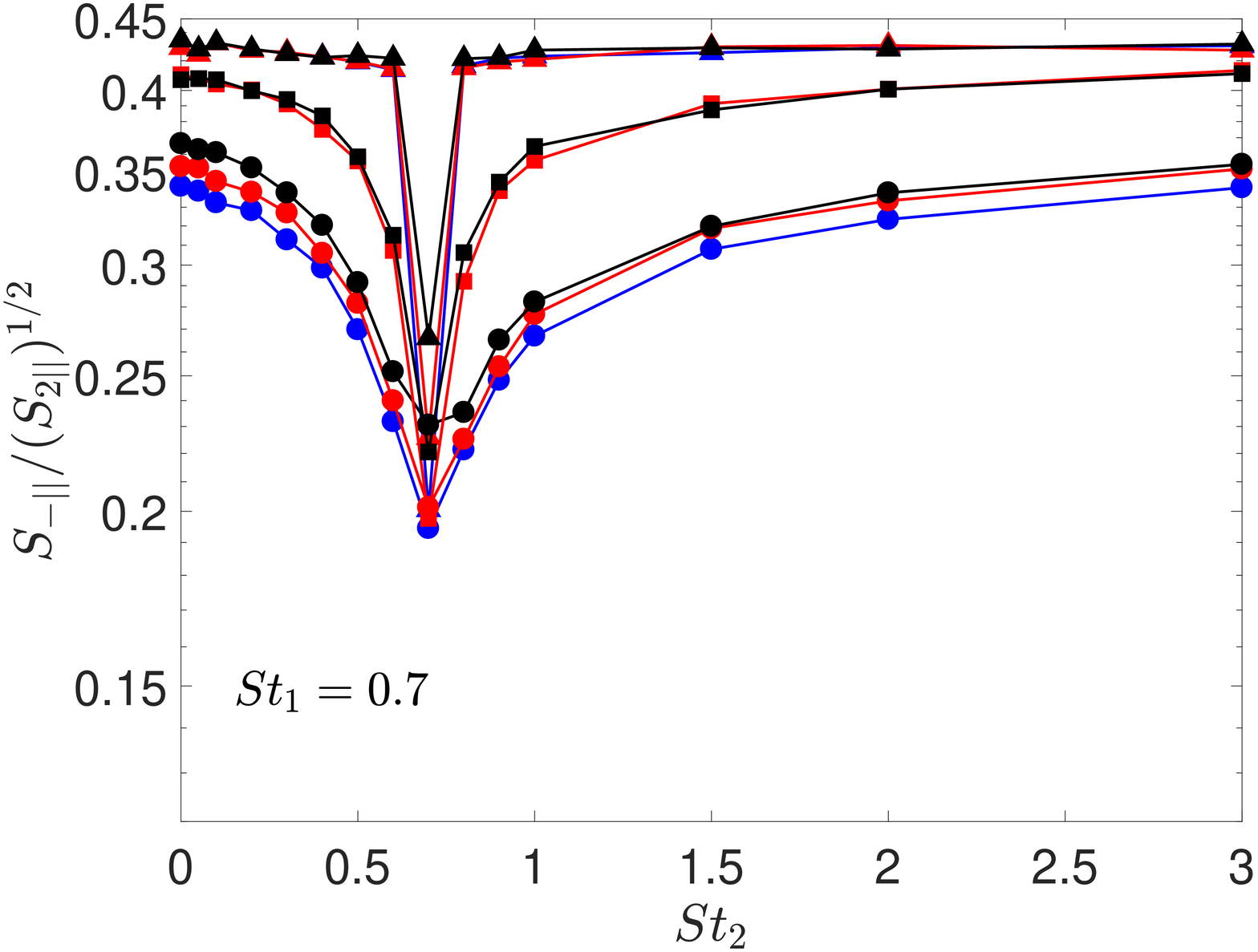}
    \caption{}
  \end{subfigure}
    \begin{subfigure}[b]{0.5\linewidth}
    \includegraphics[width=\linewidth]{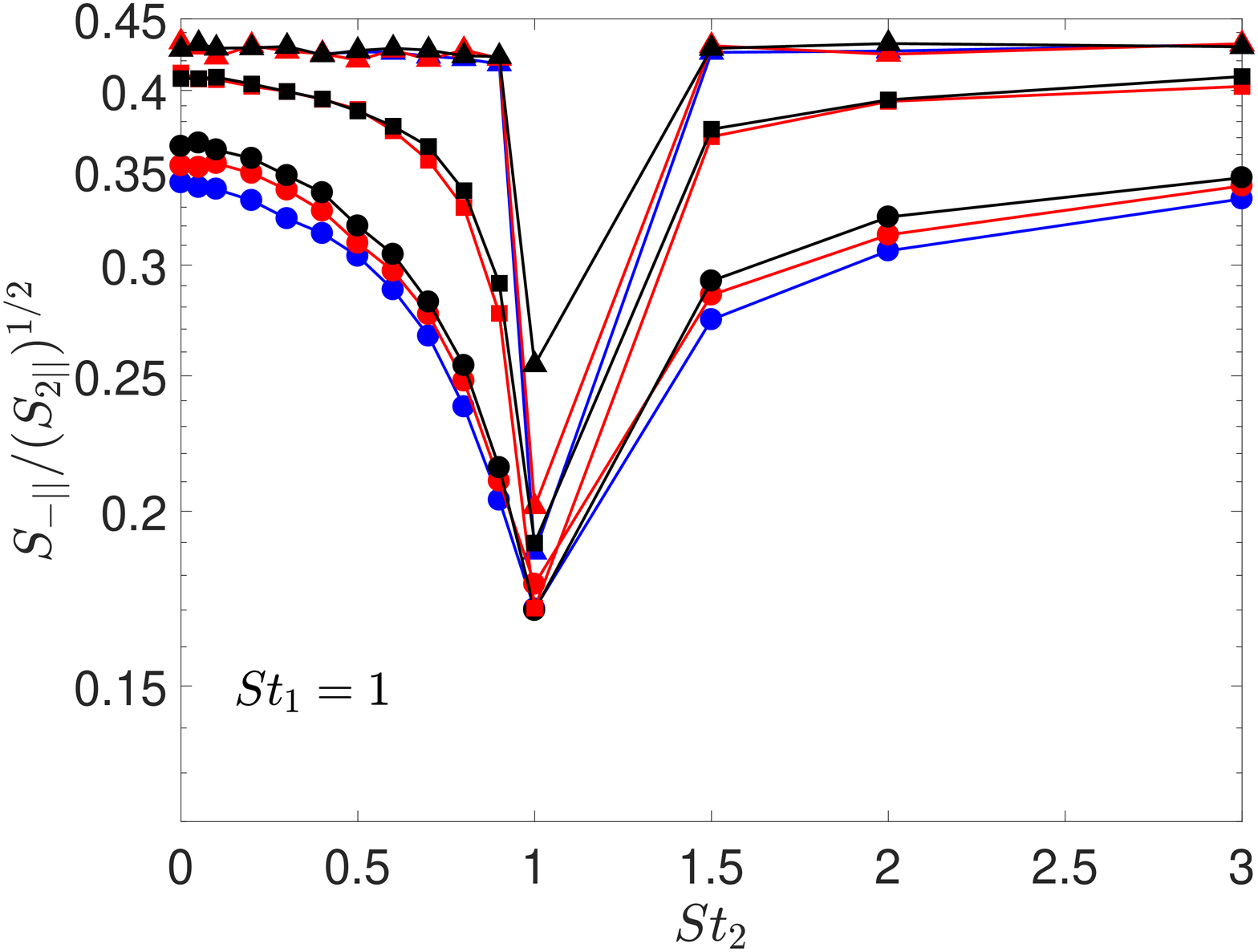}
    \caption{}
  \end{subfigure}%
    \begin{subfigure}[b]{0.5\linewidth}
    \includegraphics[width=\linewidth]{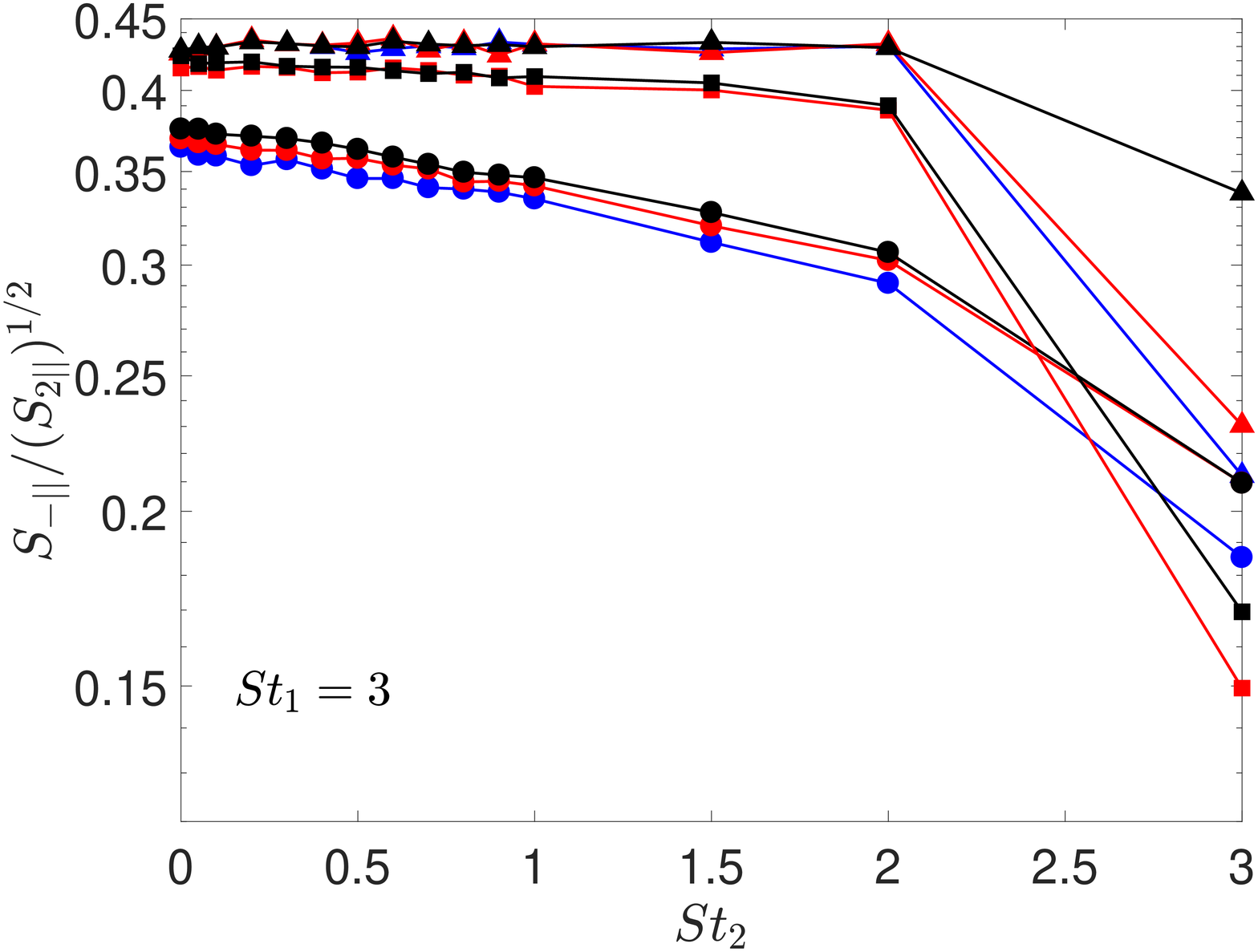}
    \caption{}
  \end{subfigure}
  \caption{Ratio between mean inward relative velocities and the standard deviation \protect\\
of the longitudinal relative velocities as a function of $St_2$, at $r/\eta=0.375$ for different $St_1$, $Fr$ and $R_\lambda$ combinations. Black, red and blue lines correspond to $R_\lambda=90$, $R_\lambda=224$ and $R_\lambda=398$, respectively and circle, square and triangle symbols denote $Fr=\infty$, $Fr=0.3$ and $Fr=0.052$, respectively.}\label{fig:Ratio_MeanInward_T_SD_RV}
\end{figure}

In order to consider the effect of $Fr$ and $R_\lambda$ on the shape of the relative velocity PDFs, in figure \ref{fig:Ratio_MeanInward_T_SD_RV} we plot the ratio $S_{-\parallel}^p/(S_{2\parallel}^p)^{1/2}$ corresponding to the ratio of mean inward longitudinal relative velocity to the square root of second-order longitudinal relative velocity structure function. These quantities are defined as $S_{-\parallel}^p\equiv\langle w^p_\parallel(t)\vert<0\rangle_r$,  $S^p_{2\parallel}(r)\equiv\langle w^p_\parallel(t)w^p_\parallel(t)\rangle_r$, where $w^p_\parallel(t)\equiv \|\boldsymbol{r}^p(t)\|^{-1}\boldsymbol{r}^p(t)\boldsymbol{\cdot}\boldsymbol{w}^p(t)$, $\langle\cdot\rangle_r$ denotes an ensemble average conditioned on $r=\|\boldsymbol{r}^p(t)\|$, and $\vert<0$ denotes that only negative values of $w^p_\parallel(t)$ contribute to the average. While other quantifies of the PDF shape could be used, we choose this measure of the PDF shape since it is of interest to the problem of particle collisions, to which we will turn our attention in the next section. The results are plotted for $St_1$ as a function of $St_2$ at three different values of $Fr$ and $R_\lambda$. For a Gaussian PDF, $S_{-\parallel}^p/(S_{2\parallel}^p)^{1/2}=1/\sqrt{2\pi}\approx0.4 $, and the results in figure \ref{fig:Ratio_MeanInward_T_SD_RV} show that the departures from the Gaussian limit are strongest for $St=O(1)$, $|\Delta St|\ll1$, and $Fr=\infty$. The dip in the curves that occurs in the monodisperse limit $|\Delta St|\to0$ show how sensitive the relative velocities are to bidispersity, and this sensitivity is enhanced as $Fr$ is decreased. This occurs because the differential sedimentation term $|\Delta St|/Fr$, though identically zero for monodisperse particles,  quickly becomes large as $|\Delta St|$ is increased if $Fr\ll1$. This emphasizes the importance of accounting for bidispersity when describing the small-scale dynamics of settling inertial particles in turbulence, such as in clouds, even if $|\Delta St|\ll1$.

\FloatBarrier
Our results also show that the departures from Gaussianity of the relative velocities become stronger as $R_\lambda$ increases, as may be expected. However, as $Fr$ is decreased, the PDFs become increasingly Gaussian. This is important for models of particle collisions in turbulence, since most of these model $S_{2\parallel}$, and then from this recover a model for the mean collision velocity $S_{-\parallel}^p$ by assuming the Gaussian relationship $S_{-\parallel}^p=\sqrt{S_{2\parallel}^p/2\pi}$ \citep[e.g., see][]{bragg14c}. Our results imply this is reasonable for $Fr\leq 0.3$ and $|\Delta St| \geq O(1)$.
  \FloatBarrier
\subsection{Particle Collisions}
We now turn our attention to the quantities that are important for particle collisions in turbulence, specifically, the RDF $g(r)$ which quantifies the spatial clustering of the particles, the collision velocity $S_{-\parallel}^p(r)$, and the collision kernel $K(r)$ \citep{sundaram4}. The collision kernel is given by $K(d)\equiv 4\pi d^2 g(d)S_{-\parallel}^p(d)$, where $d \equiv (d_1+d_2)/2$ is the collision diameter of two spherical particles with
diameters $d_1$ and $d_2$ \citep{sundaram4}. While the results so far show that $R_\lambda$ can have a strong effect on the relative motion of settling bidisperse particles, it is possible that the low-order moments characterizing the mean particle collision rates are not so sensitive to $R_\lambda$. This was found to be the case in \cite{ireland2016effectb} for settling monodisperse particles, and we now explore the bidisperse case.

\begin{figure}
  \centering
  \begin{subfigure}[b]{0.5\linewidth}
    \includegraphics[width=\linewidth]{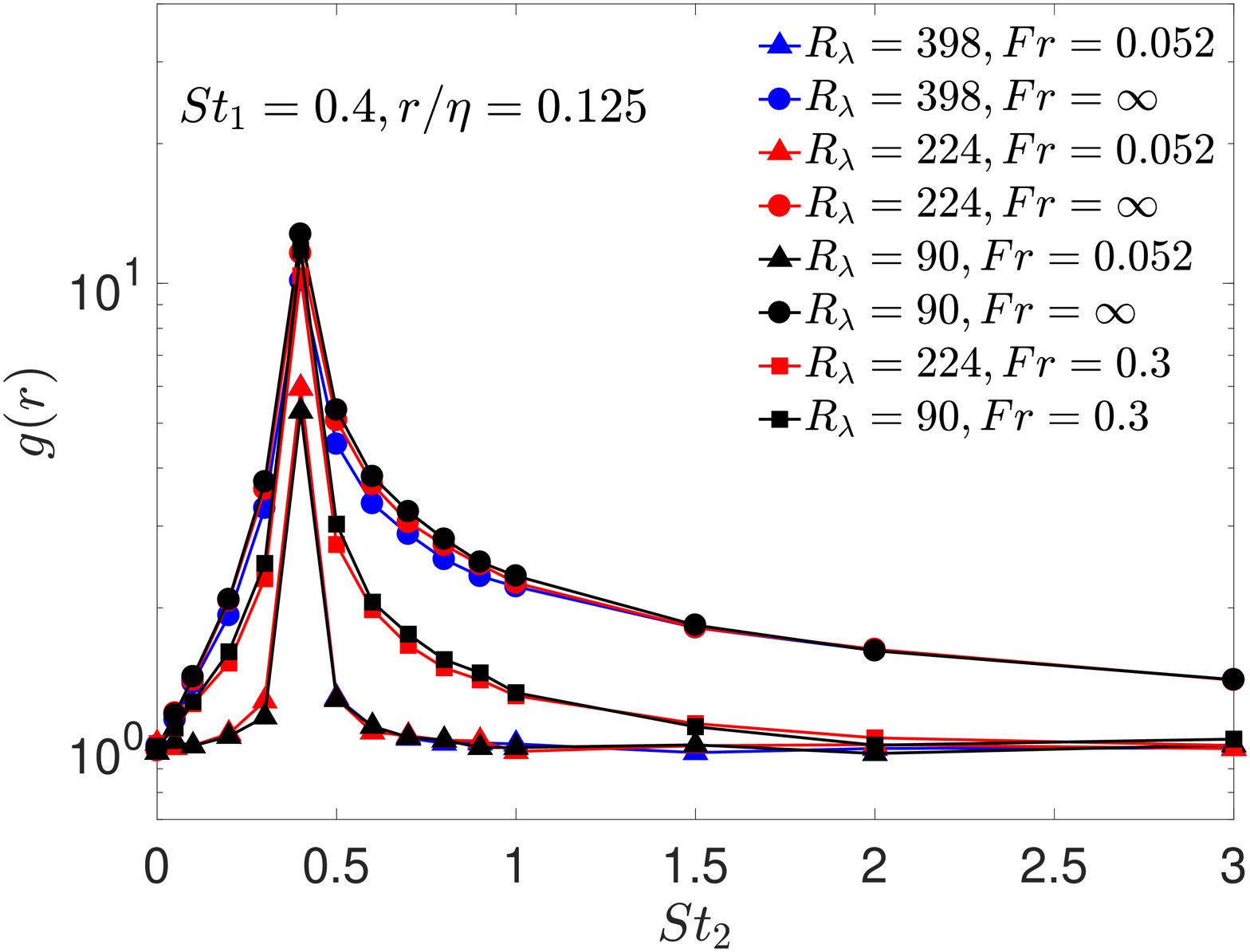}
    \caption{}
  \end{subfigure}%
    \begin{subfigure}[b]{0.5\linewidth}
    \includegraphics[width=\linewidth]{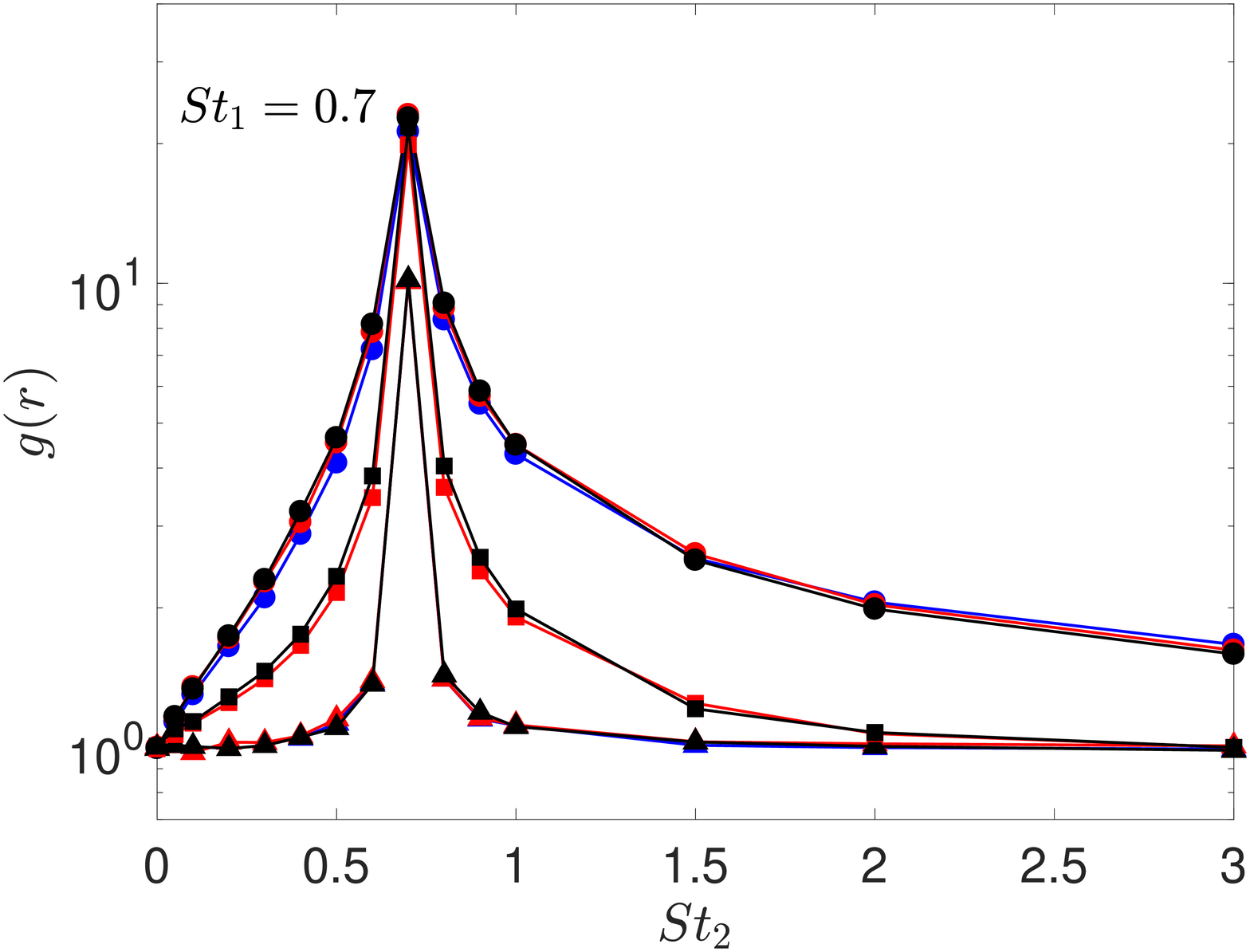}
    \caption{}
  \end{subfigure}
    \begin{subfigure}[b]{0.5\linewidth}
    \includegraphics[width=\linewidth]{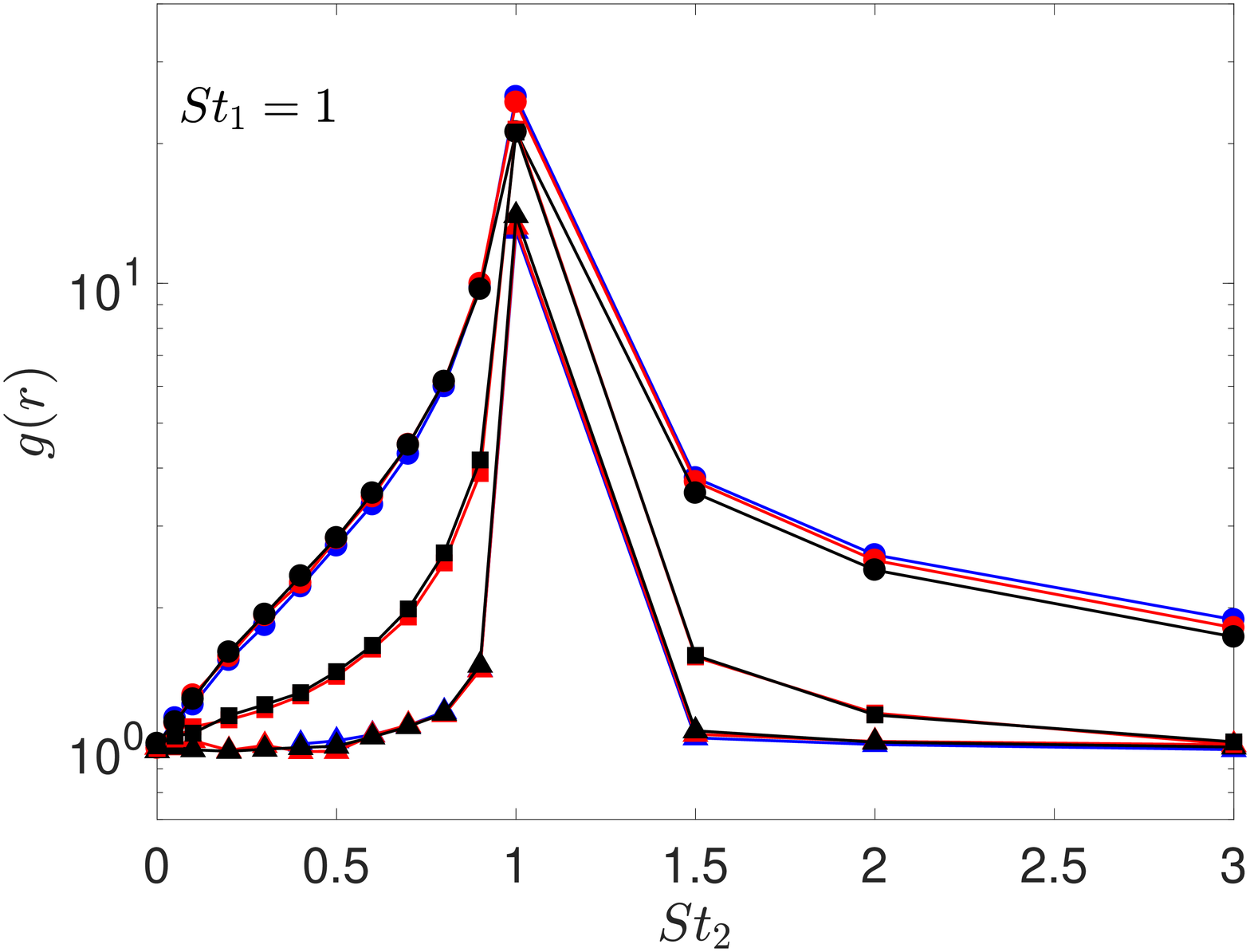}
    \caption{}
  \end{subfigure}%
    \begin{subfigure}[b]{0.5\linewidth}
    \includegraphics[width=\linewidth]{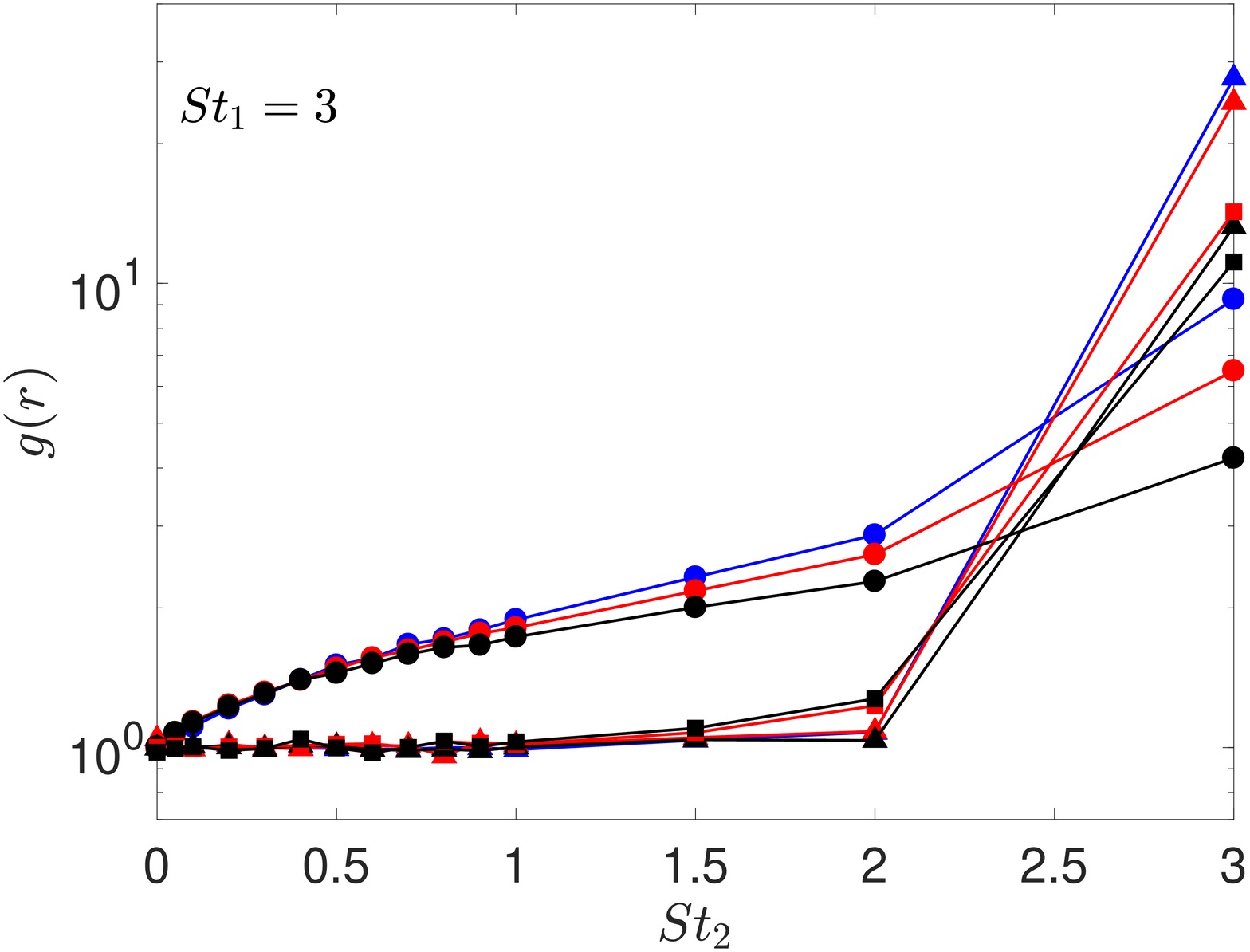}
    \caption{}
  \end{subfigure}
  \caption{Radial distribution function (RDF) at $r/\eta=0.125$, as a function of $St_2$, and for different $St_1$, $Fr$ and $R_\lambda$ combinations. Black, red and blue lines correspond to $R_\lambda=90$, $R_\lambda=224$ and $R_\lambda=398$, respectively and circle, square and triangle symbols denote $Fr=\infty$, $Fr=0.3$ and $Fr=0.052$, respectively.}\label{fig:RDF}
\end{figure}

\begin{figure}
  \centering
  \begin{subfigure}[b]{0.5\linewidth}
    \includegraphics[width=\linewidth]{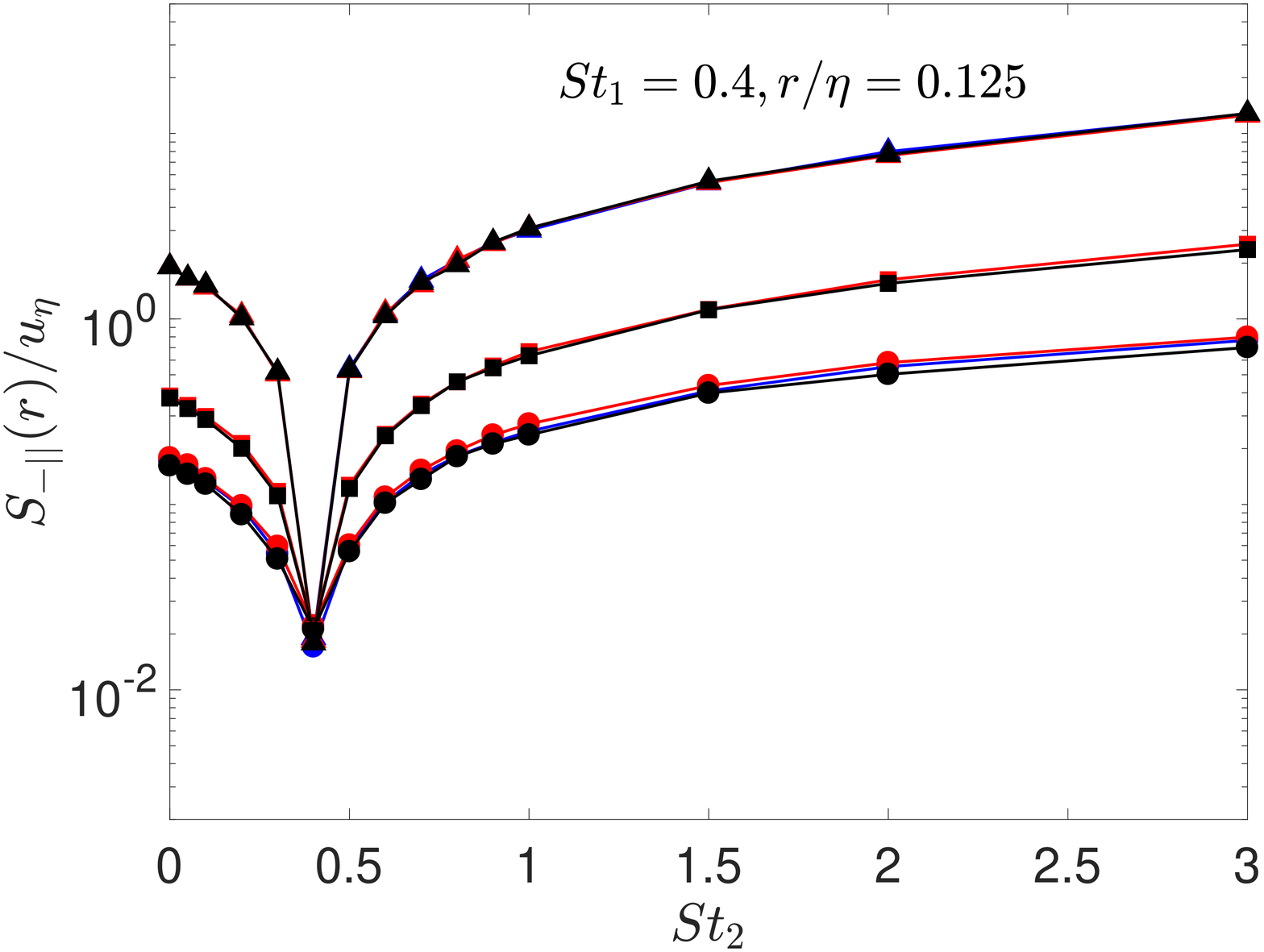}
    \caption{}
  \end{subfigure}%
    \begin{subfigure}[b]{0.5\linewidth}
    \includegraphics[width=\linewidth]{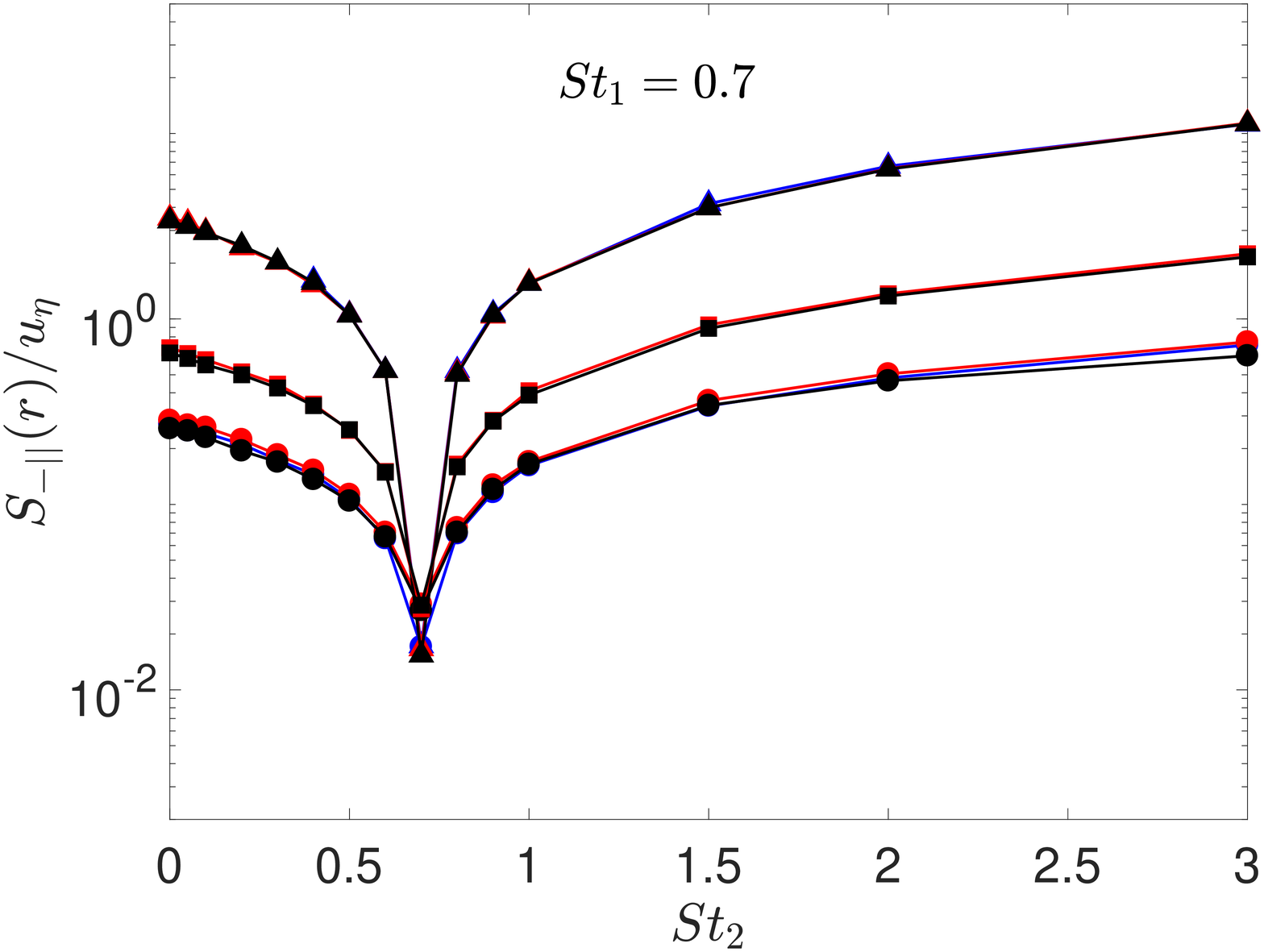}
    \caption{}
  \end{subfigure}
    \begin{subfigure}[b]{0.5\linewidth}
    \includegraphics[width=\linewidth]{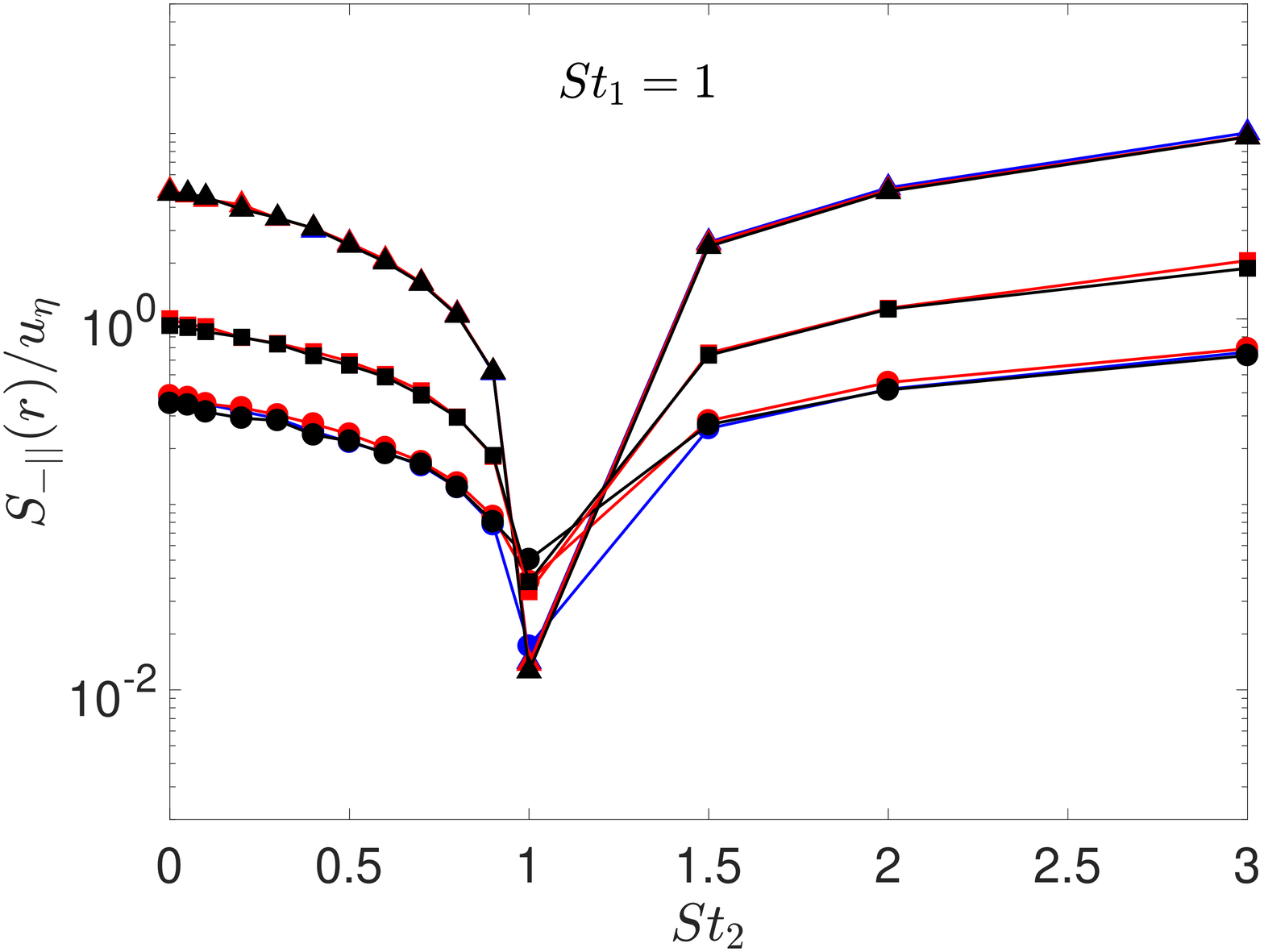}
    \caption{}
  \end{subfigure}%
    \begin{subfigure}[b]{0.5\linewidth}
    \includegraphics[width=\linewidth]{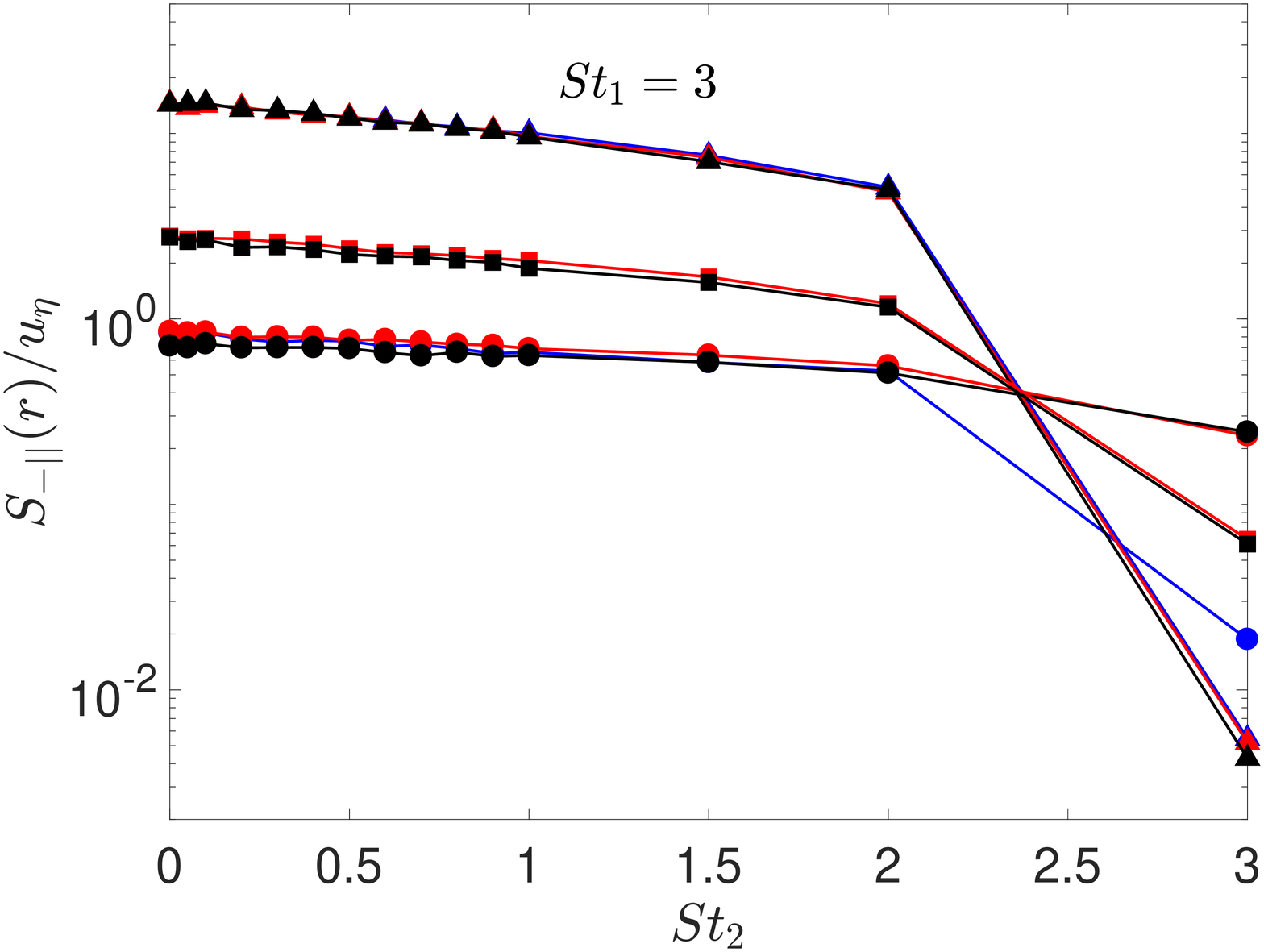}
    \caption{}
  \end{subfigure}
  \caption{Mean inward relative velocities at $r/\eta=0.125$, as a function of $St_2$, and for different $St_1$, $Fr$ and $R_\lambda$ combinations. Legend for this plot is the same as that in figure \ref{fig:RDF}. Black, red and blue lines correspond to $R_\lambda=90$, $R_\lambda=224$ and $R_\lambda=398$, respectively and circle, square and triangle symbols denote $Fr=\infty$, $Fr=0.3$ and $Fr=0.052$, respectively.}\label{fig:MeanInward_RelativeVelocity}
\end{figure}

Figure \ref{fig:RDF} shows the results for the RDF. As was shown in \cite{dhariwal2018small}, increasing bidispersity and decreasing $Fr$ both lead to the suppression of the spatial clustering of the particles, as characterized by the RDF. For $St < 1$, the RDF slightly decreases with increasing $R_\lambda$, while for $St>1$ it increases slightly with increasing $R_\lambda$. We also note that the results show that the RDF can be extremely sensitive to $|\Delta St|$, especially for $Fr\ll1$. The results in figure \ref{fig:RDF}(d) are particularly striking, showing that for $|\Delta St|\geq O(1)$, the clustering is absent for $Fr=0.052$, but then as the monodisperse limit $|\Delta St|\to 0$ is approached, the level of clustering dramatically increases, and becomes stronger than the $Fr=\infty$ case. This illustrates nicely the profound difference in the effect of gravity on the clustering of monodisperse and bidisperse inertial particles in turbulence, where in the former case it can enhance the clustering in certain parameter regimes, whereas for the latter it can dramatically suppresses the clustering.
\FloatBarrier
Figure \ref{fig:MeanInward_RelativeVelocity} shows the results for the mean collision velocity $S_{-\parallel}^p$. The results show that both increasing bidispersity and decreasing $Fr$ lead to enhancement of $S_{-\parallel}^p$, which is simply due to the enhanced contribution from the differential settling velocity. However, as for the RDF, the dependency of $S_{-\parallel}^p$ on $R_\lambda$ is very weak across the entire range $Fr=\infty$ to $Fr=0.052$, especially for $St_1,St_2\lesssim 1$. For $|\Delta St|/Fr\gg 1$, this is to be expected since in this case $S_{-\parallel}^p$ is dominated by the differential settling velocity of the particles, which is independent of the turbulence and hence independent of $R_\lambda$. In the regime $|\Delta St|/Fr\leq O(1)$, the weak dependency of both the RDF and $S_{-\parallel}^p$ on $R_\lambda$ is likely due to the fact that if $St\leq O(1)$ and $r\leq O(\eta)$, the particle-pair dynamics is dominated by the dissipation range of the turbulence \citep{bragg14d}, and also because RDF and $S_{-\parallel}^p$ are low order moments, whereas the strong effects of intermittency are mainly associated with the high-order statistics of the phase-space motion.
\FloatBarrier

\begin{figure}
  \centering
  \begin{subfigure}[b]{0.5\linewidth}
    \includegraphics[width=\linewidth]{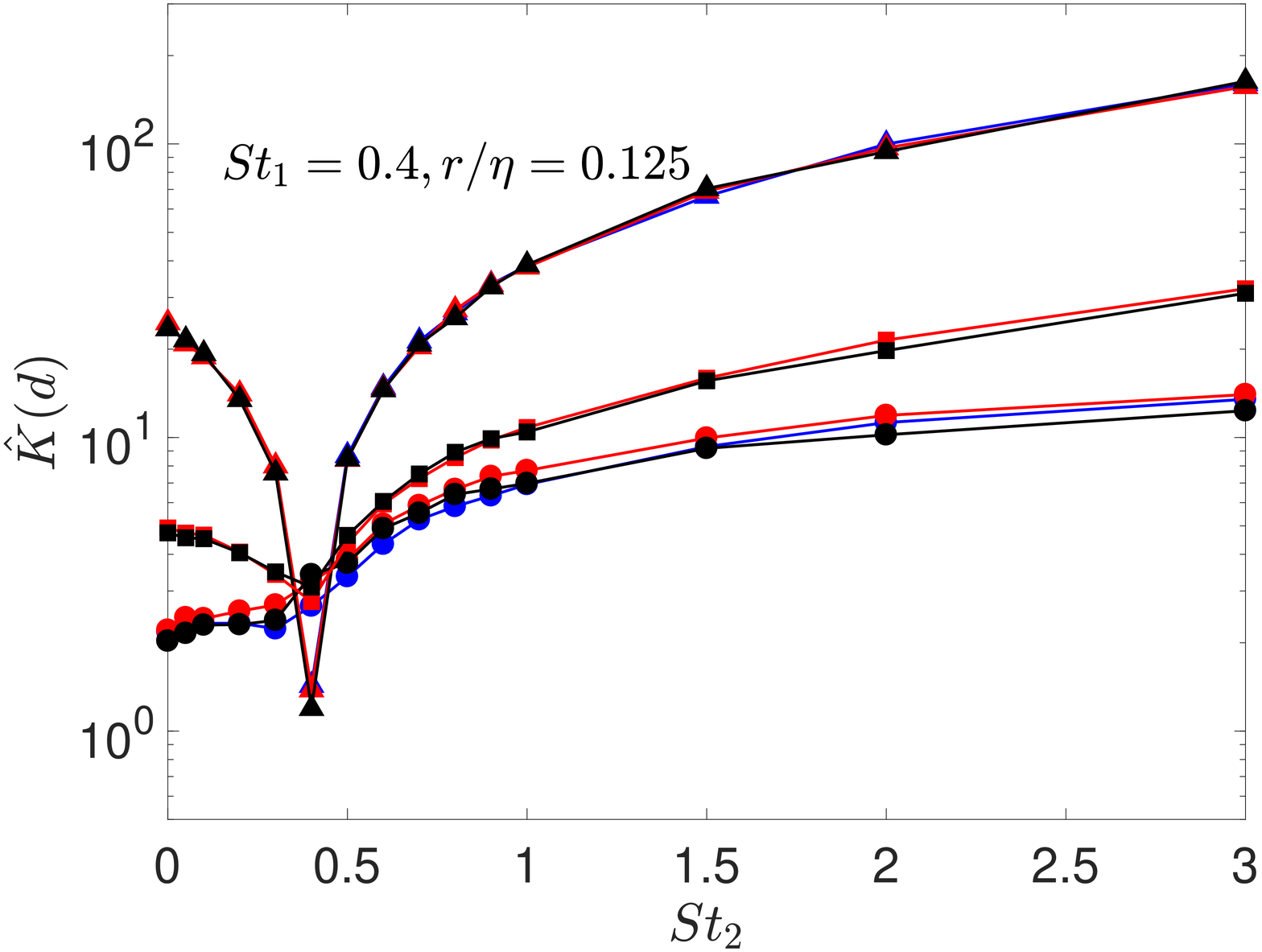}
    \caption{}
  \end{subfigure}%
    \begin{subfigure}[b]{0.5\linewidth}
    \includegraphics[width=\linewidth]{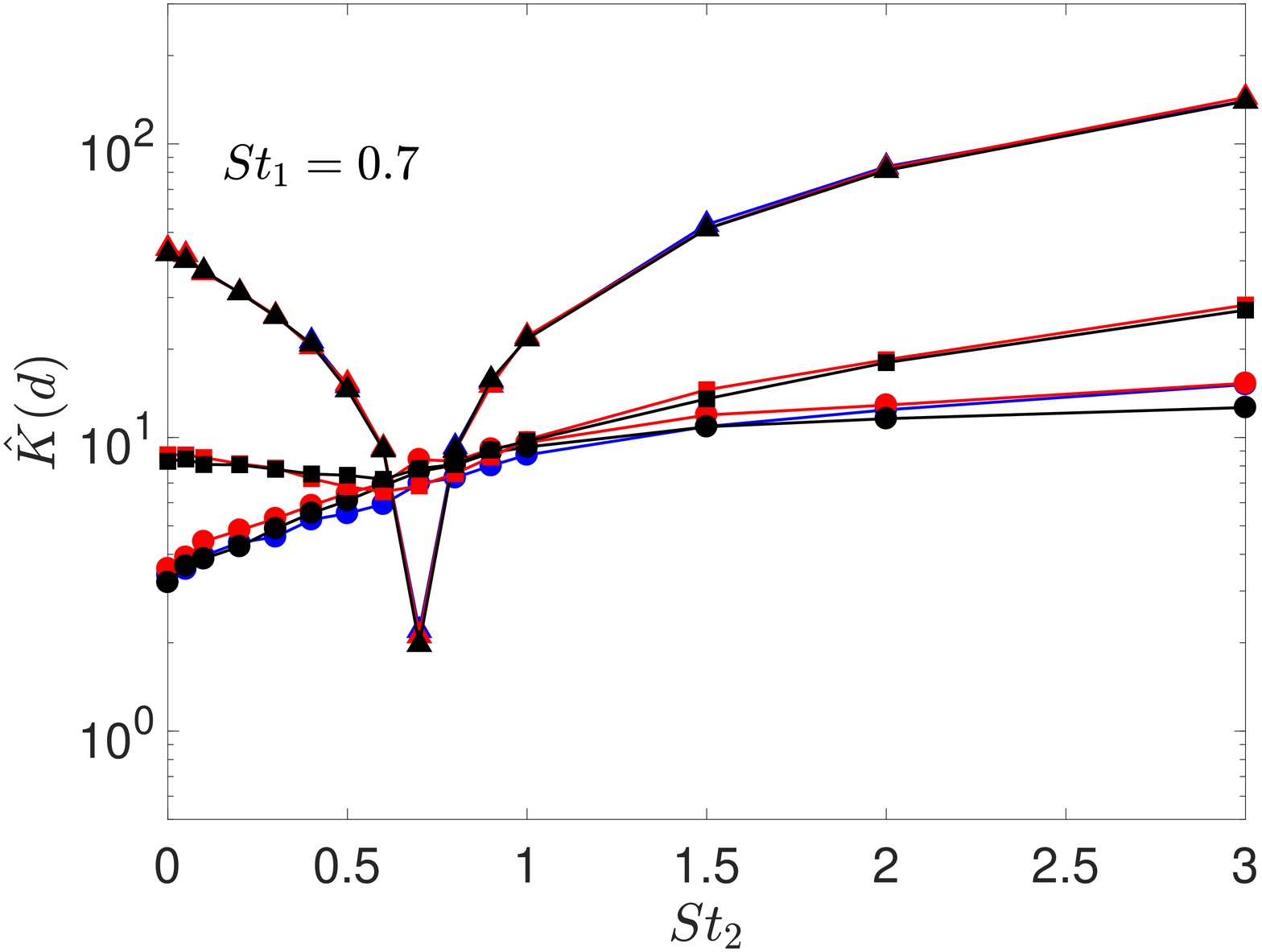}
    \caption{}
  \end{subfigure}
    \begin{subfigure}[b]{0.5\linewidth}
    \includegraphics[width=\linewidth]{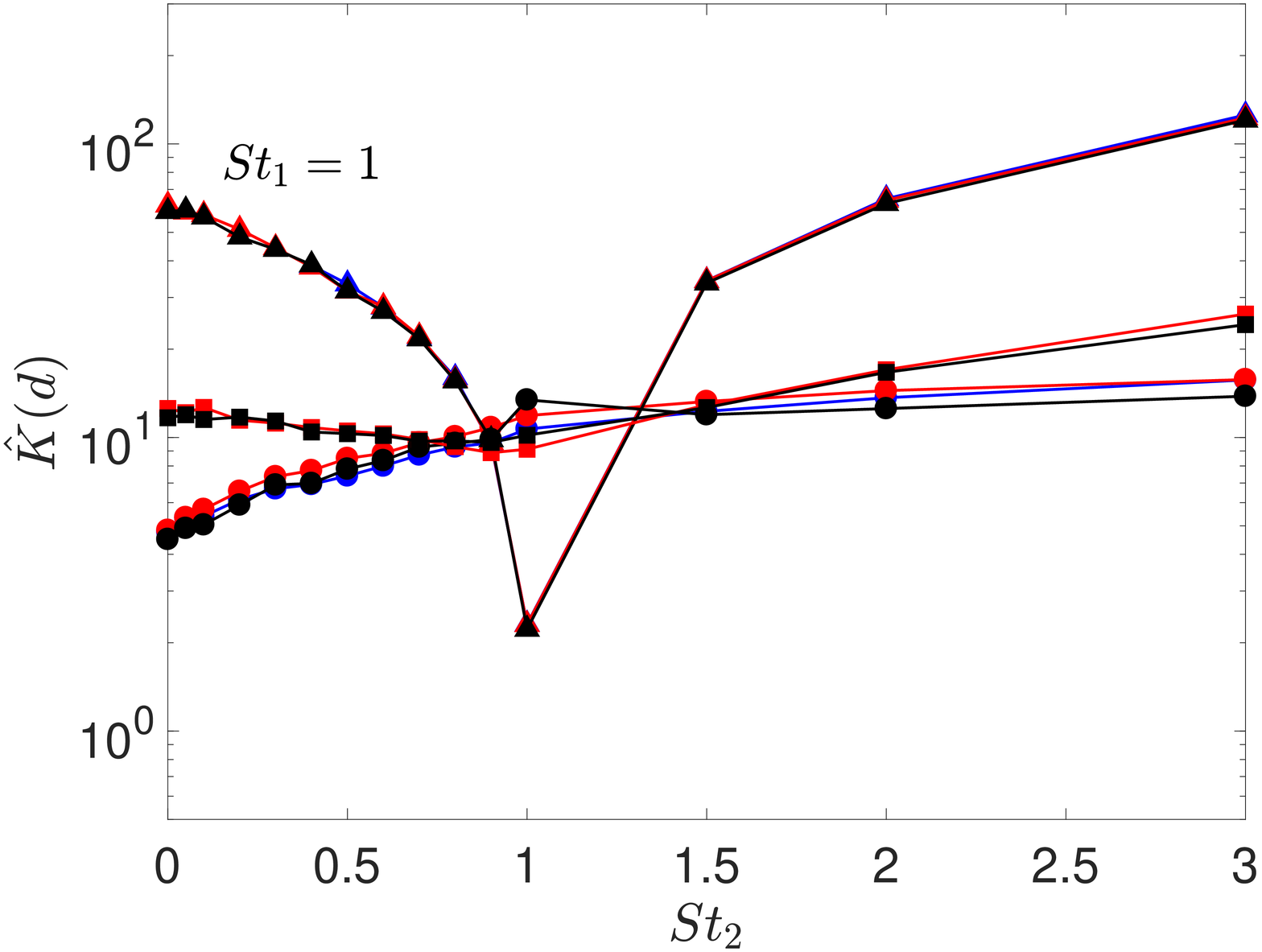}
    \caption{}
  \end{subfigure}%
    \begin{subfigure}[b]{0.5\linewidth}
    \includegraphics[width=\linewidth]{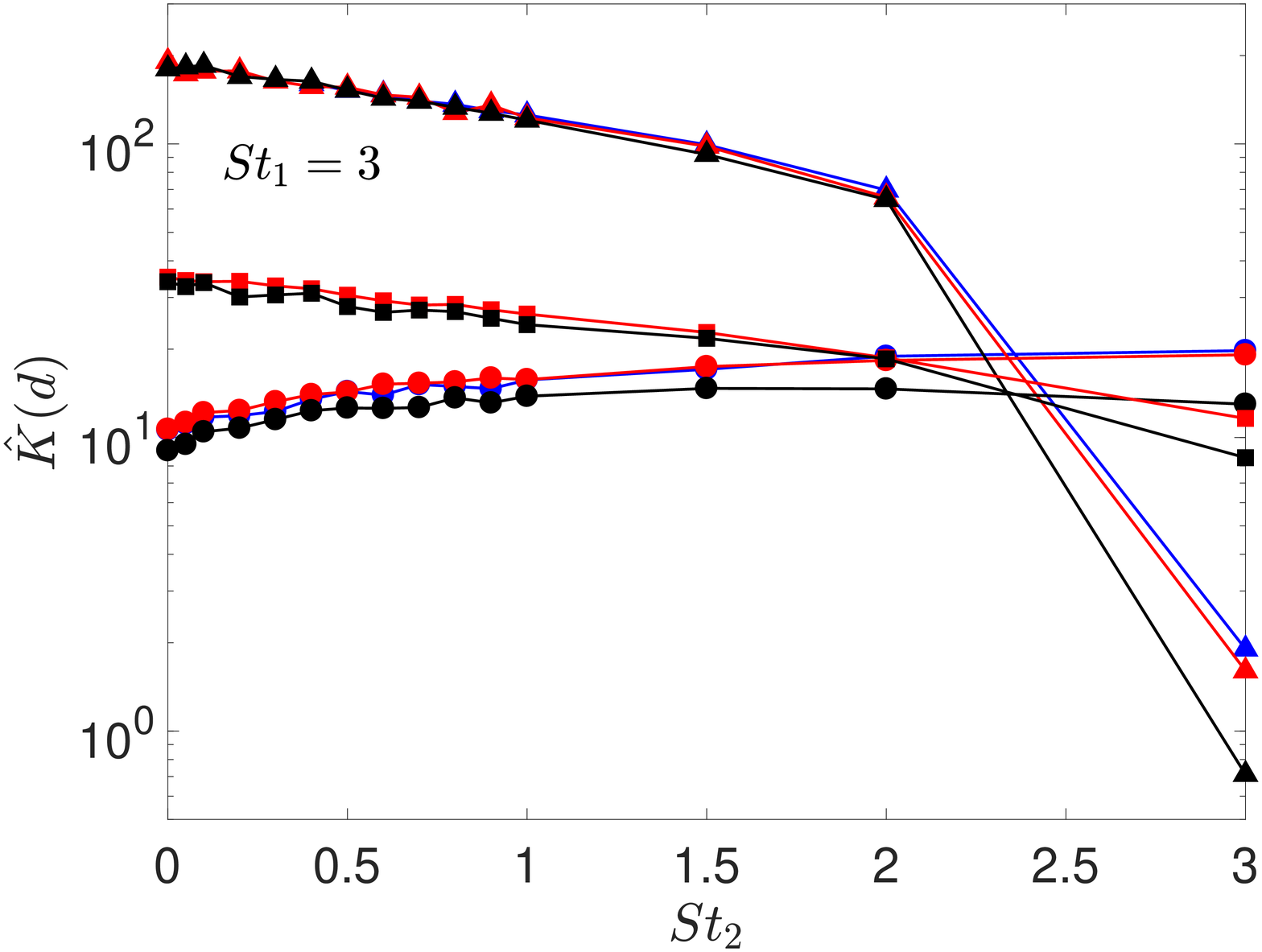}
    \caption{}
  \end{subfigure}
  \caption{Collision Kernel at $d/\eta=0.125$ as a function of $St_2$, and for different $St_1$, $Fr$ and $R_\lambda$ combinations. Legend for this plot is the same as that in figure \ref{fig:RDF}. Black, red and blue lines correspond to $R_\lambda=90$, $R_\lambda=224$ and $R_\lambda=398$, respectively and circle, square and triangle symbols denote $Fr=\infty$, $Fr=0.3$ and $Fr=0.052$, respectively.}\label{fig:Collision_Kernel}
\end{figure}

Finally, in figure \ref{fig:Collision_Kernel} we plot the normalized collision kernel $\hat{K}(d)\equiv K(d)/d^2 u_\eta$. As explained in \cite{dhariwal2018small}, we plot the results at the smallest $r/\eta$ for which our DNS data is reliable, since for bidisperse particles, the exact functional forms for  the RDF and $S_{-\parallel}^p$ are not known, and therefore we cannot justifiably extrapolate our DNS data down to the desired values of $d/\eta$. We observe that $\hat{K}$ increases both with increasing bidispersity and decreasing $Fr$, which follows because the enhancement of $S_{-\parallel}^p$ is stronger than the reduction of the RDF due to increasing bidispersity and decreasing $Fr$. As follows from the behavior of $S_{-\parallel}^p$ and the RDF, $\hat{K}$ shows a weak $R_\lambda$ dependence over the range considered here, especially for $St_1,St_2\lesssim 1$.
\FloatBarrier
\section{Conclusions}
In this paper, we have explored the influence of Froude number, $Fr$, and the Taylor Reynolds number, $R_\lambda$, on the dynamics settling, bidisperse particles in statistically stationary, isotropic turbulence using Direct Numerical Simulation (DNS). In particular, our DNS covered the ranges $0.052\le Fr \le \infty$ and $90\leq R_\lambda \leq 398$, along with particle Stokes numbers in the range $0\le St \le 3$.

We first analysed the statistics of particle accelerations since they play a key role in the relative motion of bidisperse inertial particle-pairs in turbulence. The acceleration variance and kurtosis results showed that decreasing $Fr$ enhances the inertial particle accelerations, whereas it suppresses their intermittency. Further, the Probability Density Function (PDF) of the accelerations showed that for $St>1$, the effect of $R_\lambda$ on the particle accelerations becomes more pronounced in the presence of gravity, since gravity causes the particle accelerations to be affected by a larger range of flow scales. We also observed that for $Fr\ll1$ and $St\gtrsim 1$, the acceleration PDFs become almost Gaussian, and may therefore be predicted using the asymptotic models for the acceleration variance of particles settling in turbulence given in \cite{ireland2016effectb}. 

We then studied the relative velocity of the bidisperse inertial particles. The PDF of the particle relative velocities showed that decreasing $Fr$ enhances the relative velocities of these particles in both the directions parallel to gravity (vertical) and perpendicular to gravity (horizontal), even when $St\ll1$. Most importantly, we also found that even when the particle settling velocity is large, turbulence still makes a substantial contribution to the vertical relative velocities, and increasingly so as $R_\lambda$ is increased. This behavior arises because of intermittency in the turbulence, due to which there are significant regions of the flow where the turbulent velocities are of the same order as the particle settling velocity, even though the mean-field fluctuations of the turbulence are small compared with the particle settling velocity. An important practical consequence of this is that when modeling the mixing of bidisperse particles in turbulence with $R_\lambda\ggg 1$, the effect of turbulence cannot be ignored  even when the particle settling parameter is $Sv\gg 1$ (unless only the low-order statistical properties of the mixing are of interest). Our results also show that reducing $Fr$ systematically suppresses the intermittency of the relative velocities, and in some parameter regimes the PDFs become almost Gaussian at the small-scales of the turbulence.

Finally, we examined the Radial Distribution Function (RDF) and particle collision kernels. We found that these low-order statistics are strongly dependent upon $Fr$, $St$, and the degree of bidispersity (the difference in the Stokes numbers of the particles) but are insensitive to $R_\lambda$ when $St\lesssim 1$. The latter finding is the same as was observed for monodisperse particles in \cite{ireland2016effectb}, despite the fact that the mechanisms governing the spatial clustering and collisions of monodisperse particles are in general quite different from those for bidisperse particles. These results indicate that the collisions of droplets in atmospheric clouds might be well described even by DNS with relatively low $R_\lambda$ (and more generally, for gas-solid flows where $R_\lambda\ggg1$).

It would be interesting for future experimental efforts on heavy particle motion in turbulence to test the findings presented in this paper. Experimental data for $Fr$ lower than considered in this paper is also of interest to understand how the system behaves as $Fr\to 0$, which is difficult to do in DNS due to the effects of periodic boundary conditions on the settling particles, and the associated need for large computational domains. For larger particles, the effects of non-linear drag forces (ignored in this study) on the particles can be important, and the impact of this on the motion of settling bidisperse particles in turbulence should be explored. A difficulty, however, is that studies have shown that widely used empirical laws for non-linear drag are not quantitatively accurate for settling particles in turbulence \citep{good14}, and new models are required. Future studies should also consider the full polydisperse case for inertial particles in turbulence, which is important for problems where the particle number density is sufficiently high for the binary collision assumption to fail.
\section{Acknowledgments}
This work used the Extreme Science and Engineering Discovery Environment (XSEDE), which is supported by National Science Foundation grant number ACI-1548562 \citep{xsede}. Specifically, the Comet cluster was used under allocation CTS170009.

\bibliographystyle{jfm}
\bibliography{jfm-instructions}

\begin{thebibliography}{40}
\expandafter\ifx\csname natexlab\endcsname\relax\def\natexlab#1{#1}\fi
\def\au#1{#1} \def\ed#1{#1} \def\yr#1{#1}\def\at#1{#1}\def\jt#1{\textit{#1}}
  \def\bt#1{#1}\def\bvol#1{\textbf{#1}} \def\vol#1{#1} \def\pg#1{#1}
  \def\publ#1{#1}\def\arxiv#1{#1}\def\org#1{#1}\def\st#1{\textit{#1}}

\bibitem[Ayyalasomayajula {\em et~al.\/}(2008)Ayyalasomayajula, Warhaft \&
  Collins]{sathya08a}
{\sc \au{Ayyalasomayajula, S.}, \au{Warhaft, Z.} \& \au{Collins, L.~R.}}
  \yr{2008}  \at{Modeling inertial particle acceleration statistics in
  isotropic turbulence}.  \jt{Phys. Fluids}  \bvol{20},  \pg{094104}.

\bibitem[Balachandar \& Eaton(2010)]{balachandar10}
{\sc \au{Balachandar, S.} \& \au{Eaton, J.~K.}} \yr{2010}  \at{Turbulent
  dispersed multiphase flow}.  \jt{Annu. Rev. Fluid Mech.}  \bvol{42},
  \pg{111--133}.

\bibitem[Bec {\em et~al.\/}(2006)Bec, Biferale, Boffetta, Celani, Cencini,
  Lanotte, Musacchio \& Toschi]{bec06a}
{\sc \au{Bec, J.}, \au{Biferale, L.}, \au{Boffetta, G.}, \au{Celani, A.},
  \au{Cencini, M.}, \au{Lanotte, A.~S.}, \au{Musacchio, S.} \& \au{Toschi, F.}}
  \yr{2006}  \at{Acceleration statistics of heavy particles in turbulence}.
  \jt{J. Fluid Mech.}  \bvol{550},  \pg{349--358}.

\bibitem[Bec {\em et~al.\/}(2007)Bec, Biferale, Cencini, Lanotte, Musacchio \&
  Toschi]{bec07}
{\sc \au{Bec, J.}, \au{Biferale, L.}, \au{Cencini, M.}, \au{Lanotte, A.~S.},
  \au{Musacchio, S.} \& \au{Toschi, F.}} \yr{2007}  \at{Heavy particle
  concentration in turbulence at dissipative and inertial scales}.  \jt{Phys.
  Rev. Lett.}  \bvol{98},  \pg{084502}.

\bibitem[Bec {\em et~al.\/}(2014)Bec, Homann \& Ray]{bec14b}
{\sc \au{Bec, J\'er\'emie}, \au{Homann, Holger} \& \au{Ray, Samriddhi~Sankar}}
  \yr{2014}  \at{Gravity-driven enhancement of heavy particle clustering in
  turbulent flow}.  \jt{Phys. Rev. Lett.}  \bvol{112},  \pg{184501}.

\bibitem[Borghi \& Anselmet(2013)]{borghi2013turbulent}
{\sc \au{Borghi, Roland} \& \au{Anselmet, Fabien}} \yr{2013} {\em Turbulent
  multiphase flows with heat and mass transfer\/}.  \publ{John Wiley \& Sons}.

\bibitem[Bragg \& Collins(2014{\natexlab{{\em a\/}}})]{bragg14b}
{\sc \au{Bragg, A.D.} \& \au{Collins, L.R.}} \yr{2014{\natexlab{{\em a\/}}}}
  \at{New insights from comparing statistical theories for inertial particles
  in turbulence: {I}. spatial distribution of particles.}  \jt{New J. Phys.}
  \bvol{16},  \pg{055013}.

\bibitem[Bragg \& Collins(2014{\natexlab{{\em b\/}}})]{bragg14c}
{\sc \au{Bragg, A.D.} \& \au{Collins, L.R.}} \yr{2014{\natexlab{{\em b\/}}}}
  \at{New insights from comparing statistical theories for inertial particles
  in turbulence: {II}. relative velocities of particles.}  \jt{New J. Phys.}
  \bvol{16},  \pg{055014}.

\bibitem[Bragg {\em et~al.\/}(2015)Bragg, Ireland \& Collins]{bragg14d}
{\sc \au{Bragg, A.~D.}, \au{Ireland, P.~J.} \& \au{Collins, L.~R.}} \yr{2015}
  \at{On the relationship between the non-local clustering mechanism and
  preferential concentration}.  \jt{Journal of Fluid Mechanics}  \bvol{780},
  \pg{327--343}.

\bibitem[Brennen \& Brennen(2005)]{brennen2005fundamentals}
{\sc \au{Brennen, Christopher~Earls} \& \au{Brennen, Christopher~E}} \yr{2005}
  {\em Fundamentals of multiphase flow\/}.  \publ{Cambridge university press}.

\bibitem[Chun {\em et~al.\/}(2005)Chun, Koch, Rani, Ahluwalia \&
  Collins]{chun05}
{\sc \au{Chun, J.}, \au{Koch, D.~L.}, \au{Rani, S.}, \au{Ahluwalia, A.} \&
  \au{Collins, L.~R.}} \yr{2005}  \at{Clustering of aerosol particles in
  isotropic turbulence}.  \jt{J. Fluid Mech.}  \bvol{536},  \pg{219--251}.

\bibitem[Dhariwal \& Bragg(2018)]{dhariwal2018small}
{\sc \au{Dhariwal, Rohit} \& \au{Bragg, Andrew~D}} \yr{2018}  \at{Small-scale
  dynamics of settling, bidisperse particles in turbulence}.  \jt{Journal of
  Fluid Mechanics}  \bvol{839},  \pg{594--620}.

\bibitem[Frisch(1995)]{frisch}
{\sc \au{Frisch, Uriel}} \yr{1995} {\em Turbulence: {T}he Legacy of {A}. {N}.
  {K}olmogorov\/}.  \publ{Cambridge University Press}.

\bibitem[Good {\em et~al.\/}(2014)Good, Ireland, Bewley, Bodenschatz, Collins
  \& Warhaft]{good14}
{\sc \au{Good, G.~H.}, \au{Ireland, P.~J.}, \au{Bewley, G.~P.},
  \au{Bodenschatz, E.}, \au{Collins, L.~R.} \& \au{Warhaft, Z.}} \yr{2014}
  \at{Settling regimes of inertial particles in isotropic turbulence}.
  \jt{Journal of Fluid Mechanics}  \bvol{759}.

\bibitem[Gustavsson \& Mehlig(2016)]{gustavsson16}
{\sc \au{Gustavsson, K.} \& \au{Mehlig, B.}} \yr{2016}  \at{Statistical models
  for spatial patterns of heavy particles in turbulence}.  \jt{Advances in
  Physics}  \bvol{65}~(1),  \pg{1--57}.

\bibitem[Gustavsson {\em et~al.\/}(2012)Gustavsson, Meneguz, Reeks \&
  Mehlig]{gustavsson12}
{\sc \au{Gustavsson, K.}, \au{Meneguz, E.}, \au{Reeks, M.} \& \au{Mehlig, B.}}
  \yr{2012}  \at{Inertial-particle dynamics in turbulent flows: caustics,
  concentration fluctuations and random uncorrelated motion}.  \jt{NJP}
  \bvol{14},  \pg{115017}.

\bibitem[Gustavsson {\em et~al.\/}(2014)Gustavsson, Vajedi \&
  Mehlig]{gustavsson14}
{\sc \au{Gustavsson, K.}, \au{Vajedi, S.} \& \au{Mehlig, B.}} \yr{2014}
  \at{Clustering of particles falling in a turbulent flow}.  \jt{Phys. Rev.
  Lett.}  \bvol{112},  \pg{214501}.

\bibitem[Hanafizadeh {\em et~al.\/}(2014)Hanafizadeh, Momenifar, Geimassi \&
  Ghanbarzadeh]{hanafizadeh2014void}
{\sc \au{Hanafizadeh, Pedram}, \au{Momenifar, Mohammadreza}, \au{Geimassi,
  A~Nouri} \& \au{Ghanbarzadeh, S}} \yr{2014}  \at{Void fraction and wake
  analysis of a gas-liquid two-phase cross-flow}.  \jt{Multiphase Science and
  Technology}  \bvol{26}~(4).

\bibitem[Ireland {\em et~al.\/}(2016{\natexlab{{\em a\/}}})Ireland, Bragg \&
  Collins]{ireland2016effecta}
{\sc \au{Ireland, Peter~J}, \au{Bragg, Andrew~D} \& \au{Collins, Lance~R}}
  \yr{2016{\natexlab{{\em a\/}}}}  \at{The effect of reynolds number on
  inertial particle dynamics in isotropic turbulence. part 1. simulations
  without gravitational effects}.  \jt{Journal of Fluid Mechanics}  \bvol{796},
   \pg{617--658}.

\bibitem[Ireland {\em et~al.\/}(2016{\natexlab{{\em b\/}}})Ireland, Bragg \&
  Collins]{ireland2016effectb}
{\sc \au{Ireland, Peter~J}, \au{Bragg, Andrew~D} \& \au{Collins, Lance~R}}
  \yr{2016{\natexlab{{\em b\/}}}}  \at{The effect of reynolds number on
  inertial particle dynamics in isotropic turbulence. part 2. simulations with
  gravitational effects}.  \jt{Journal of Fluid Mechanics}  \bvol{796},
  \pg{659--711}.

\bibitem[Ireland {\em et~al.\/}(2013)Ireland, Vaithianathan, Sukheswalla, Ray
  \& Collins]{ireland2013highly}
{\sc \au{Ireland, Peter~J}, \au{Vaithianathan, T}, \au{Sukheswalla, Parvez~S},
  \au{Ray, Baidurja} \& \au{Collins, Lance~R}} \yr{2013}  \at{Highly parallel
  particle-laden flow solver for turbulence research}.  \jt{Computers \&
  Fluids}  \bvol{76},  \pg{170--177}.

\bibitem[Kolev \& Kolev(2005)]{kolev2005multiphase}
{\sc \au{Kolev, Nikolay~Ivanov} \& \au{Kolev, Nikolay~I}} \yr{2005} {\em
  Multiphase flow dynamics: Fundamentals\/}.  \publ{Springer}.

\bibitem[{La Porta} {\em et~al.\/}(2001){La Porta}, Voth, Crawford, Alexander
  \& Bodenschatz]{laporta01}
{\sc \au{{La Porta}, A.}, \au{Voth, G.~A.}, \au{Crawford, A.~M.},
  \au{Alexander, J.} \& \au{Bodenschatz, E.}} \yr{2001}  \at{Fluid particle
  accelerations in fully developed turbulence}.  \jt{Nature}  \bvol{409},
  \pg{1017--1019}.

\bibitem[Maxey(1987)]{maxey87}
{\sc \au{Maxey, M.~R.}} \yr{1987}  \at{The gravitational settling of aerosol
  particles in homogeneous turbulence and random flow fields}.  \jt{J. Fluid
  Mech.}  \bvol{174},  \pg{441--465}.

\bibitem[Maxey \& Riley(1983)]{maxey1983equation}
{\sc \au{Maxey, Martin~R} \& \au{Riley, James~J}} \yr{1983}  \at{Equation of
  motion for a small rigid sphere in a nonuniform flow}.  \jt{The Physics of
  Fluids}  \bvol{26}~(4),  \pg{883--889}.

\bibitem[Momenifar {\em et~al.\/}(2015)Momenifar, Akhavan-Behabadi, Nasr \&
  Hanafizadeh]{momenifar2015effect}
{\sc \au{Momenifar, MR}, \au{Akhavan-Behabadi, MA}, \au{Nasr, M} \&
  \au{Hanafizadeh, P}} \yr{2015}  \at{Effect of lubricating oil on flow boiling
  characteristics of r-600a/oil inside a horizontal smooth tube}.  \jt{Applied
  Thermal Engineering}  \bvol{91},  \pg{62--72}.

\bibitem[Pan \& Padoan(2010)]{pan10}
{\sc \au{Pan, L.} \& \au{Padoan, P.}} \yr{2010}  \at{Relative velocity of
  inertial particles in turbulent flows}.  \jt{J. Fluid Mech.}  \bvol{661},
  \pg{73--107}.

\bibitem[Pan {\em et~al.\/}(2014)Pan, Padoan \& Scalo]{pan14}
{\sc \au{Pan, Liubin}, \au{Padoan, Paolo} \& \au{Scalo, John}} \yr{2014}
  \at{Turbulence-induced relative velocity of dust particles. ii. the
  bidisperse case}.  \jt{The Astrophysical Journal}  \bvol{791}~(1),  \pg{48}.

\bibitem[Parishani {\em et~al.\/}(2015)Parishani, Ayala, Rosa, Wang \&
  Grabowski]{parishani2015effects}
{\sc \au{Parishani, H}, \au{Ayala, O}, \au{Rosa, B}, \au{Wang, L-P} \&
  \au{Grabowski, WW}} \yr{2015}  \at{Effects of gravity on the acceleration and
  pair statistics of inertial particles in homogeneous isotropic turbulence}.
  \jt{Physics of Fluids}  \bvol{27}~(3),  \pg{033304}.

\bibitem[Pinsky {\em et~al.\/}(2007)Pinsky, Khain \&
  Shapiro]{pinsky2007collisions}
{\sc \au{Pinsky, MB}, \au{Khain, AP} \& \au{Shapiro, M}} \yr{2007}
  \at{Collisions of cloud droplets in a turbulent flow. part iv: Droplet
  hydrodynamic interaction}.  \jt{Journal of the atmospheric sciences}
  \bvol{64}~(7),  \pg{2462--2482}.

\bibitem[Pope(2000)]{pope}
{\sc \au{Pope, S.~B.}} \yr{2000} {\em Turbulent Flows\/}.  \publ{New York:
  Cambridge University Press}.

\bibitem[Prosperetti \& Tryggvason(2009)]{prosperetti2009computational}
{\sc \au{Prosperetti, Andrea} \& \au{Tryggvason, Gr{\'e}tar}} \yr{2009} {\em
  Computational methods for multiphase flow\/}.  \publ{Cambridge university
  press}.

\bibitem[Salazar \& Collins(2009)]{salazar09}
{\sc \au{Salazar, J. P. L.~C.} \& \au{Collins, L.~R.}} \yr{2009}
  \at{Two-particle dispersion in isotropic turbulent flows}.  \jt{Annu. Rev.
  Fluid Mech.}  \bvol{41},  \pg{405--432}.

\bibitem[Shaw(2003)]{shaw2003particle}
{\sc \au{Shaw, Raymond~A}} \yr{2003}  \at{Particle-turbulence interactions in
  atmospheric clouds}.  \jt{Annual Review of Fluid Mechanics}  \bvol{35}~(1),
  \pg{183--227}.

\bibitem[Sundaram \& Collins(1997)]{sundaram4}
{\sc \au{Sundaram, S.} \& \au{Collins, L.~R.}} \yr{1997}  \at{Collision
  statistics in an isotropic, particle-laden turbulent suspension {I}. {D}irect
  numerical simulations}.  \jt{J. Fluid Mech.}  \bvol{335},  \pg{75--109}.

\bibitem[Toschi \& Bodenschatz(2009)]{toschi2009lagrangian}
{\sc \au{Toschi, Federico} \& \au{Bodenschatz, Eberhard}} \yr{2009}
  \at{Lagrangian properties of particles in turbulence}.  \jt{Annual review of
  fluid mechanics}  \bvol{41},  \pg{375--404}.

\bibitem[Towns {\em et~al.\/}(2014)Towns, Cockerill, Dahan, Foster, Gaither,
  Grimshaw, Hazlewood, Lathrop, Lifka, Peterson, Roskies, Scott \&
  Wilkins-Diehr]{xsede}
{\sc \au{Towns, J.}, \au{Cockerill, T.}, \au{Dahan, M.}, \au{Foster, I.},
  \au{Gaither, K.}, \au{Grimshaw, A.}, \au{Hazlewood, V.}, \au{Lathrop, S.},
  \au{Lifka, D.}, \au{Peterson, G.~D.}, \au{Roskies, R.}, \au{Scott, J.~R.} \&
  \au{Wilkins-Diehr, N.}} \yr{2014}  \at{Xsede: Accelerating scientific
  discovery}.  \jt{Computing in Science \& Engineering}  \bvol{16}~(5),
  \pg{62--74}.

\bibitem[Wang \& Maxey(1993)]{wang93}
{\sc \au{Wang, L.~P.} \& \au{Maxey, M.~R.}} \yr{1993}  \at{Settling velocity
  and concentration distribution of heavy particles in homogeneous isotropic
  turbulence}.  \jt{J. Fluid Mech.}  \bvol{256},  \pg{27--68}.

\bibitem[Wilkinson \& Mehlig(2005)]{wilkinson05}
{\sc \au{Wilkinson, M.} \& \au{Mehlig, B.}} \yr{2005}  \at{Caustics in
  turbulent aerosols}.  \jt{Europhys. Lett.}  \bvol{71},  \pg{186--192}.

\bibitem[Woittiez {\em et~al.\/}(2009)Woittiez, Jonker \&
  Portela]{woittiez2009combined}
{\sc \au{Woittiez, Eric~JP}, \au{Jonker, Harm~JJ} \& \au{Portela, Lu{\'\i}s~M}}
  \yr{2009}  \at{On the combined effects of turbulence and gravity on droplet
  collisions in clouds: a numerical study}.  \jt{Journal of the atmospheric
  sciences}  \bvol{66}~(7),  \pg{1926--1943}.

\end{thebibliography}

\end{document}